\documentclass[3p,preprint,12pt]{elsarticle}
\usepackage{tabulary,xcolor}
\usepackage{amsfonts,amsmath,amssymb}
\usepackage{graphicx}
\usepackage{epstopdf}
\usepackage{subfig}
\usepackage{bm}
\usepackage{slashed}
\usepackage{caption}
\usepackage{float}
\usepackage{braket}

\newcommand{\bfb}{{\bf b}_{\perp}}

\newcommand{\bfp}{{\bf p}_{\perp}}
\newcommand{\bfn}{{\bf n}}
\newcommand{\bfS}{{\bf S}_{\perp}}

\newcommand{\bea}{\begin{eqnarray}}
\newcommand{\eea}{\end{eqnarray}}
\def\be{\begin{equation}}
\def\ee{\end{equation}}
\def\ba{\begin{array}}
\def\ea{\end{array}}
\def\bc{\begin{center}}
\def\ec{\end{center}}
\makeatletter
\let\save@ps@pprintTitle\ps@pprintTitle
\def\ps@pprintTitle{\save@ps@pprintTitle\gdef\@oddfoot{\footnotesize\itshape \null\hfill\today}}
\def\hlinewd#1{%
  \noalign{\ifnum0=`}\fi\hrule \@height #1%
  \futurelet\reserved@a\@xhline}

\AtBeginDocument{\ifNAT@numbers \biboptions{sort&compress}\fi}
\makeatother

\usepackage{ifluatex}
\ifluatex
\usepackage{fontspec}
\defaultfontfeatures{Ligatures=TeX}
\usepackage[]{unicode-math}
\unimathsetup{math-style=TeX}
\else
\usepackage[utf8]{inputenc}
\fi
\ifluatex\else\usepackage{stmaryrd}\fi

\usepackage{url,multirow,morefloats,floatflt,cancel,textcomp,tfrupee}
\usepackage{pifont}
\usepackage[nointegrals]{wasysym}
\makeatletter

\AtBeginDocument{
\expandafter\ifx\csname eqalign\endcsname\relax
\def\eqalign#1{\null\vcenter{\def\\{\cr}\openup\jot\m@th
  \ialign{\strut$\displaystyle{##}$\hfil&$\displaystyle{{}##}$\hfil
      \crcr#1\crcr}}\,}
\fi
}

\let\lt=<
\let\gt=>
\def\processVert{\ifmmode|\else\textbar\fi}

\@ifundefined{subparagraph}{
\def\subparagraph{\@startsection{paragraph}{5}{2\parindent}{0ex plus 0.1ex minus 0.1ex}%
{0ex}{\normalfont\small\itshape}}%
}{}

\newcommand\role[1]{\unskip}
\newcommand\aucollab[1]{\unskip}

\@ifundefined{tsGraphicsScaleX}{\gdef\tsGraphicsScaleX{1}}{}
\@ifundefined{tsGraphicsScaleY}{\gdef\tsGraphicsScaleY{.9}}{}
\def\checkGraphicsWidth{\ifdim\Gin@nat@width>\textwidth
	\tsGraphicsScaleX\textwidth\else\Gin@nat@width\fi}

\def\checkGraphicsHeight{\ifdim\Gin@nat@height>.9\textheight
	\tsGraphicsScaleY\textheight\else\Gin@nat@height\fi}

\def\fixFloatSize#1{\@ifundefined{processdelayedfloats}{\setbox0=\hbox{\includegraphics{#1}}\ifnum\wd0<\columnwidth\relax\renewenvironment{figure*}{\begin{figure}}{\end{figure}}\fi}{}}
\let\ts@includegraphics\includegraphics

\def\inlinegraphic[#1]#2{{\edef\@tempa{#1}\edef\baseline@shift{\ifx\@tempa\@empty0\else#1\fi}\edef\tempZ{\the\numexpr(\numexpr(\baseline@shift*\f@size/100))}\protect\raisebox{\tempZ pt}{\ts@includegraphics{#2}}}}

\AtBeginDocument{\def\includegraphics{\@ifnextchar[{\ts@includegraphics}{\ts@includegraphics[width=\checkGraphicsWidth,height=\checkGraphicsHeight,keepaspectratio]}}}

\def\URL#1#2{\@ifundefined{href}{#2}{\href{#1}{#2}}}

\def\UrlOrds{\do\*\do\-\do\~\do\'\do\"\do\-}%
\g@addto@macro{\UrlBreaks}{\UrlOrds}
\makeatother

\journal{Nucl. Phys. B.}
\begin{document}

\begin{frontmatter}
	
\title{Wigner distributions for an electron}
\author{Narinder Kumar}
\address{Department of Physics, Indian Institute of Technology Kanpur, Kanpur-208016, India}
\ead{narinder@iitk.ac.in}
\author{Chandan Mondal}
\address{Institute for Modern Physics, Chinese Academy of Sciences, Lanzhou-730000, China}
\ead{mondal@impcas.ac.in}
\begin{abstract}
We study the Wigner distributions for a physical electron, which reveal the multidimensional images of the electron. The physical electron is considered as a composite system of a bare electron and photon.
The Wigner distributions for unpolarized, longitudinally polarized and transversely polarized electron are presented in transverse momentum plane as well as in impact-parameter plane. The spin-spin correlations
between the bare electron and the physical electron are discussed.
We also evaluate all the leading twist generalized transverse momentum distributions (GTMDs) for electron.
\end{abstract}

\end{frontmatter}
\section{Introduction}\label{intro}
What is the shape of an electron? Most probably our answer is spherical but electron is a point particle having no underlying substructure. In quantum electrodynamics (QED) electron is considered as an elementary field. When we talk about the electron ``structure", what we are actually doing is probing the fluctuations in the quantum theory. According to quantum theory, one can consider the fluctuations of electron into electron-photon pair i.e., $e \rightarrow e\gamma \rightarrow e$ with same quantum number. The virtual photon further broke up into virtual electrons and positrons with all possible combinations. Thus, as a result bare electron is no more an isolated particle but it is surrounded by virtual cloud of electrons, positrons and photons. Therefore, bare electron becomes a dressed electron and one can consider electrons, positrons and photons as composite particles in the original electron.

The structure of the electron can be revealed when virtual cloud interacts with a probe,  the parton content of the electron is resolved. Recently Wigner distributions have been used as powerful tools to understand the multi-dimensional images of the hadrons. Ji \cite{Ji:2003ak} first introduced the Wigner distributions in quantum chromodynamics (QCD), commonly known as phase-space distributions. These distributions reduce to generalized parton distributions (GPDs) \cite{Diehl:2003ny}  and transverse momentum distributions (TMDs) after certain phase-space reductions. GPDs and TMDs are measurable in high energy experiments (For review on these distributions and experiments to measure them see \cite{Ji:1998pc,Goeke:2001tz,Mulders:1995dh,Boer:1997nt}). GPDs provide us the method for spatial imaging of nucleon \cite{Burkardt:2000za,Burkardt:2002hr,Burkardt:2005td,Diehl:2005jf, Pasquini:2007xz} by using the definition of impact-parameter dependent parton distribution functions (ipdpdfs) which resolve the correlation between the parton distribution in impact-parameter plane and longitudinal momentum of different parton and target polarizations. On the other side, TMDs contains three dimensional information regarding the spin-spin and spin-orbit correlations in momentum plane \cite{Goeke:2005hb,Meissner:2007rx,Lorce:2011zta}.

Therefore, complete understanding of internal structure can be gained by combining the position and momentum distribution called as Wigner distribution. Wigner distributions contain the one-body information of partons.
Wigner distributions have been used in many areas of physics for example in quantum molecular dynamics, quantum information, heavy ion collision and signal analysis \cite{Balazs:1983hk,Hillery:1983ms,LEE1995147} and measurable in some experiments \cite{Banaszek:1999ya,Smithey:1993zz,Vogel:1989zz}. Wigner distributions have been studied in different models. e.g.,  light-cone chiral quark soliton model \cite{Lorce:2011kd}, light-front dressed quark model \cite{Mukherjee:2014nya}, light-cone spectator model\cite{Liu:2015eqa}, AdS/QCD inspired quark-diquark model \cite{Chakrabarti:2016yuw,Chakrabarti:2017teq} etc. 
Wigner distributions are also related with generalized parton correlation functions (GPCFs) \cite{Meissner:2009ww} and integrating over the light-cone energy of the quark provide us the generlaized transverse momentum distributions (GTMDs) \cite{Lorce:2011dv,Echevarria:2016mrc,Kanazawa:2015zpa}which have been widely studied in light-front quark model and AdS/ QCD inspired quark model \cite{Lorce:2011kd,Chakrabarti:2017teq,Chakrabarti:2016yuw} models. GTMDs are also called as mother distributions as they give rise to GPDs and TMDs. Therefore, Wigner distributions are important tools to understand the internal structure of nucleons. 
A nice attempt has been made to understand the structure of electron in QED in Ref.\cite{Miller:2014vla} where several properties such as form factors, generalized parton distributions, transverse densities, Wigner distributions and the angular momentum content have been evaluated for the electron-photon component of the electron wave function. The transverse shape of electron has also been discussed in Ref. \cite{Hoyer:2009sg}.\\
 From last many years, it was believed that the quantum mechanical version of the Wigner distributions cannot be measured in experiments. But recent series of papers put this conception to an end. Recently, it was shown that GTMDs of the gluons can be accessed by diffractive di-jet production in deep-inelastic lepton-nucleus scattering \cite{Hatta:2016aoc,Hatta:2016dxp,Ji:2016jgn} and in virtual photon-nucleus quasi-elastic scattering \cite{Zhou:2016rnt}. A saturation model carrying numerical calculations on the gluons at small $x$ is also presented in Ref. \cite{Zhou:2016rnt,Hagiwara:2016kam}. But for the first time, a physical process which can give access to quark GTMDs by entering the exclusive pion-nucleus double Drell-Yan process is presented in Ref. \cite{Bhattacharya:2017bvs}. An upcoming experimental facility Electron Ion Collider \cite{Accardi:2012qut} in USA  and JLab 12 GeV \cite{Dudek:2012vr} upgrade is gearing up to measure these distributions . In addition to this, azimuthal Wigner distributions for twisted photons is also measured \cite{Mirhosseini}.

In the present work, we investigate the Wigner distribution of an electron, which provide the multidimensional images on the distribution of QED partons in the dressed electron. We consider the self-consistent model of spin-$\frac{1}{2}$ composite system based on the quantum fluctuations of electron in QED \cite{Brodsky:2000ii}. We represent the spin-$\frac{1}{2}$ composite system with a mass $M$ consist of spin-$\frac{1}{2}$ fermion (electron) and spin-1 vector boson (photon) with respective masses $m$ and $\lambda$.
This model is also act as a guideline to nucleon strcture because electron-photon component of the physical electron can be used as a component of the nucleon wave function consists of a quark and a vector diquark and therefore it probes the structure of nucleon and widely used for the calculations of gravitational form factors and spin and orbital angualr momentum of a composite relativistic system \cite{Brodsky:2000ii}, GPDs \cite{Brodsky:2006ku,Kumar:2015fta}, ipdpdfs \cite{Kumar:2015tpa}, charge and magnetization densities \cite{Kumar:2014coa}. Recently this approach has been used to understand the electron TMDs in momentum plane \cite{Bacchetta:2015qka}. Using the overlap of light-front wave functions (LFWFs) we evaluate the Wigner distributions for electron dressed with another electron having different polarizations (unpolarized, longitudinal polarized and transversely polarized). The QED-based LFWFs can be useful since they are frame-independent and satisfy properties such as $J^z$ conservation.
 The spin-spin correlations for longitudinally and transversely polarized spin-$\frac{1}{2}$ composite system and fermion constituent, transversely polarized  spin-$\frac{1}{2}$ composite system with longitudinally polarized fermion constituent and  longitudinally polarized spin-$\frac{1}{2}$ composite system with transversely polarized fermion constituent are also investigated. Further, all the leading twist GTMDs for a physical electron are evaluated from the Wigner distributions.
 Presently we do not have any experimental data over the distribution of QED partons in both transverse position and monetum space and therefore a direct comparison of our calcuclations is not possbile.  Experimental facilities like ACME collaboration  \cite{baron, hudson} is putting efforts to understand the shape of electron by means of atomic studies i.e., measuring the charge distributions and electric dipole moment (EDM) \cite{wesley}. They improved the upper limits for electric dipole moment of the electron and focused on the non-spherical shape of electron. New limits obtained there suggests the deviation of the shape of electron from spherical symmetry. Within the standard model,  a non-spherical shape of the electron arises naturally via the quantum fluctuation in QED.\\
The manuscript is organized as follows. In section \ref{model}, we discuss the QED model. In section \ref{wigner}, we discuss the Wigner distributions and present the analytic expressions for the different Wigner distribution functions. Results and discussion are given in section \ref{impact}. In section \ref{spin_spin}, we discuss the spin-spin correlations and in section \ref{gtmd}, we calculate all the leading twist GTMDs. Transverse shape of the electron is given in section \ref{shape-electron}. Conclusions are drawn in section \ref{con}.

\vskip0.2in
\noindent
\section {QED model}\label{model}
We evaluate the results for the Wigner distribution of the physical electron by considering it as a two particle state (electron and photon). The two particle  Fock state for an electron with $J^z=+\frac{1}{2}$ has four possible combinations \cite{Brodsky:2000ii}:
\begin{eqnarray*}
&&\Ket {\Psi^{\uparrow}(P^+, {\bf P_\perp}= {\bf 0_\perp})}
= \int \frac{dx\ d^2 {\bf p_\perp}}{\sqrt{x(1-x)}16 \pi^3}\Bigg[\psi^{\uparrow}_{+\frac{1}{2}+1}(x,{\bf p_\perp}) \Ket{ +\frac{1}{2}+1; x\ P^+, {\bf p_\perp}}\nonumber\\
\end{eqnarray*}
\begin{eqnarray}
&&+ \psi^{\uparrow}_{+\frac{1}{2}-1}(x,{\bf p_\perp})\Ket{ +\frac{1}{2}-1; x\ P^+, {\bf p_\perp}} + \psi^{\uparrow}_{-\frac{1}{2}+1}(x,{\bf p_\perp})\Ket{ -\frac{1}{2}+1; x\ P^+, {\bf p_\perp}}\nonumber\\
&&+\psi^{\uparrow}_{-\frac{1}{2}-1}(x,{\bf p_\perp})\Ket{- \frac{1}{2}-1; x\ P^+, {\bf p_\perp}}\Bigg].
\label{electron_up}
\end{eqnarray}
The wavefunctions can be represented as
\begin{eqnarray}
\psi^{\uparrow}_{+\frac{1}{2}+1}(x,{\bf p_\perp})&=& - \sqrt{2} \frac{(-p^1+i p^2)}{x(1-x)}\varphi,\nonumber\\
\psi^{\uparrow}_{+\frac{1}{2}-1}(x,{\bf p_\perp})&=& - \sqrt{2} \frac{(+p^1+i p^2)}{(1-x)}\varphi,\nonumber\\
\psi^{\uparrow}_{-\frac{1}{2}+1}(x,{\bf p_\perp})&=& - \sqrt{2} \Bigg(M-\frac{m}{x}\Bigg) \varphi,\nonumber\\
\psi^{\uparrow}_{-\frac{1}{2}-1}(x,{\bf p_\perp})&=& 0.
\end{eqnarray}
Similarly, two particle Fock's state for an electron with $J^z=-\frac{1}{2}$ also has four possible combinations:
\begin{eqnarray}
&&\Ket {\Psi^{\downarrow}(P^+, {\bf P_\perp}= {\bf 0_\perp})} = \int \frac{dx\ d^2 {\bf p_\perp}}{\sqrt{x(1-x)}16 \pi^3}\Bigg[\psi^{\downarrow}_{+\frac{1}{2}+1}(x,{\bf p_\perp})\Ket{ +\frac{1}{2}+1; x\ P^+, {\bf p_\perp}}\nonumber\\
&&+ \psi^{\downarrow}_{+\frac{1}{2}-1}(x,{\bf p_\perp})\Ket{ +\frac{1}{2}-1; x\ P^+, {\bf p_\perp}} + \psi^{\downarrow}_{-\frac{1}{2}+1}(x,{\bf p_\perp})\Ket{ -\frac{1}{2}+1; x\ P^+, {\bf p_\perp}}\nonumber\\
&&+\psi^{\downarrow}_{-\frac{1}{2}-1}(x,{\bf p_\perp})\Ket{- \frac{1}{2}-1; x\ P^+, {\bf p_\perp}}\Bigg],
\label{electron}
\end{eqnarray}
where
\begin{eqnarray}
\psi^{\downarrow}_{+\frac{1}{2}+1}(x,{\bf p_\perp})&=& 0,\nonumber\\
\psi^{\downarrow}_{+\frac{1}{2}-1}(x,{\bf p_\perp})&=& - \sqrt{2} \Bigg(M-\frac{m}{x}\Bigg) \varphi,\nonumber\\
\psi^{\downarrow}_{-\frac{1}{2}+1}(x,{\bf p_\perp})&=& - \sqrt{2} \frac{(-p^1+i p^2)}{(1-x)}\varphi, \nonumber\\
\psi^{\downarrow}_{-\frac{1}{2}-1}(x,{\bf p_\perp})&=& - \sqrt{2}\frac{(+p^1+i p^2)}{x(1-x)}\varphi,
\end{eqnarray}
with
\begin{equation}
\varphi(x,{\bf p_\perp})=\frac{e}{\sqrt{1-x}} \frac{1}{M^2-\frac{{\bf p_\perp^2}+m^2}{x}+ \frac{{\bf p_\perp^2}+\lambda^2}{1-x}}.
\label{electron_down}
\end{equation}
This one loop model is self consistent since it has correct correlation of different Fock's components of the state as given by light-front eigen value equation.
\section{Wigner Distributions}\label{wigner}
Wigner distribution in the light-front framework is defined as \cite{Lorce:2011kd,Meissner:2009ww,Meissner:2008ay,Lorce:2011ni}
\begin{eqnarray}
\rho^{[\Gamma]}({\bf b_\perp},{\bf p_\perp},x;S)
= \int \frac{d^2 \Delta_\perp}{(2 \pi)^2} e^{-i {\bf \Delta_\perp} \cdot {\bf b_\perp}}
 W^{[\Gamma]}(\Delta_\perp, {\bf p_\perp},x;S),
\end{eqnarray}
where
\begin{eqnarray}
W^{[\Gamma]}(\Delta_\perp, {\bf p_\perp},x;S) = \int \frac{dz^- d^2 z_\perp}{2(2 \pi)^3} e^{i p \cdot z}\Bra{P'';S} \bar{\psi}(-z/2) \Gamma \mathcal{W}_{[-\frac{z}{2},\frac{z}{2}]}\psi(z/2)\Ket{P';S},
\label{wigner-operator}
\end{eqnarray}
with $\Gamma$, for example $\gamma^+, \gamma^{+}\gamma^{5}, i \sigma^{j +}\gamma_5$ and $S$ is the spin of the composite system. We have defined the $P'=(P^+, P'^-, \frac{\Delta_\perp}{2})$ and $P''=(P^+, P''^-,- \frac{\Delta_\perp}{2})$ are the initial and final momentum of the composite system.
From Eq. \ref{wigner-operator}, we can write the quark-quark correlator by assuming different operators ($\gamma^+, \gamma^{+}\gamma^{5}, i \sigma^{1 +}\gamma_5, i \sigma^{2 +} \gamma_5$).
Wigner distribution is a five-dimensional$(b_x, b_y, p_x, p_y, x)$ quantity. To obtain the phase-space distribution, we will integrate it over $x$ so that the distribution is in impact-parameter and transverse momentum plane and therefore called as transverse Wigner distribution. In this work, we present the Wigner distribution in impact-parameter plane with fixed transverse momentum and in transverse momentum plane with fixed value of impact-parameter. One can also obtain the TMDs and ipdpdfs from the Wigner distributions. Wigner distributions can be presented in mixed plane i.e., by integrating over a transverse momentum and an impact-parameter along two transverse direction
\begin{eqnarray}
\tilde{\rho}(b_x, p_y, x)&=& \int db_y\ dp_x \ \rho({\bf b_\perp},{\bf p_\perp},x), \nonumber\\
\bar{\rho}(b_y, p_x, x)&=& \int db_x\ dp_y \ \rho({\bf b_\perp},{\bf p_\perp},x).
\end{eqnarray}
These mixing distributions are real distributions and the remaining variables are not violated by the uncertainty principle. They describe the correlation of impact-parameter and transverse momentum in perpendicular directions. Following, we have presented the results for the Wigner distribution by combining the various polarization configurations of unpolarized (U), longitudinal polarized (L) and transversely polarized (T) physical electron and internal electron. Among 16 independent twist-two Wigner distributions, we define the unpolarized Wigner distribution
\begin{eqnarray}
\rho_{UU}({\bf b_\perp},{\bf p_\perp},x)= \frac{1}{2} \Big[\rho^{[\gamma^+]}({\bf b_\perp},{\bf p_\perp},x;+\hat{e}_z)+\rho^{[\gamma^+]}({\bf b_\perp},{\bf p_\perp},x;-\hat{e}_z)\Big],
\end{eqnarray}
the unpolarized-longitudinally polarized Wigner distribution
\begin{eqnarray}
\rho_{UL}({\bf b_\perp},{\bf p_\perp},x)&=& \frac{1}{2} \Big[\rho^{[\gamma^+ \gamma_5]}({\bf b_\perp},{\bf p_\perp},x;+\hat{e}_z) +\rho^{[\gamma^+ \gamma_5]}({\bf b_\perp},{\bf p_\perp},x;-\hat{e}_z)\Big],
\label{ul}
\end{eqnarray}
the unpolarized-transversely polarized Wigner distribution
\begin{eqnarray}
\rho^{j}_{UT}({\bf b_\perp},{\bf p_\perp},x)&=& \frac{1}{2} \Big[\rho^{[i\sigma^{+j} \gamma_5]} ({\bf b_\perp},{\bf p_\perp},x;+\hat{e}_z) +\rho^{[i\sigma^{+j} \gamma_5]}({\bf b_\perp},{\bf p_\perp},x;-\hat{e}_z)\Big],
\label{ut}
\end{eqnarray}
the longitudinal-unpolarized Wigner distribution
\begin{eqnarray}
\rho_{LU}({\bf b_\perp},{\bf p_\perp},x)&=&\frac{1}{2}\Big[\rho^{[\gamma^+]} ({\bf b_\perp},{\bf p_\perp},x;+\hat{e}_z) -\rho^{[\gamma^+]} ({\bf b_\perp},{\bf p_\perp},x;-\hat{e}_z)\Big],
\label{lu}
\end{eqnarray}
the longitudinal polarized Wigner distribution
\begin{eqnarray}
\rho_{LL}({\bf b_\perp},{\bf p_\perp},x)&=&\frac{1}{2} \Big[\rho^{[\gamma^+ \gamma_5]} ({\bf b_\perp},{\bf p_\perp},x;+\hat{e}_z)-\rho^{[\gamma^+ \gamma_5]} ({\bf b_\perp},{\bf p_\perp},x;-\hat{e}_z)\Big],
\label{ll}
\end{eqnarray}
the longitudinal-transversely polarized Wigner distribution
\begin{eqnarray}
\rho^{j}_{LT}({\bf b_\perp},{\bf p_\perp},x)&=& \Big[\rho^{[i \sigma^{+j} \gamma_5]} ({\bf b_\perp},{\bf p_\perp},x;+\hat{e}_z) -\rho^{[i \sigma^{+j} \gamma_5]} ({\bf b_\perp},{\bf p_\perp},x;-\hat{e}_z)\Big],
\label{lt}
\end{eqnarray}
the transverse-unpolarized Wigner distribution
\begin{eqnarray}
\rho^i_{TU}({\bf b_\perp},{\bf p_\perp},x)&=& \frac{1}{2} \Big[\rho^{[\gamma^{+} ]} ({\bf b_\perp},{\bf p_\perp},x;+\hat{e}_i)  - \rho^{[\gamma^{+}]} ({\bf b_\perp},{\bf p_\perp},x;-\hat{e}_i)\Big],
\label{tu}
\end{eqnarray}
the transverse-longitudinal Wigner distribution
\begin{eqnarray}
\rho^i_{TL}({\bf b_\perp},{\bf p_\perp},x)&=& \frac{1}{2} \Big[\rho^{[\gamma^{+} \gamma_5 ]} ({\bf b_\perp},{\bf p_\perp},x;+\hat{e}_i)  - \rho^{[\gamma^{+} \gamma_5]} ({\bf b_\perp},{\bf p_\perp},x;-\hat{e}_i)\Big],
\label{tl}
\end{eqnarray}
the transverse Wigner distribution 
\begin{eqnarray}
\rho_{TT}({\bf b_\perp},{\bf p_\perp},x)&=& \frac{1}{2} \delta_{ij} \Big[\rho^{[i \sigma^{+j} \gamma_5]}({\bf b_\perp},{\bf p_\perp},x;+\hat{e}_i)  - \rho^{[i \sigma^{+j} \gamma_5]}({\bf b_\perp},{\bf p_\perp},x;-\hat{e}_i)\Big],
\label{tt}
\end{eqnarray}
and the pretzelous  Wigner distribution
\begin{eqnarray}
\rho^{\perp}_{TT}({\bf b_\perp},{\bf p_\perp},x)&=& \frac{1}{2} \epsilon_{ij} \Big[\rho^{[i \sigma^{+j} \gamma_5]}({\bf b_\perp},{\bf p_\perp},x;+\hat{e}_i)  - \rho^{[i \sigma^{+j} \gamma_5]}({\bf b_\perp},{\bf p_\perp},x;-\hat{e}_i)\Big].
\label{pret}
\end{eqnarray}
Here, first subscript in $\rho$ represent the polarization state of physical electron and second subscript represents the polarization state of internal electron.
Using Eqs. \ref{electron_up} and \ref{electron_down} in Eq. \ref{wigner-operator}, the correlator function $W^{[\Gamma]}(\Delta_\perp, {\bf p_\perp},x;S)$ can be expressed in terms of LFWFs as
\bea
 W^{[\gamma^+]}_{s s'}({\bf \Delta_\perp},{\bf p_\perp},x)=\frac{1}{16 \pi^3} \sum_{\lambda^{'}_{1}, \lambda_1, \lambda_2} \ \Psi^{* s'}_{\lambda^{'}_{1} \lambda_2} \ \chi^{\dagger}_{\lambda^{'}_{1}} \ \chi_{\lambda_1}\ \Psi^{s}_{\lambda_1 \lambda_2},
\label{unpol}
\eea
\bea
W^{[\gamma^+ \gamma_5]}_{s s'}({\bf \Delta_\perp},{\bf p_\perp},x)=\frac{1}{16 \pi^3} \sum_{\lambda^{'}_{1}, \lambda_1, \lambda_2} \ \Psi^{* s'}_{\lambda^{'}_{1} \lambda_2} \ \chi^{\dagger}_{\lambda^{'}_{1}} \ \sigma_3 \ \chi_{\lambda_1}\ \Psi^{s}_{\lambda_1 \lambda_2},
\label{long-pol}
\eea
\bea
W^{[i \sigma^{+j} \gamma_5]}_{s s'}({\bf \Delta_\perp},{\bf p_\perp},x)=\frac{1}{16 \pi^3} \sum_{\lambda^{'}_{1}, \lambda_1, \lambda_2} \ \Psi^{* s'}_{\lambda^{'}_{1} \lambda_2} \ \chi^{\dagger}_{\lambda^{'}_{1}}\ \sigma_j \ \chi_{\lambda_1} \Psi^{s}_{\lambda_1 \lambda_2},
\label{trans-pol}
\eea
where $\sigma_i$ are the Pauli spin matrices. GTMDs in unpolarized, longitudinally polarized and transversely polarized cases can be represent in the form of LFWFs by Eqs. \ref{unpol},\ \ref{long-pol} and \ref{trans-pol}.
Using the overlap of LFWFs, we can explicitly represent the expression for the different Wigner distributions as
\begin{eqnarray}
\rho_{UU}({\bf b_\perp},{\bf p_\perp})
&=&\frac{4 e^2}{2(2\pi)^2 16 \pi^3} \int d\Delta_x d\Delta_y \int dx \cos(\Delta_x b_x+ \Delta_y b_y)\nonumber\\
&\times&\Big[\frac{1+x^2}{x^2(1-x)^2}\Big({\bf p_\perp^2}-\frac{(1-x)^2}{4}{\bf \Delta_\perp^2}\Big)
+\Big(M-\frac{m}{x}\Big)^2 \Big]\varphi^\dagger({\bf p''_\perp}) \varphi({\bf p'_\perp}),
\label{rhouu}
\end{eqnarray}
\begin{eqnarray}
\rho_{LU}({\bf b_\perp},{\bf p_\perp})
&=&\frac{4 \ e^2}{2(2\pi)^2 16 \pi^3} \int d\Delta_x d\Delta_y \int dx \sin(\Delta_x b_x+ \Delta_y b_y)\nonumber\\
&\times&\frac{(\Delta_x p_y- \Delta_y p_x) }{x^2 (1-x)}(x^2-1)
\varphi^\dagger({\bf p''_\perp})\varphi({\bf p'_\perp}),
\label{rholu}
\end{eqnarray}
\begin{eqnarray}
\rho^{1}_{UT}({\bf b_\perp},{\bf p_\perp})
&=&\frac{-4 e^2}{2(2\pi)^2 16 \pi^3} \int d\Delta_x d\Delta_y \int dx \sin(\Delta_x b_x+ \Delta_y b_y)\nonumber\\
&\times&\frac{\Delta_y}{x}\Big(M-\frac{m}{x}\Big) \varphi^\dagger({\bf p''_\perp}) \varphi({\bf p'_\perp}),
\label{rhout}
\end{eqnarray}
\begin{eqnarray}
\rho_{LL}({\bf b_\perp},{\bf p_\perp})
&=& \frac{4\ e^2}{2(2\pi)^2 16 \pi^3} \int d\Delta_x d\Delta_y \int dx \cos(\Delta_x b_x+ \Delta_y b_y)\Big[\frac{(1+x^2)}{x^2(1-x)^2} \nonumber\\
&\times&\Big({\bf p_\perp^2}-\frac{(1-x)^2}{4}{\bf\Delta_\perp^2}\Big)
-\Big(M- \frac{m}{x}\Big)^2 \Big] \varphi^\dagger({\bf p''_\perp}) \varphi({\bf p'_\perp}),
\label{rholl}
 \end{eqnarray}
 \begin{eqnarray}
\rho^{1}_{LT}({\bf b_\perp},{\bf p_\perp})
&=&\frac{-4 \ e^2}{2 (2\pi)^2 \ 16 \pi^3} \int d\Delta_x d\Delta_y \int dx \cos(\Delta_x b_x+ \Delta_y b_y)\nonumber\\
&\times&\frac{p_x}{x(1-x)}\Big(M-\frac{m}{x}\Big)
\varphi^\dagger({\bf p''_\perp}) \varphi({\bf p'_\perp}),
\label{rholt}
\end{eqnarray}
\begin{eqnarray}
\rho^{1}_{TU}({\bf b_\perp},{\bf p_\perp})
&=& - \frac{4 e^2}{2 (2\pi)^2 16 \pi^3}\int d\Delta_x d\Delta_y \int dx  \cos(\Delta_x b_x + \Delta_y b_y)\nonumber\\
&\times&\Delta_x\Big(M-\frac{m}{x}\Big)
\varphi^\dagger({\bf p''_\perp}) \varphi({\bf p'_\perp}),
\label{rhotu}
 \end{eqnarray}
 \begin{eqnarray}
\rho^{1}_{TL}({\bf b_\perp},{\bf p_\perp})
&=&- \frac{4 e^2}{2  (2\pi)^216 \pi^3} \int d\Delta_x d\Delta_y \int dx \sin(\Delta_x b_x + \Delta_y b_y)\nonumber\\
&\times&p_y\Big(M- \frac{m}{x}\Big)
 \varphi^\dagger({\bf p''_\perp}) \varphi({\bf p'_\perp}),
\label{rhotl}
\end{eqnarray}
for $i=j=1 $ and $i=j=2$, 
\begin{eqnarray}
\rho_{TT}({\bf b_\perp},{\bf p_\perp})
&=& \frac{4 e^2}{2  (2\pi)^216\pi^3} \int d\Delta_x \ d\Delta_y \int dx \cos(\Delta_x b_x+\Delta_y b_y)\nonumber\\
&\times&\frac{1}{x(1-x)^2}\Big({\bf p^2_\perp}- \frac{(1-x)^2}{4} {\bf \Delta^2_\perp}\Big)
\varphi^\dagger({\bf p''_\perp}) \varphi({\bf p'_\perp}),
 \label{rhott}
 \end{eqnarray}
for $i=1,j=2$  and for $i=2, j=1,$
\be
\rho_{TT}^{\perp}({\bf b_\perp},{\bf p_\perp})=0,
\ee
where
\begin{eqnarray}
{\bf p'_\perp}&=&{\bf p_\perp}- (1-x) \frac{{\bf \Delta_\perp}}{2},\nonumber\\
{\bf p''_\perp}&=&{\bf p_\perp}+ (1-x) \frac{{\bf \Delta_\perp}}{2},
\end{eqnarray}
and
\begin{eqnarray*}
\varphi({\bf p'_\perp})
&=& \frac{e}{\sqrt{1-x}}\frac{x(1-x)}{{\bf p'^2_\perp}- M^2 x (1-x)+m^2(1-x)+ \lambda^2 x},\nonumber\\
\varphi^\dagger({\bf p''_\perp})
&=& \frac{e}{\sqrt{1-x}}\frac{x(1-x)}{{\bf p''^2_\perp}- M^2 x (1-x)+m^2(1-x)+ \lambda^2 x}.
\end{eqnarray*}
\begin{figure}
\begin{minipage}[c]{0.98\textwidth}
\small{(a)}
\includegraphics[width=7cm,clip]{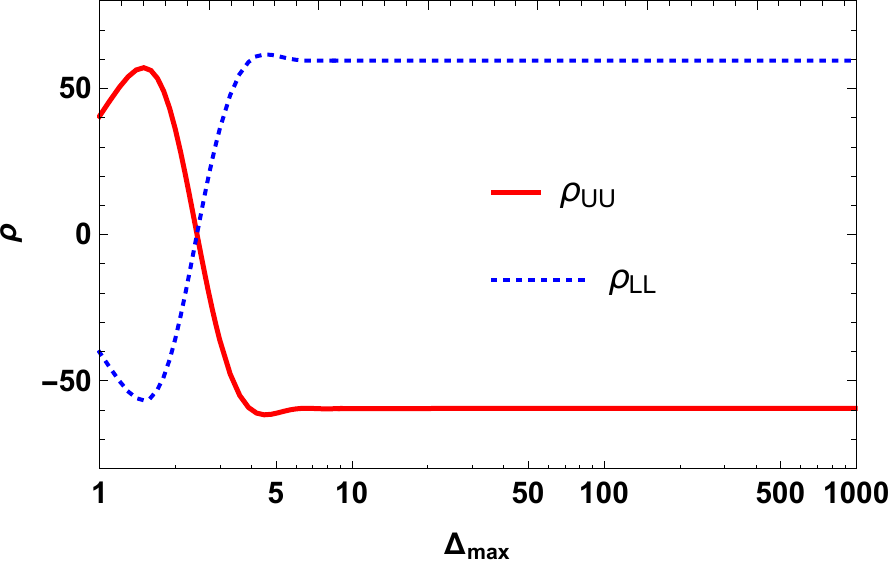}
\hspace{0.1cm}%
\small{(b)}\includegraphics[width=7cm,clip]{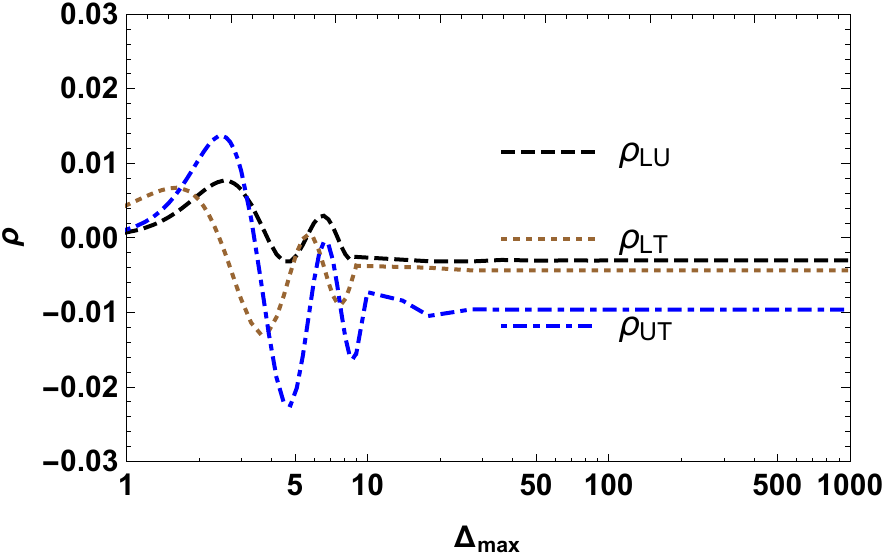}
\end{minipage}
\begin{minipage}[c]{0.98\textwidth}
\small{(c)}
\includegraphics[width=7cm,clip]{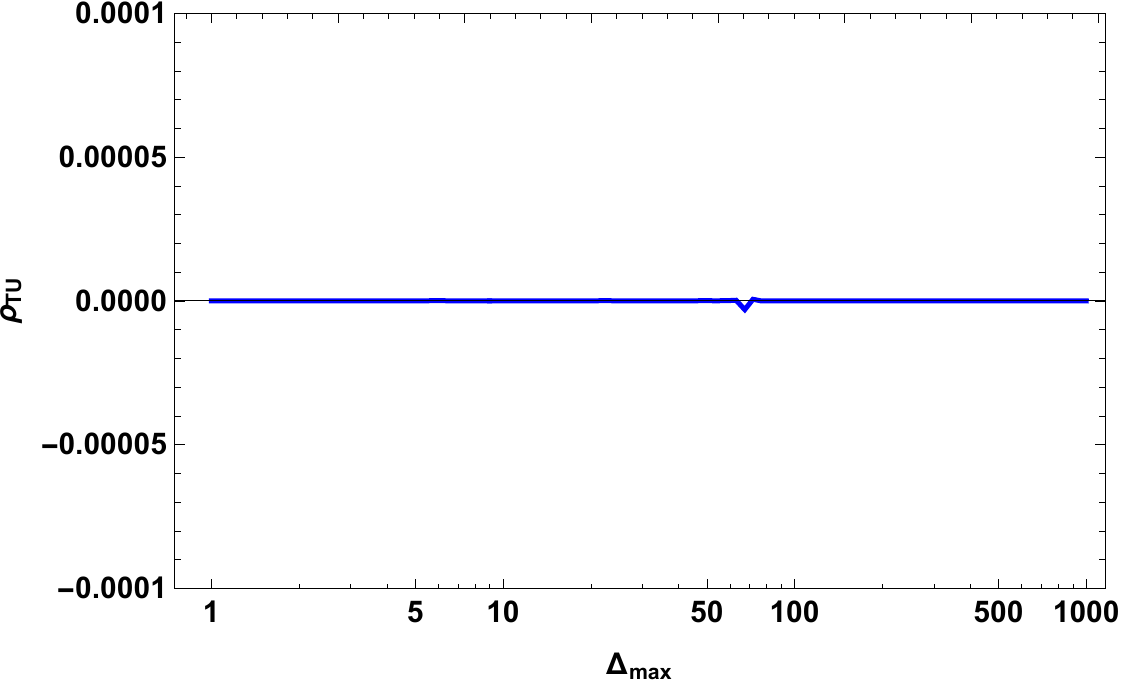}
\hspace{0.1cm}%
\small{(d)}\includegraphics[width=7cm,clip]{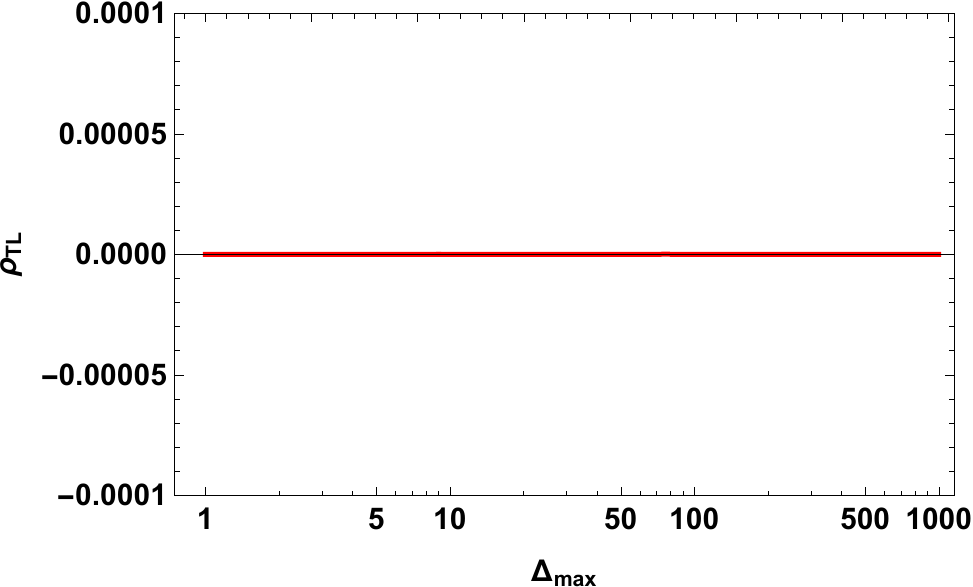}
\end{minipage}
\caption{(color online) Plots of (a) $\rho_{UU}, \rho_{LL}$; (b) $ \rho_{TT}$, $\rho_{LT}$, $\rho_{LU}$ and $\rho_{UT}$; (c) $\rho_{TU}$; (d) $\rho_{TL}$ for fixed values of $b_x=0.7 ~MeV^{-1}, b_y=0.9~MeV^{-1}, p_x=0.9 ~MeV$ and $p_y=0.8~MeV$ to show convergence using Levin's integration method \cite{levin1,levin2}. }
  \label{numerical-strategy}
\end{figure}
\begin{figure}[hbtp]
\centering
\includegraphics[width=7cm,clip]{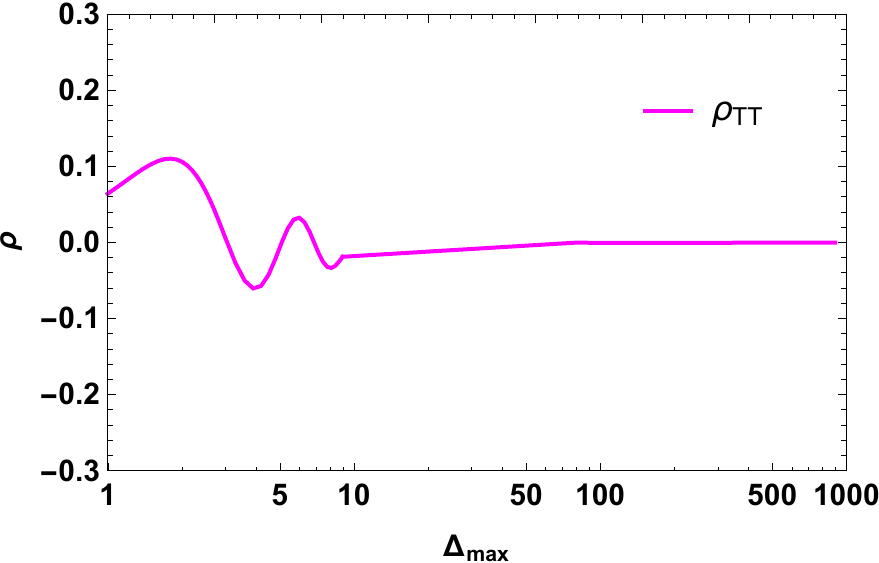}
\caption{(color online) Plot of $\rho_{TT}$ for fixed values of $b_x=0.7 MeV^{-1}, b_y= 0.7 MeV^{-1}, p_x = 0.9 MeV$ and $p_y = 0.8 MeV$ to show convergence using Levin's integration method.}
\end{figure}

\begin{figure}[htbp]
\centering
\begin{minipage}[c]{0.98\textwidth}
\small{(a)}
\includegraphics[width=4.5cm]{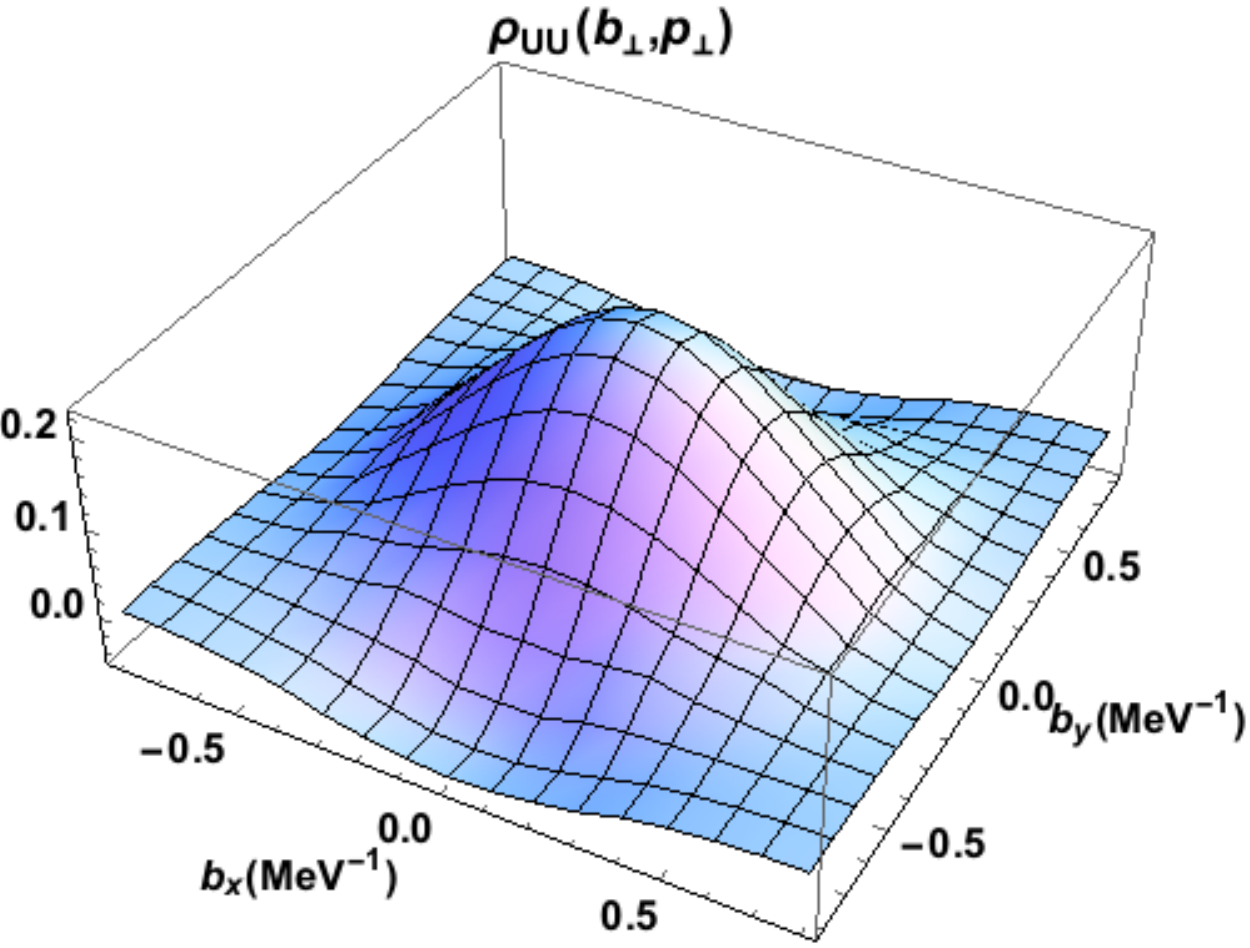}
\hspace{0.1cm}%
\small{(b)}\includegraphics[width=4.5cm]{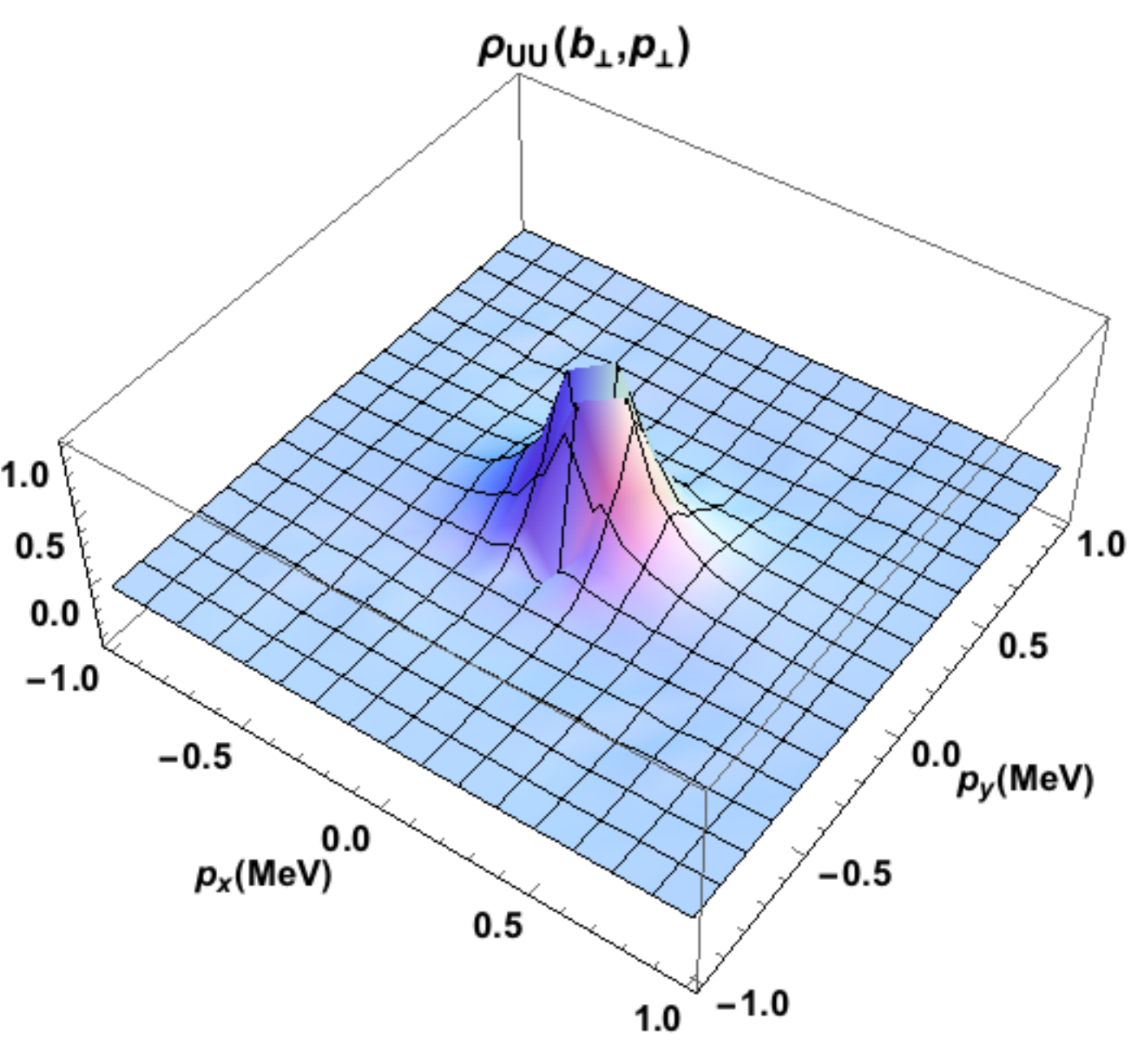}
\hspace{0.1cm}%
\small{(c)}\includegraphics[width=4.5cm]{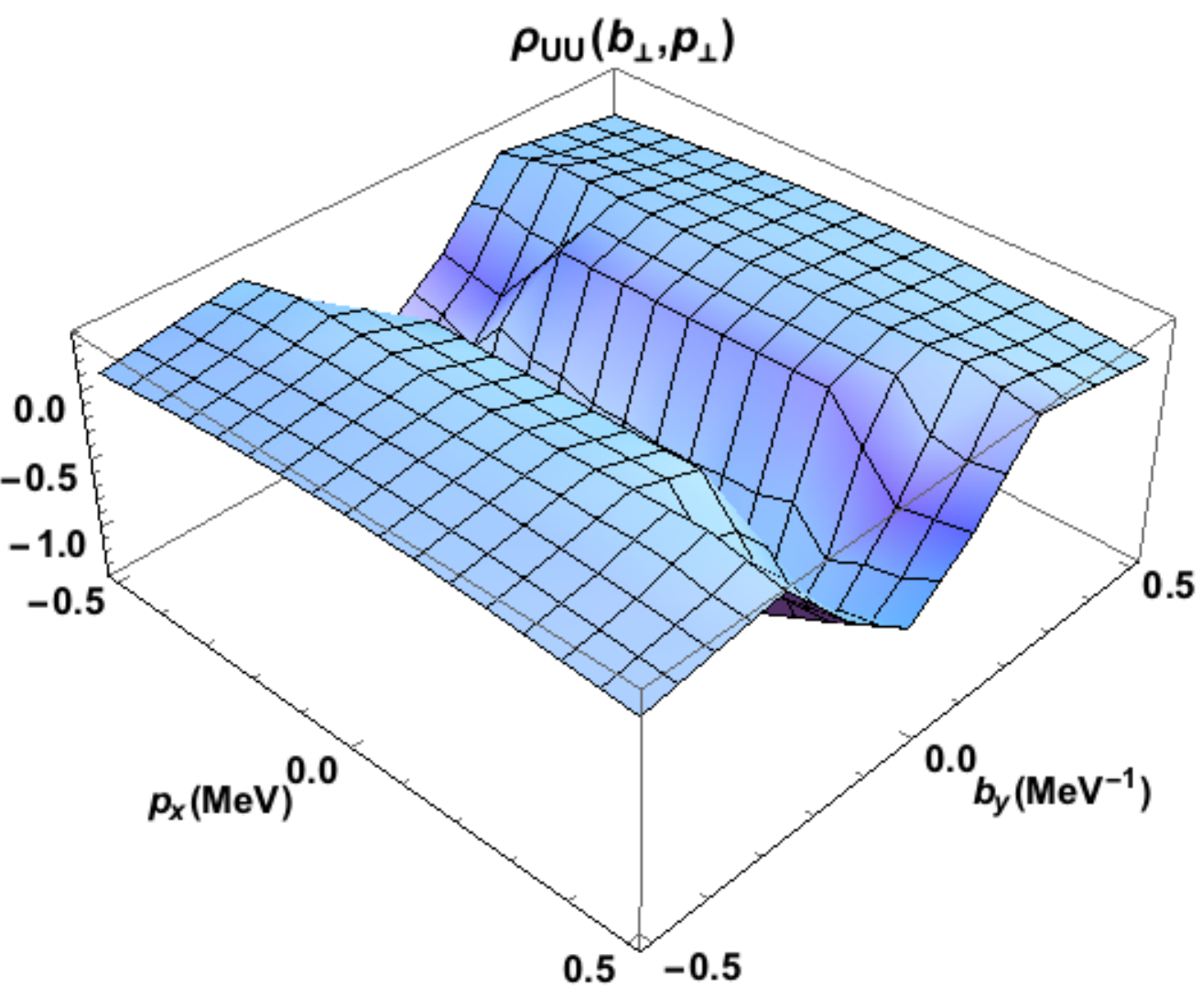}
\\
\small{(d)}
\includegraphics[width=4.5cm]{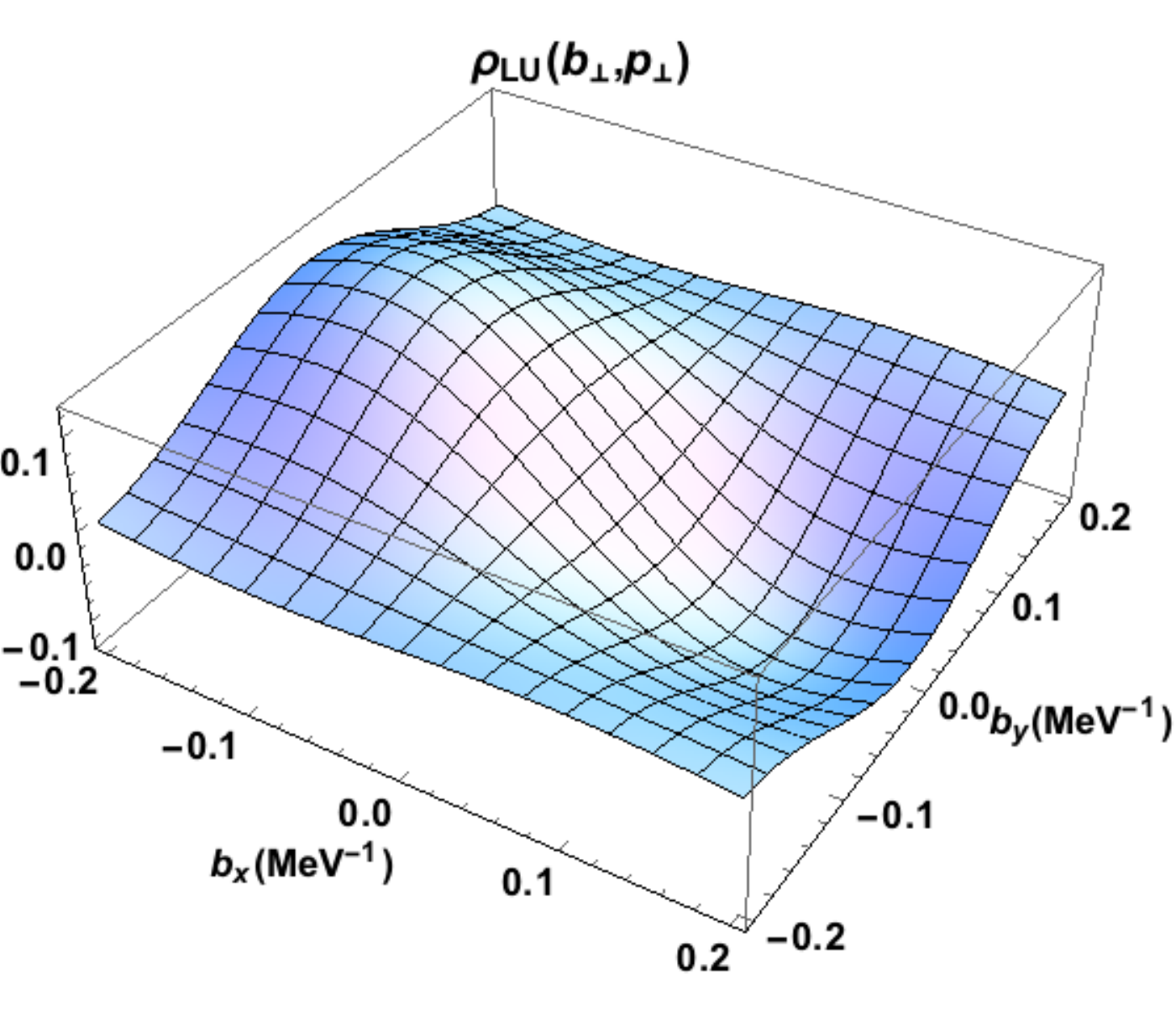}
\hspace{0.1cm}%
\small{(e)}\includegraphics[width=4.5cm]{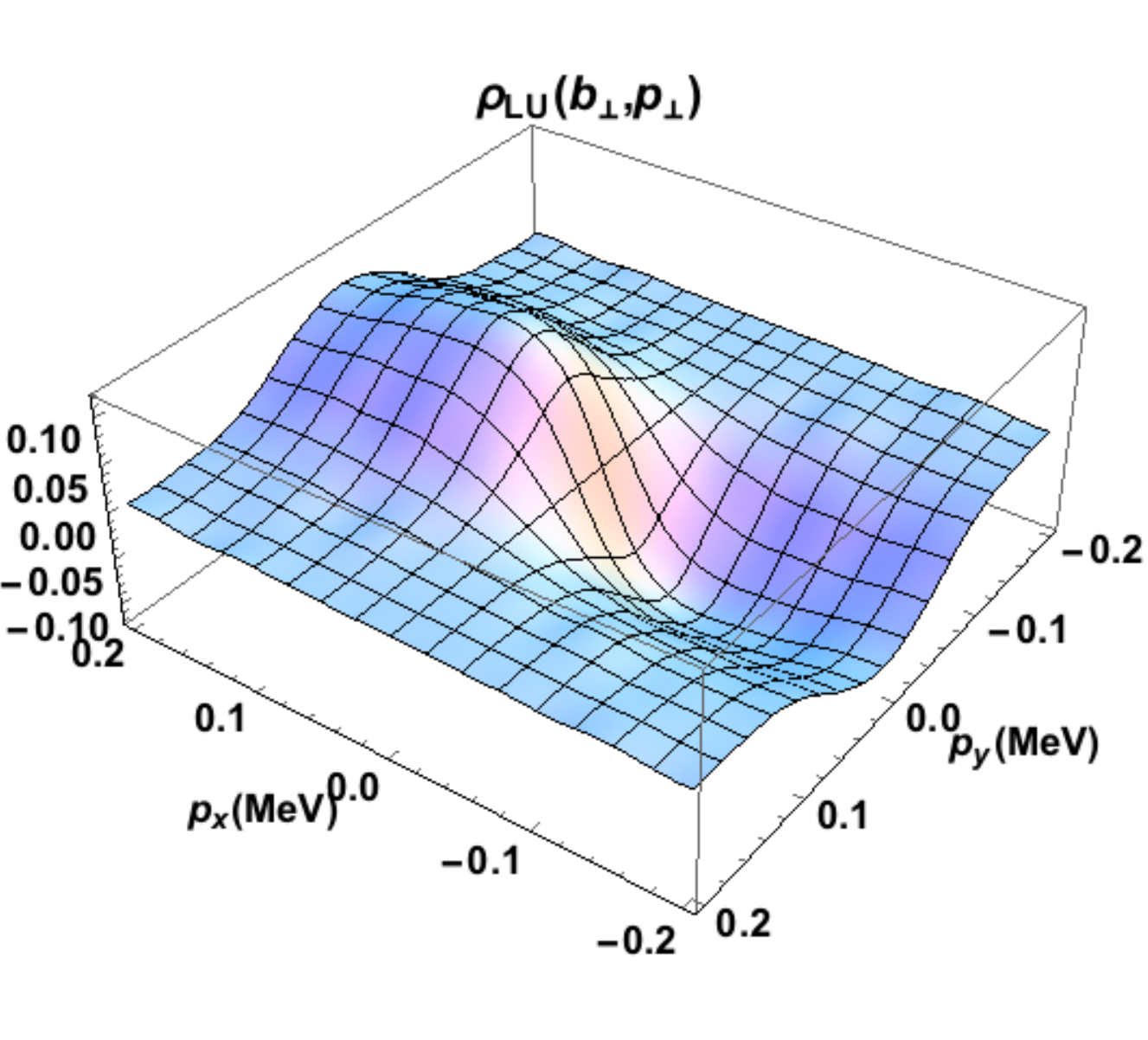}
\hspace{0.1cm}%
\small{(f)}\includegraphics[width=4.5cm]{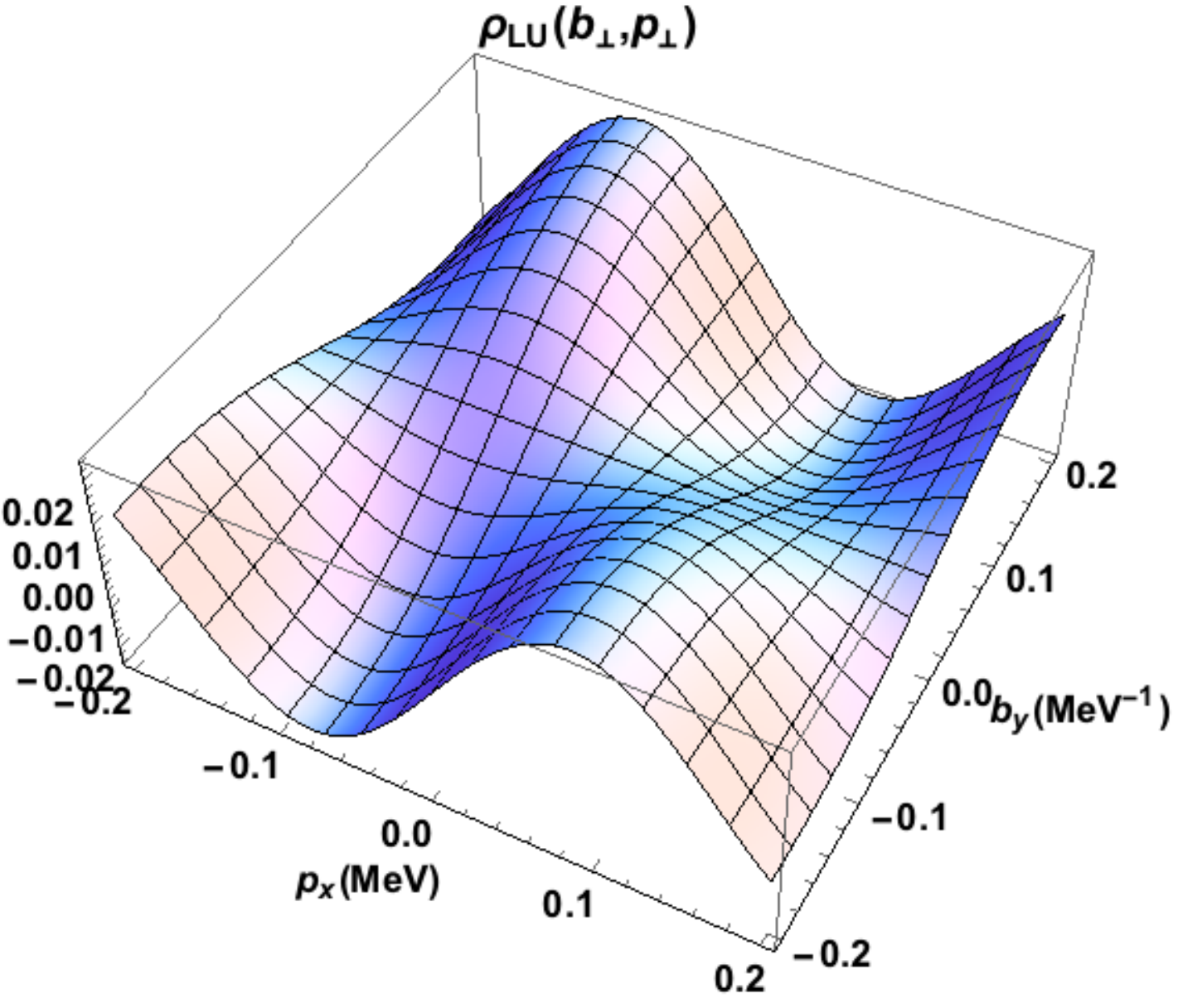}
\end{minipage}
\caption{(color online)  Plots of Wigner distribution $\rho_{UU}(\bfb,\bfp)$ and $\rho_{LU}(\bfb,\bfp)$ for physical electron in impact-parameter plane with fixed transverse momentum ${\bf p}_\perp= 0.8 ~MeV$ $\hat{e}_x$ (left panel), in momentum plane with fixed impact-parameter ${\bf b}_\perp= 0.8 ~MeV^{-1}$ $\hat{e}_x$ (middle panel) and in mixed plane (right panel). The upper panel represents $\rho_{UU}$ and the lower panel is for $\rho_{LU}$.}
  \label{rhoUU_LU}
\end{figure}
\begin{figure}
\begin{minipage}[c]{0.98\textwidth}
\small{(a)}
\includegraphics[width=4.5cm,clip]{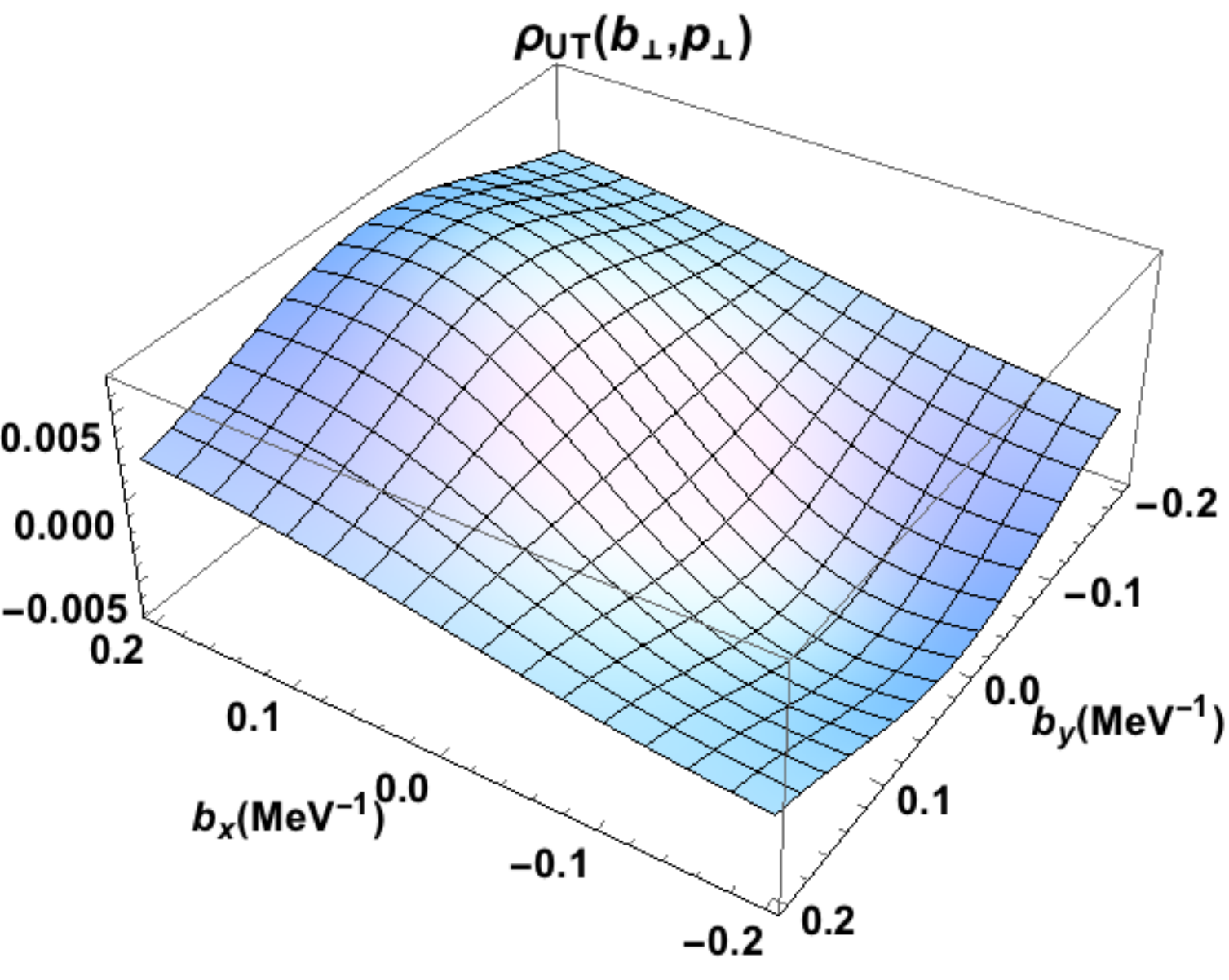}
\hspace{0.1cm}%
\small{(b)}\includegraphics[width=5cm,clip]{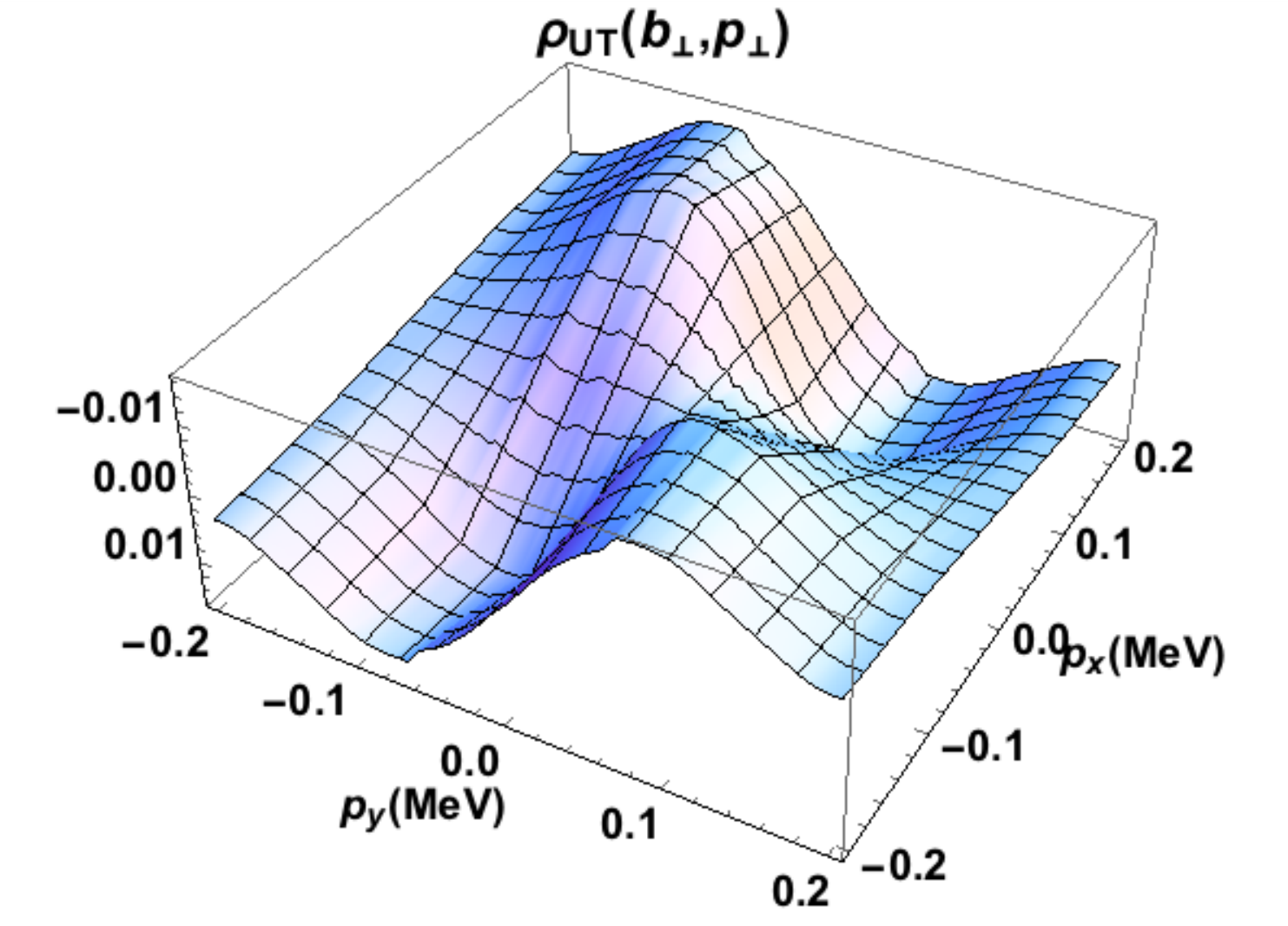}
\hspace{0.1cm}%
\small{(c)}\includegraphics[width=4.6cm,clip]{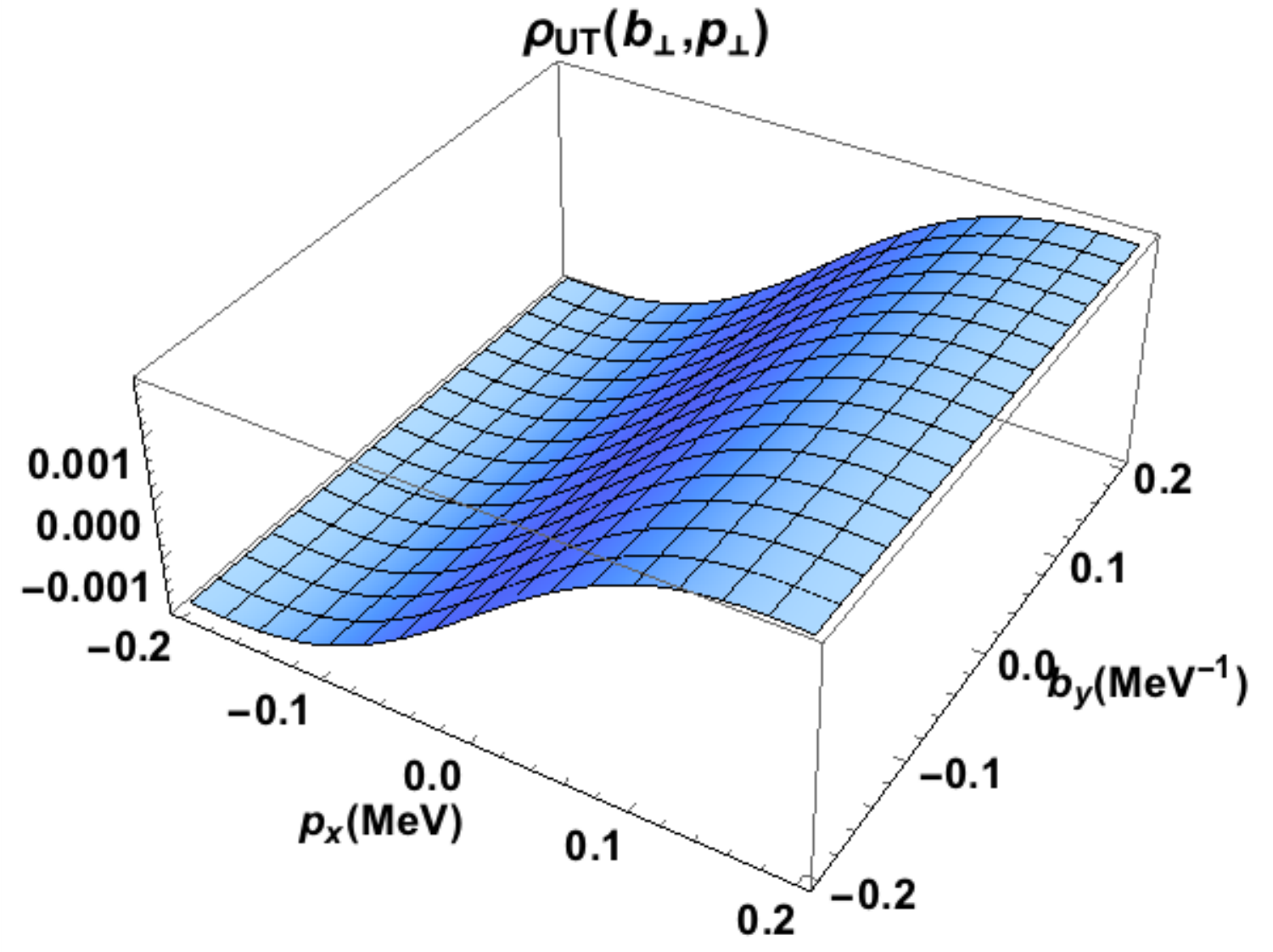}
\end{minipage}
\caption{(color online)  Plots of Wigner distribution $\rho_{UT}(\bfb,\bfp)$ for physical electron in impact-parameter plane with fixed transverse momentum ${\bf p}_\perp= 0.8 ~MeV$ $\hat{e}_x$ (left panel), in momentum plane with fixed impact-parameter ${\bf b}_\perp= 0.8 ~MeV^{-1}$ $\hat{e}_x$ (middle panel) and in mixed plane (right panel). The transverse polarization of the bare electron or the physical electron is taken along $x$-direction.}
\label{rhoUT_TU}
\end{figure}
\begin{figure}
\centering
\begin{minipage}[c]{0.98\textwidth}
\small{(a)}
\includegraphics[width=4.5cm,clip]{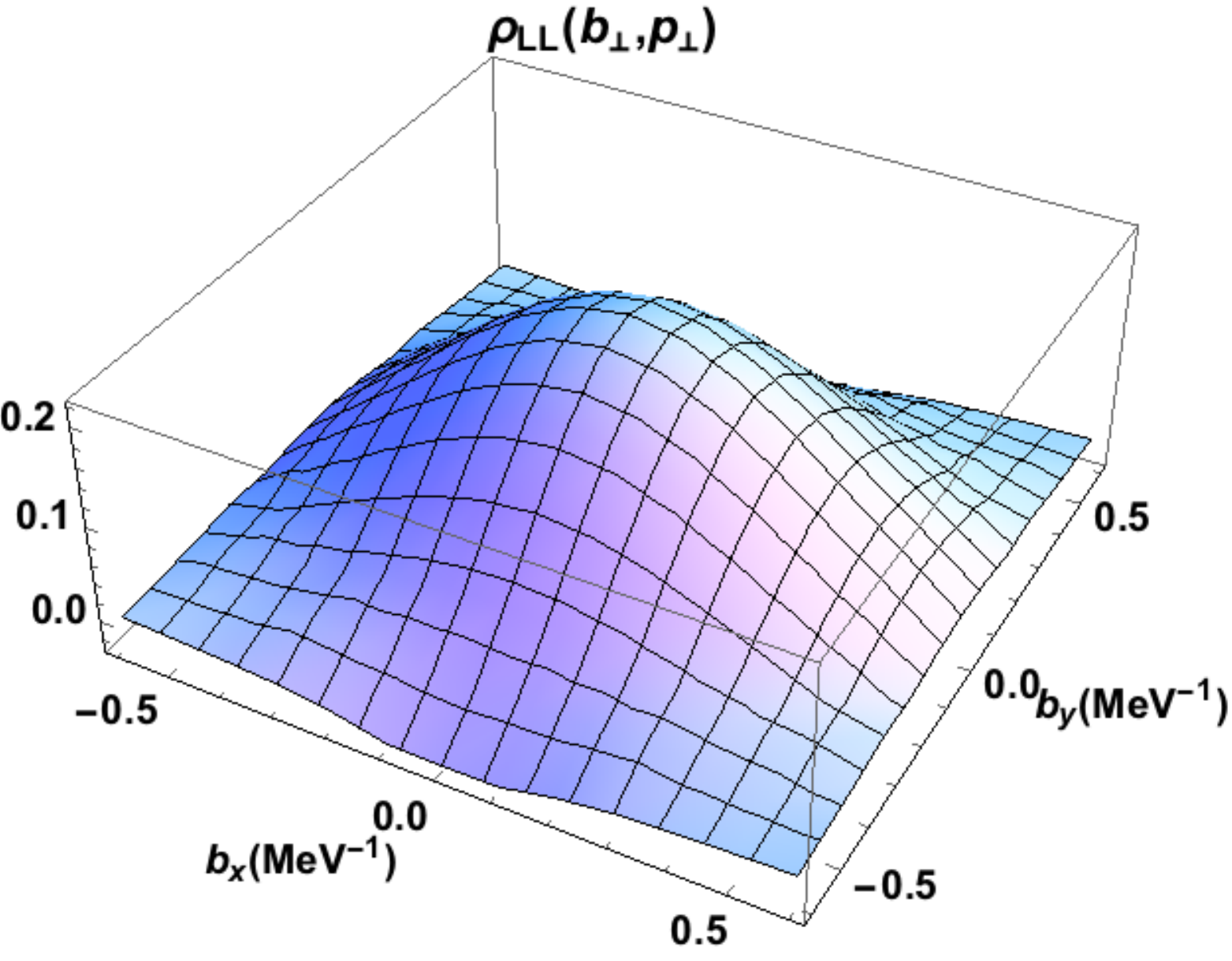}
\hspace{0.1cm}%
\small{(b)}\includegraphics[width=4.5cm,clip]{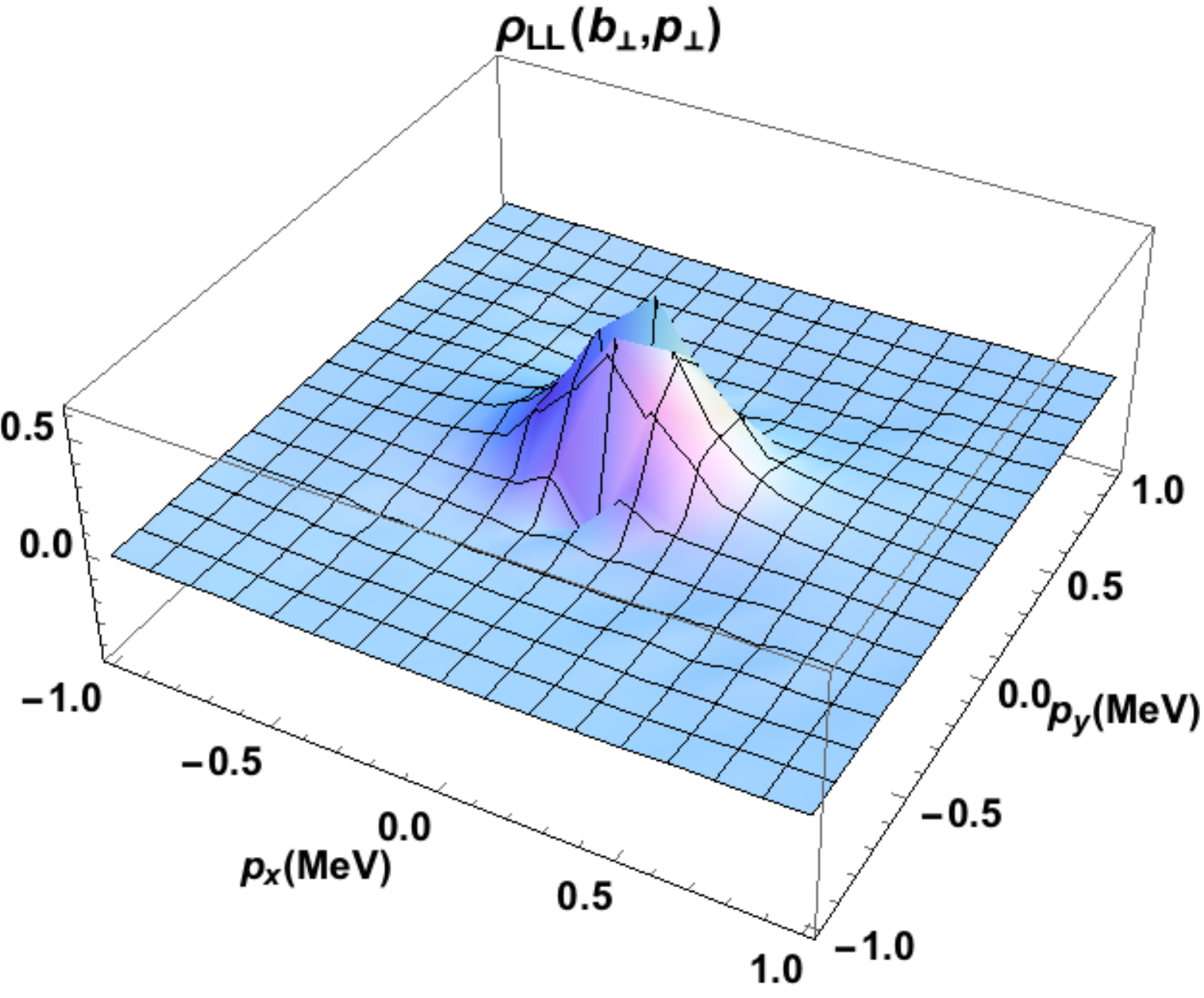}
\hspace{0.1cm}%
\small{(c)}\includegraphics[width=4.5cm,clip]{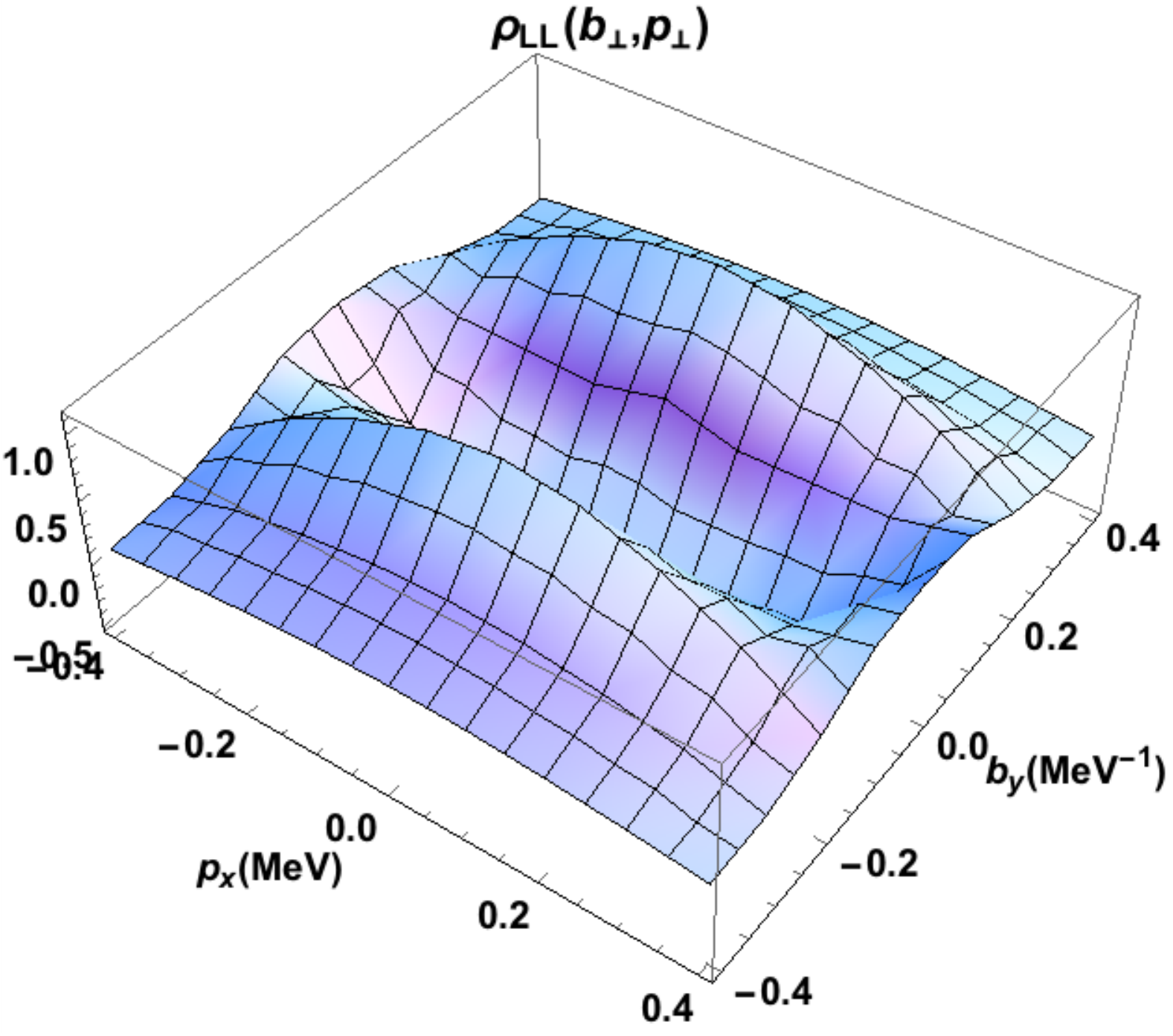}
\\
\small{(d)}
\includegraphics[width=4.5cm,clip]{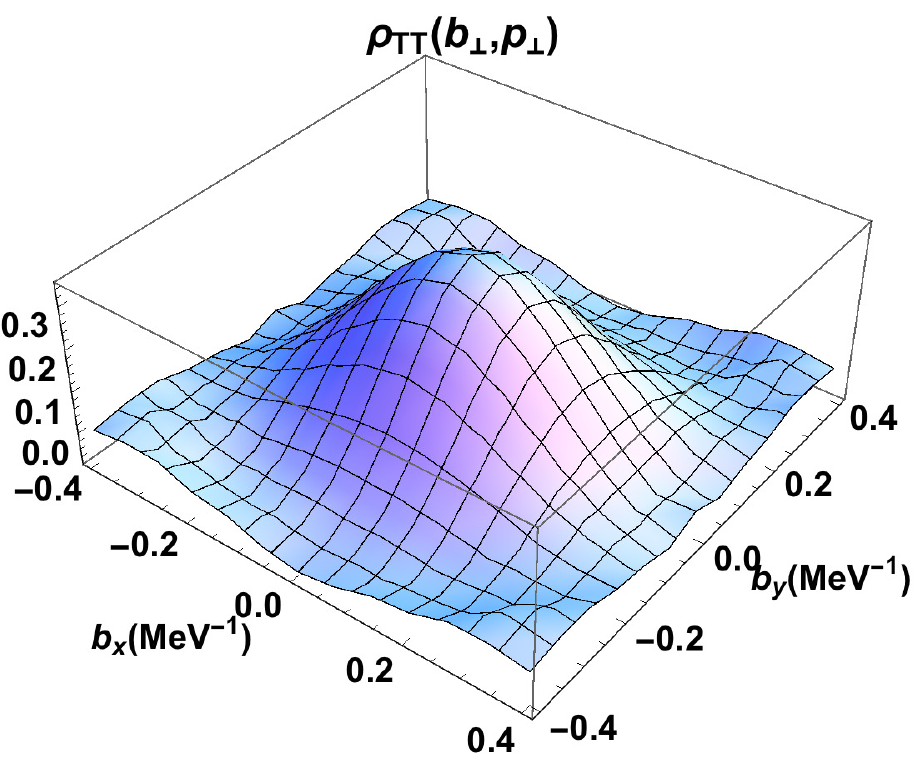}
\hspace{0.1cm}%
\small{(e)}\includegraphics[width=4.5cm,clip]{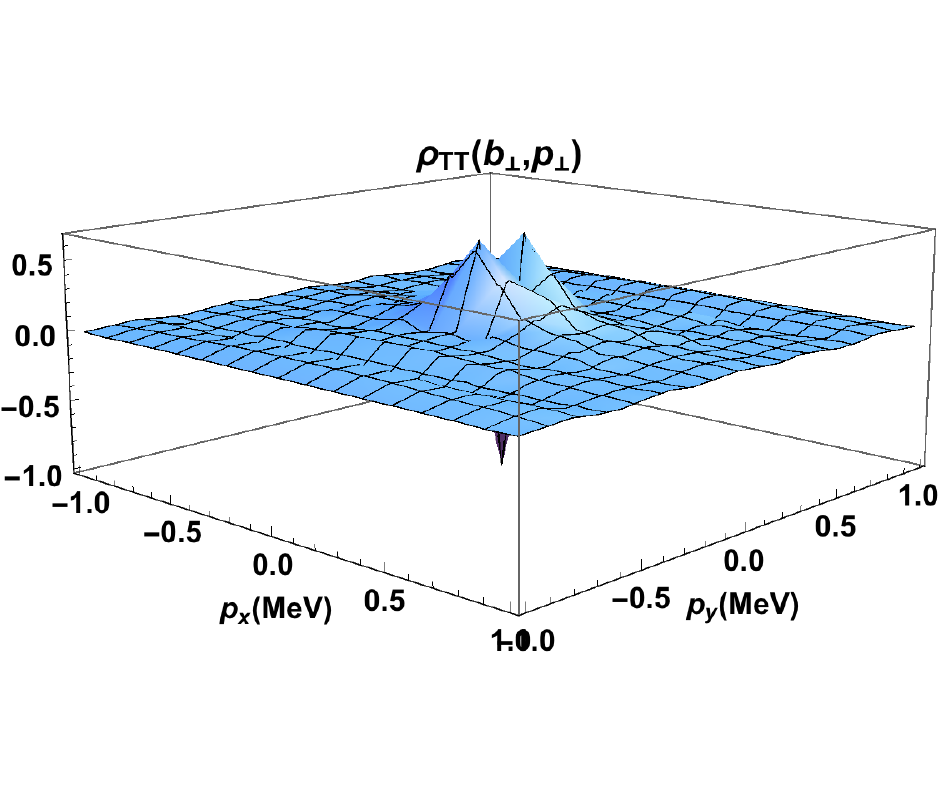}
\hspace{0.1cm}%
\small{(f)}\includegraphics[width=4.5cm,clip]{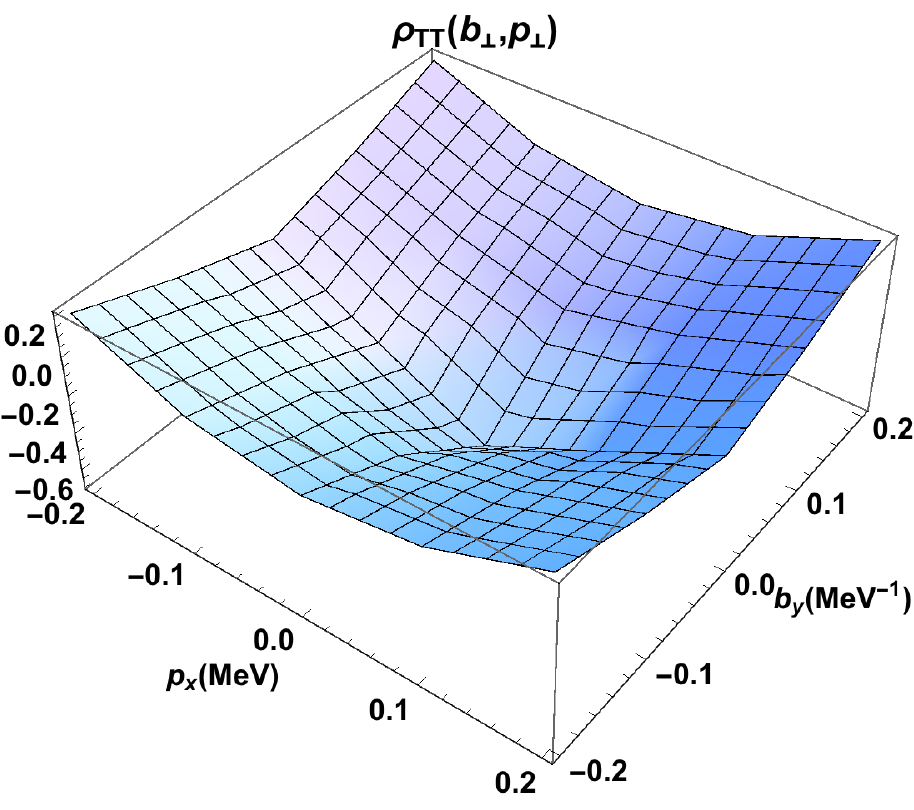}
\end{minipage}
\caption{(color online)  Plots of Wigner distribution $\rho_{LL}(\bfb,\bfp)$ and $\rho_{TT}(\bfb,\bfp)$ for physical electron in impact-parameter plane with fixed transverse momentum ${\bf p}_\perp= 0.8 ~MeV$ $\hat{e}_x$ (left panel), in momentum plane with fixed impact-parameter ${\bf b}_\perp= 0.8 ~MeV^{-1}$ $\hat{e}_x$ (middle panel) and in mixed plane (right panel). The upper and panel represent $\rho_{LL}$ and $\rho_{TT}$, respectively. For $\rho_{TT}$, the transverse polarizations of both the bare and the physical electron are along $y$-direction. }
  \label{rhoLL_TT}
\end{figure}
\begin{figure}
\centering
\begin{minipage}[c]{0.98\textwidth}
\small{(a)}
\includegraphics[width=4.5cm,clip]{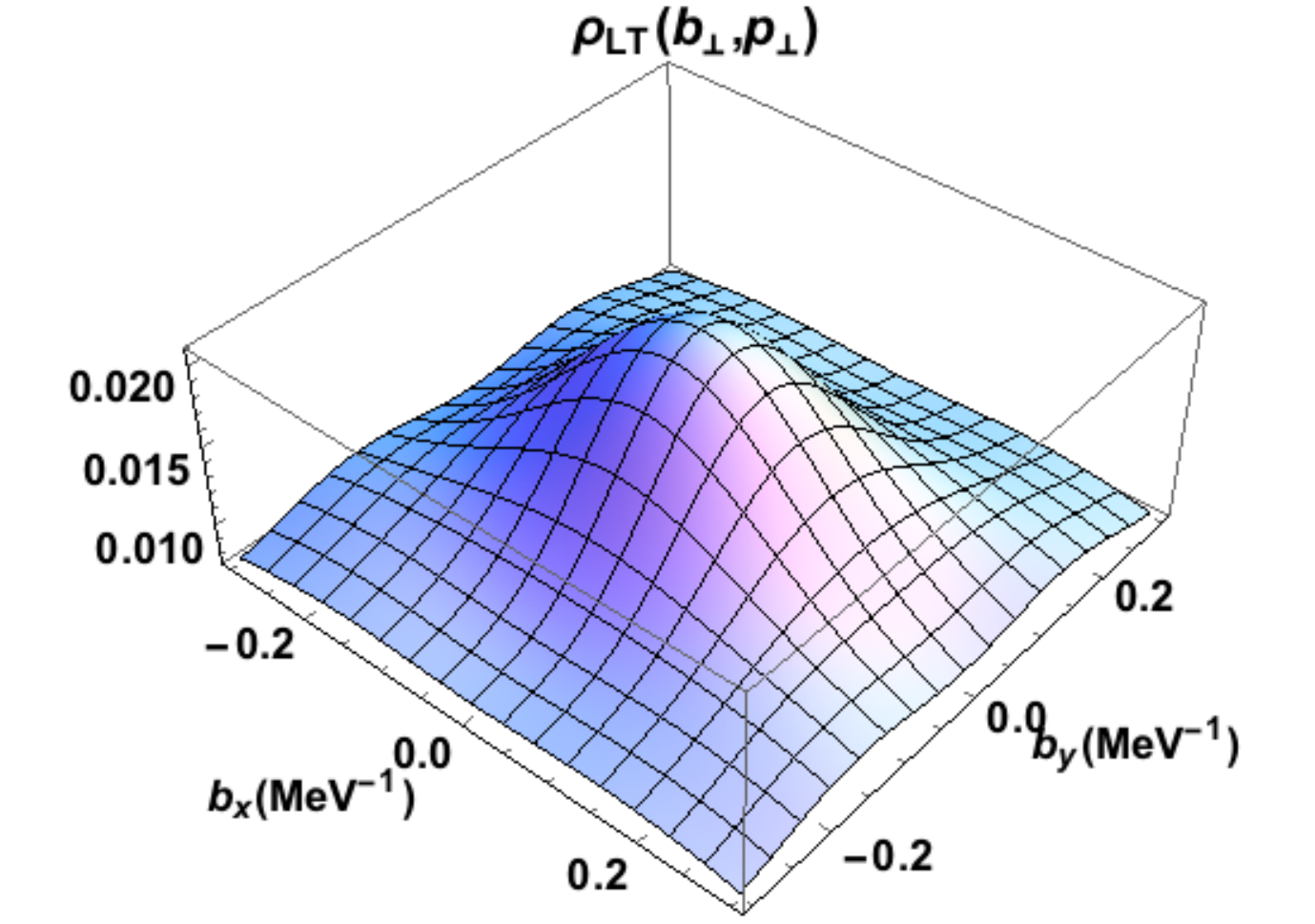}
\hspace{0.1cm}%
\small{(b)}\includegraphics[width=4.5cm,clip]{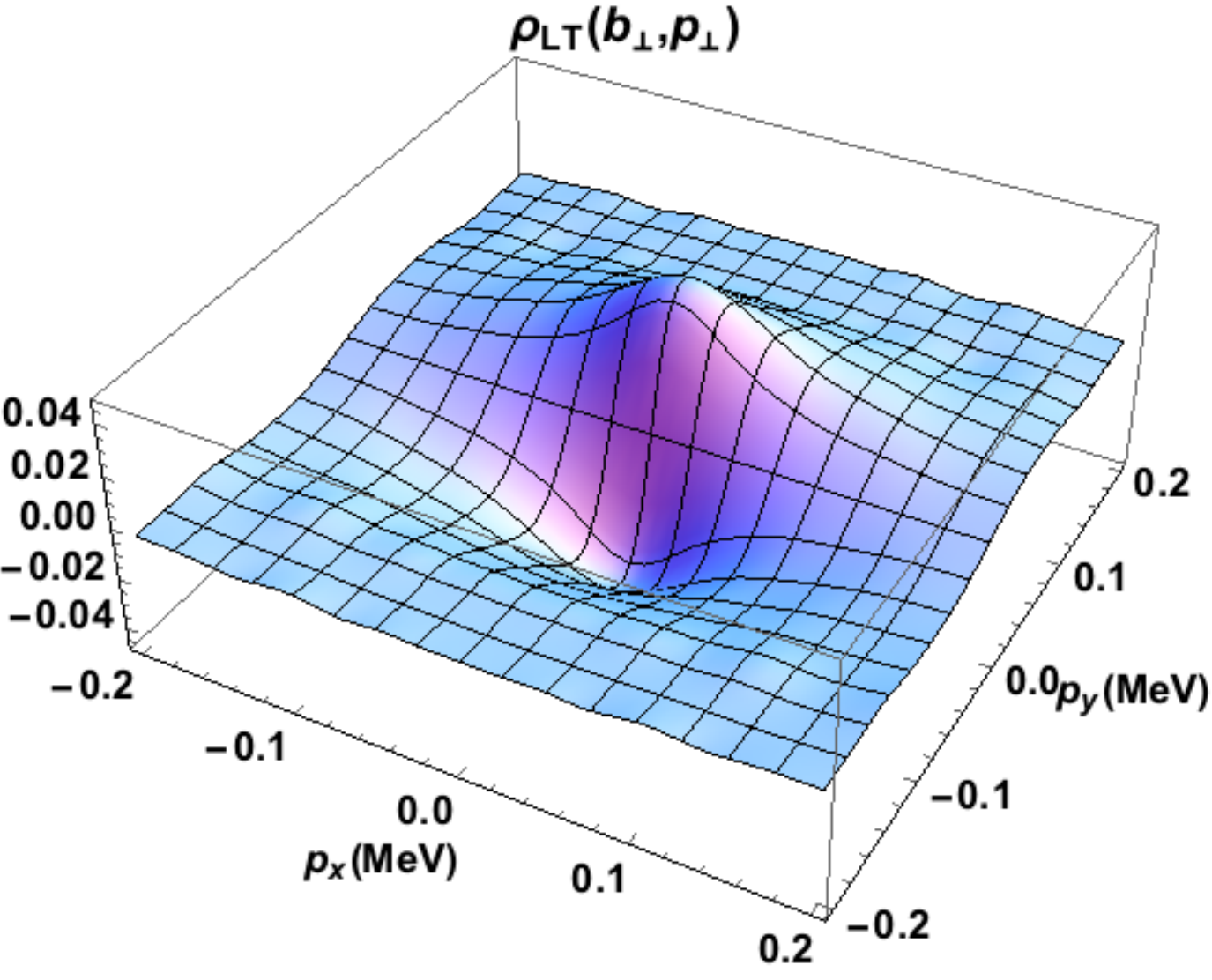}
\hspace{0.1cm}%
\small{(c)}\includegraphics[width=4.5cm,clip]{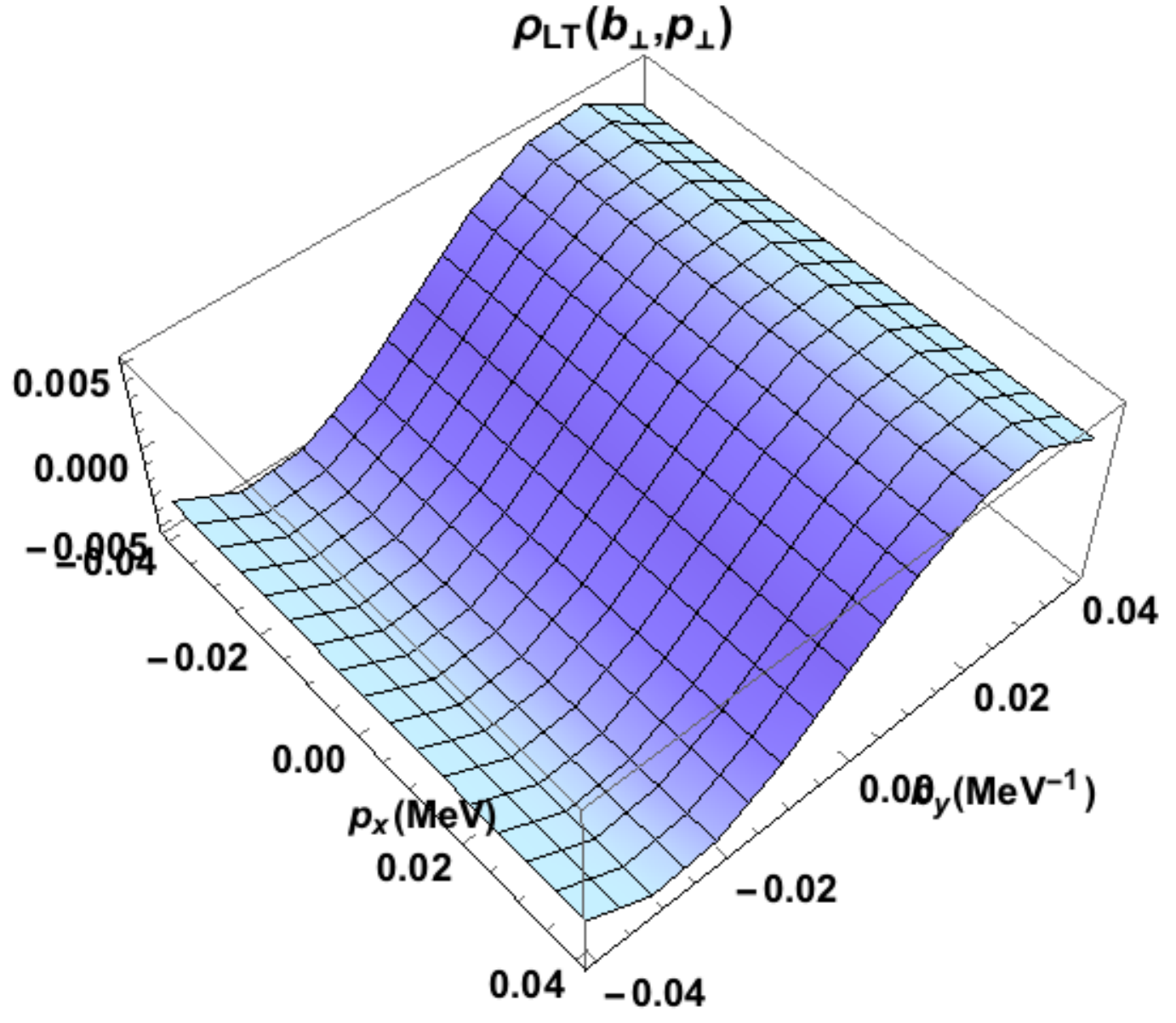}
\end{minipage}
\caption{(color online)  Plots of Wigner distribution $\rho_{LT}(\bfb,\bfp)$  for physical electron in impact-parameter space with fixed transverse momentum ${\bf p}_\perp= 0.8 ~MeV$ $\hat{e}_x$ and ${\bf b}_\perp= 0.8 ~MeV^{-1}$ for $\rho_{LT}$ . The transverse polarization of the bare electron or the physical electron is taken along $x$-direction.}
  \label{rhoLT_TL}
\end{figure}

\section{Numerical results and discussion}\label{impact}
\ 
In this section, we discuss the numerical results for Wigner distributions of electron with different polarizations configurations. The Wigner distributions in the pure transverse plane are obtained by integrating them over $x$.
To get the complete contribution correctly at $x=1$, one needs to include the contribution from single particle Fock state expansion of the physical electron state  i.e. $| e^- \rangle= |e^- \rangle + | e^- \gamma \rangle$. The form of  the single particle contribution to Wigner distribution function is $\rho_{single-particle}(\bfb, \bfp, x)\sim\delta(1-x) \delta^2(\bfb)\delta^2(\bfp)$~\cite{Brodsky:2000ii,levin}. This corresponds that the single particle carries all the longitudinal momentum at $\bfb = 0$ and the transverse momentum $\bfp$ is also zero. In the present work, we fix the non-zero value of $\bfb$ in $\bfp$ plane and vice-versa, which makes $\rho_{single-particle}(\bfb, \bfp, x)= 0$, thus, the Wigner distributions do not get the contribution from the single particle component.
The LFWFs include the three mass parameters $M$, $m$ and $\lambda$ where $M$ is the mass of physical electron, $m$ is the bare electron mass and $\lambda$ is the  photon mass. The value of masses are taken as $M=0.51 MeV$, $m=0.5 MeV$ and $\lambda=0.02 MeV$. The non-zero mass of photon is used to regulate the infrared divergence \cite{Miller:2014vla}. In order to obtain the results in impact-parameter plane $(b_x-b_y)$, one needs to fix the value of transverse momentum and integrate over $x$. Here, we perform the numerical integration by setting the higher limit of $x=0.95$.
Similarly, we need to fix the value of transverse position in order to obtain the results in momentum plane $(p_x-p_y)$. In the case of mixed plane representation one can integrate over $p_x-b_y$ or $p_y-b_x$ plane to get the distributions. In both cases i.e. (in momentum and mixed plane) we again impose the same higher cut-off on the integration over $x$.

Further, the integration in Eqs.(\ref{rhouu}-\ref{rhott}) is set over $\Delta_\perp$ having range from -$\infty$ to $\infty$. We perform the numerical integration by applying a suitable cut off on the $\Delta_\perp$ ($\Delta_{max}$). We employ the Levin's integration technique \cite{levin,levin1,levin2} to handle the oscillatory behaviour of integrands. The $\Delta_{max}$ dependence of the Wigner distributions are shown in Fig. \ref{numerical-strategy} upto 1000 $MeV$. It can be noticed that $\rho_{UU}$ and $\rho_{LL}$ is constant after the $\Delta_{max}=6~ MeV$ and hence this represents the cut-off on $\Delta_\perp$ integration. Similarly for the other Wigner distributions, the cut-off on $\Delta_\perp$ integration is $\Delta_{max}=20 ~MeV$. The Wigner distributions $\rho_{TU}$ and $\rho_{TL}$ vanish in this QED model. Recently, the Wigner distributions of a quark dressed with gluon have been studied in light-front dressed quark model \cite{More:2017zqq}, where the authors have also used the same cut-off procedure over the $\Delta_\perp$ integration to handle the oscillatory behaviour of the integrands.

We show the first Mellin moment of the Wigner distributions $\rho_{UU}(x,\bfb,\bfp)$ and $\rho_{LU}(x,\bfb,\bfp)$ for a composite system of bare electron and photon in Fig.\ref{rhoUU_LU} which represent the transverse phase-space distribution of the unpolarized and longitudinally polarized physical electron when the constituent bare electron is unpolarized.  Fig.\ref{rhoUU_LU}(a) and Fig.\ref{rhoUU_LU}(d) show the distributions $\rho_{UU}(\bfb,\bfp)$ and $\rho_{LU}(\bfb,\bfp)$ in impact-parameter plane
 respectively with fixed transverse momentum $\bfp$ along $\hat{x}$ for $p_x=0.8~MeV$ whereas the variation of the distributions in the transverse momentum plane are shown in Fig.\ref{rhoUU_LU}(b) and Fig.\ref{rhoUU_LU}(e) with fixed  impact-parameter $\bfb$ along $\hat{x}$ for $b_x=0.8~MeV^{-1}$. The mixing distributions of $\rho_{UU}$ and $\rho_{LU}$ are shown in Fig.\ref{rhoUU_LU}(c) and Fig.\ref{rhoUU_LU}(f), respectively. For the mixed plane representation, we present the results in $(p_x,b_y)$ plane for both $\rho_{UU}$ and $\rho_{LU}$. 
 In the present model, we obtain $\rho_{UL}=\rho_{LU}$, therefore results are not presented for $\rho_{UL}$. One can notice that the distributions $\rho_{UU}$ in transverse momentum plane as well as in impact-parameter plane are circularly symmetric but in momentum plane, the peak of the distribution is in negative direction. The distributions $\rho_{LU}$ in both the impact-parameter and the momentum planes exhibit dipolar patterns due to privilege direction  but the peaks of the dipolar pattern are opposite to each other. We observe the  quadrupole structure for $\rho_{LU}$ in mixed plane representation. Such patterns are also observed in recent study of quark Wigner distributions in light-front dressed quark model where a quark is dressed with gluon \cite{More:2017zqq}. This model is analogous to the present QED model. In the GPDs limit, $\rho_{UU}$ reduces to the GPD $H$ in impact-parameter plane which by integrating over $x$ provides charge density of the unpolarized physical electron. In the TMDs limit, $\rho_{UU}$ reduces to the unpolarized TMD $f_1$ which gives the electron density in momentum plane.
The Wigner distribution $\rho_{LU}$ is connected with the Fourier transform of the GTMD $\mathcal{F}_{1,4}$ \cite{Chakrabarti:2016yuw} and the multipole structure is due to the presence of factor $\epsilon^{ij}_\perp p_\perp^i \frac{\partial}{\partial b_\perp^j}$ which usually break the right-left symmetry. Futher they are also used to study the orbital angular momentum issues \cite{Lorce:2011kd,Chakrabarti:2016yuw,Lorce:2011ni}. For longitudinal-unpolarized Wigner distribution, no twist-2 TMDs or IPDs are related and therefore they will vanish at the TMD/IPD limit.
  Concentrating on the physical interpretation, these distributions reflects the correlation between the electron spin and orbital angular momentum
\begin{equation}
C_z= \int dx d^2p_\perp d^2b_\perp (\vec{b}_\perp \times \vec{p}_\perp)_z \rho_{UL}(\vec{b}_\perp, \vec{p}_\perp, x),
\end{equation}
if $C_z > 0$, electron spin and OAM is parallel to each other and if $C_z < 0$ then electron spin and OAM are anti-parallel to each other. Further, OAM can be calculated from the $F_{1,4}$
 GTMD as
 \begin{equation}
 l_z= - \int dx d^2p_\perp \frac{\vec{p}_\perp^2}{M^2} F_{1,4}(x,0,\vec{p}_\perp^2,0,0).
 \end{equation}

 Wigner distributions do not have a probabilistic interpretation due to uncertainty principle. We integrate over $p_y$ and $b_x$ dependence which in result giving us the probability densities and correlating $p_x$ and $b_y$, and this correlation is not violated by the uncertainty principle. As Wigner distributions have only quasi-probabilistic interpretation, a original probabilistic interpretation can be obtained by integrating over ${\bf b_\perp}$ and ${\bf p_\perp}$ which reduce them to TMDs and GPDs respectively.
In Fig. \ref{rhoUT_TU}, we present the results for the Wigner distribution $\rho_{UT}(\bfb,\bfp)$. 
They describe the distributions when the bare electron is transversely polarized in an unpolarized physical electron. In impact-parameter plane, $\rho_{UT}$ shows the dipole behaviour whereas in momentum plane it shows the quadruple structure. In the IPD limit, $\rho_{UT}$ reduces to $\tilde{H}_T$ together with some other
distributions and in the TMD limit it reduces to $h_1^\perp$ (Boer-Mulder's function) whereas former corresponds to T-even part and latter corresponds to T-odd part.
In the mixed plane representation, $\rho_{UT}$ gives the dipole structure.
The physical interpretation of $\rho_{UT}$ is connected with the spin structure.    $\rho_{UT}$ only corresponds the correlation between the the transverse polarization of the bare electron with its perpendicular transverse coordinate.
It is clear from Eq.(\ref{rhout}) that this distribution vanishes when electron intrinsic transverse coordinate is parallel to polarization i.e. electron transverse spin has no correlation with its parallel transverse coordinate.
In Fig. \ref{rhoLL_TT}, we show the longitudinal and transverse Wigner distributions i.e. $\rho_{LL}$ and $\rho_{TT}$ which describe the distributions when both the bare and physical electrons are longitudinally and transversely  polarized, respectively. In impact-parameter plane, we fix the transverse momentum ${\bf p_\perp}=p_\perp \hat{e_x}$ with $p_\perp=0.8 MeV$ and $\rho_{LL}$  and $\rho_{TT}$ shows a circularly symmetric behavior as similar as $\rho_{UU}$. For the distributions in momentum plane, we fix impact-parameter ${\bf b_\perp}=b_\perp \hat{e_x}$ with $b_\perp=0.8 MeV^{-1}$ and observe that both $\rho_{LL}$ and $\rho_{TT}$ again exhibit a  similar nature as $\rho_{UU}$ and the peak is in negative direction in momentum plane. In Fig. \ref{rhoLL_TT}(c) and (f), we show the distributions in the mixed plane for $\rho_{LL}$ and $\rho_{TT}$ respectively. In the TMD limit, transverse Wigner distribution reduces to transversity distribution $h_1$ when the polarization of the bare electron is parallel to the physical electron. Here, we choose the polarizations along $\hat{x}$-axis. Further, the bare electron and the physical electron can be polarized perpendicular to each other which corresponds to pretzelous Wigner distribution. The pretzelous Wigner distribution vanishes in this QED model. However in QCD spectator model \cite{Liu:2015eqa} it is nonzero but it also vanishes in the dressed quark model \cite{More:2017zqq}.
\begin{figure}
\centering
\begin{minipage}[c]{0.98\textwidth}
\small{(a)}
\includegraphics[width=4.5cm,clip]{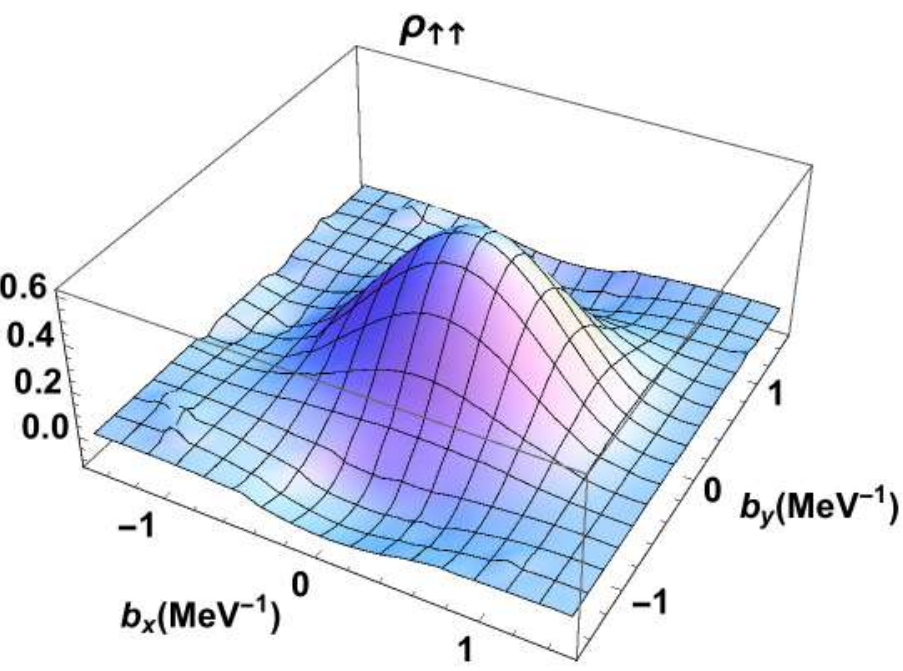}
\hspace{0.1cm}%
\small{(b)}\includegraphics[width=4.5cm,clip]{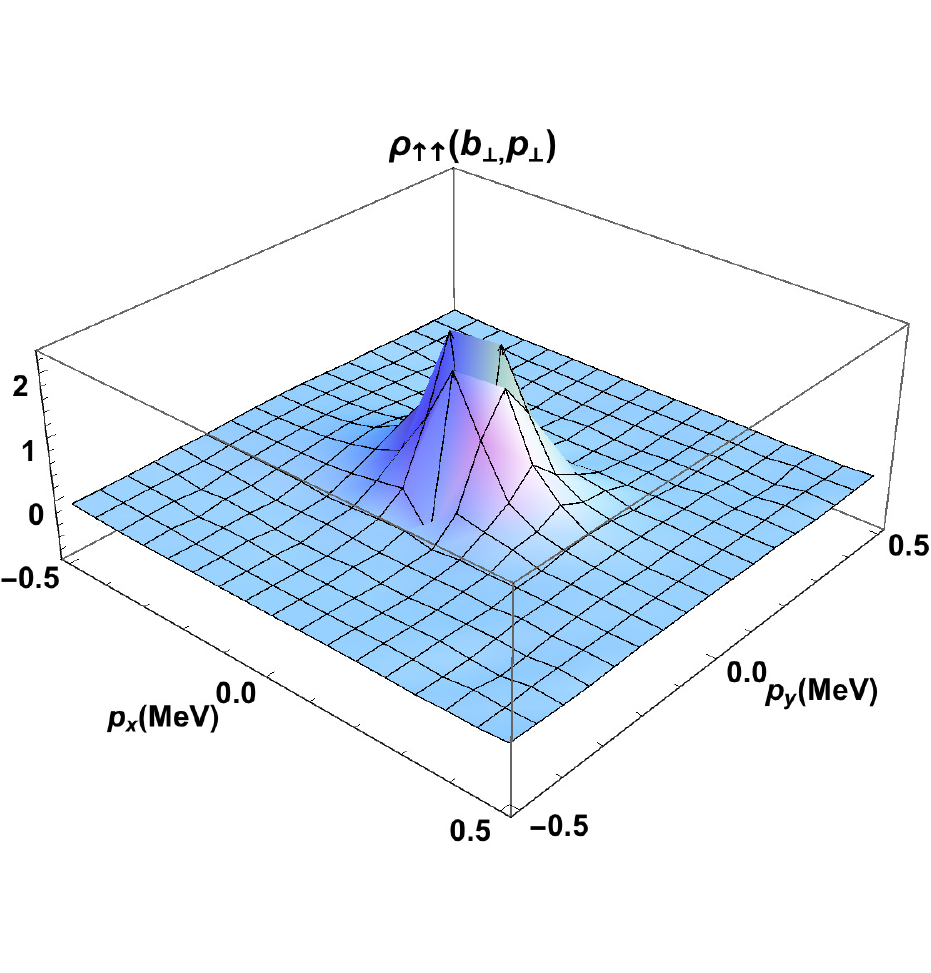}
\hspace{0.1cm}%
\small{(c)}\includegraphics[width=4.5cm,clip]{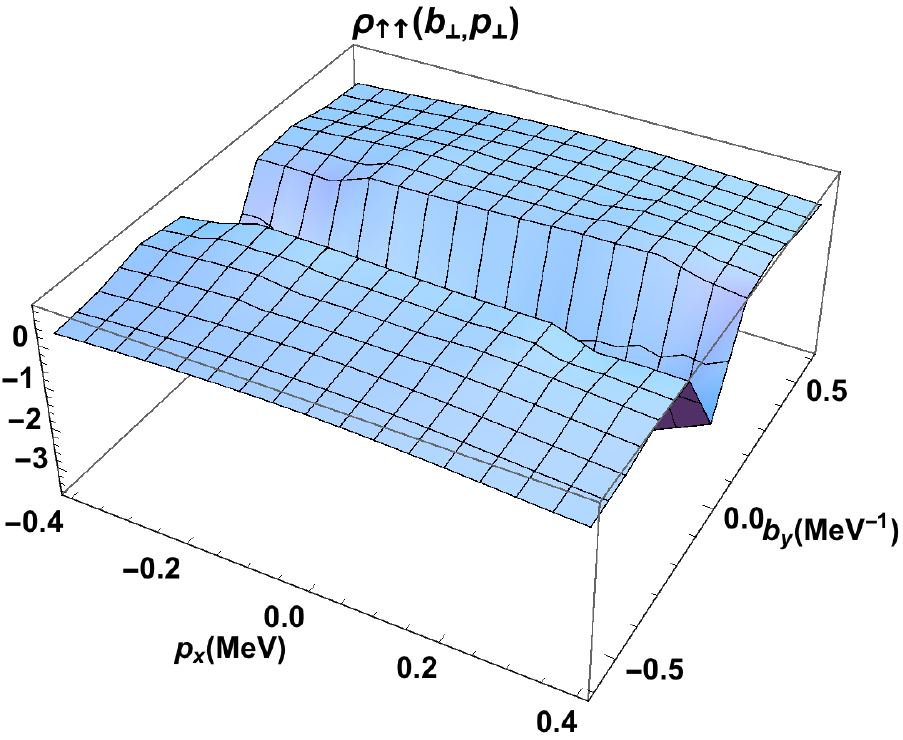}
\\
\small{(d)}
\includegraphics[width=4.5cm,clip]{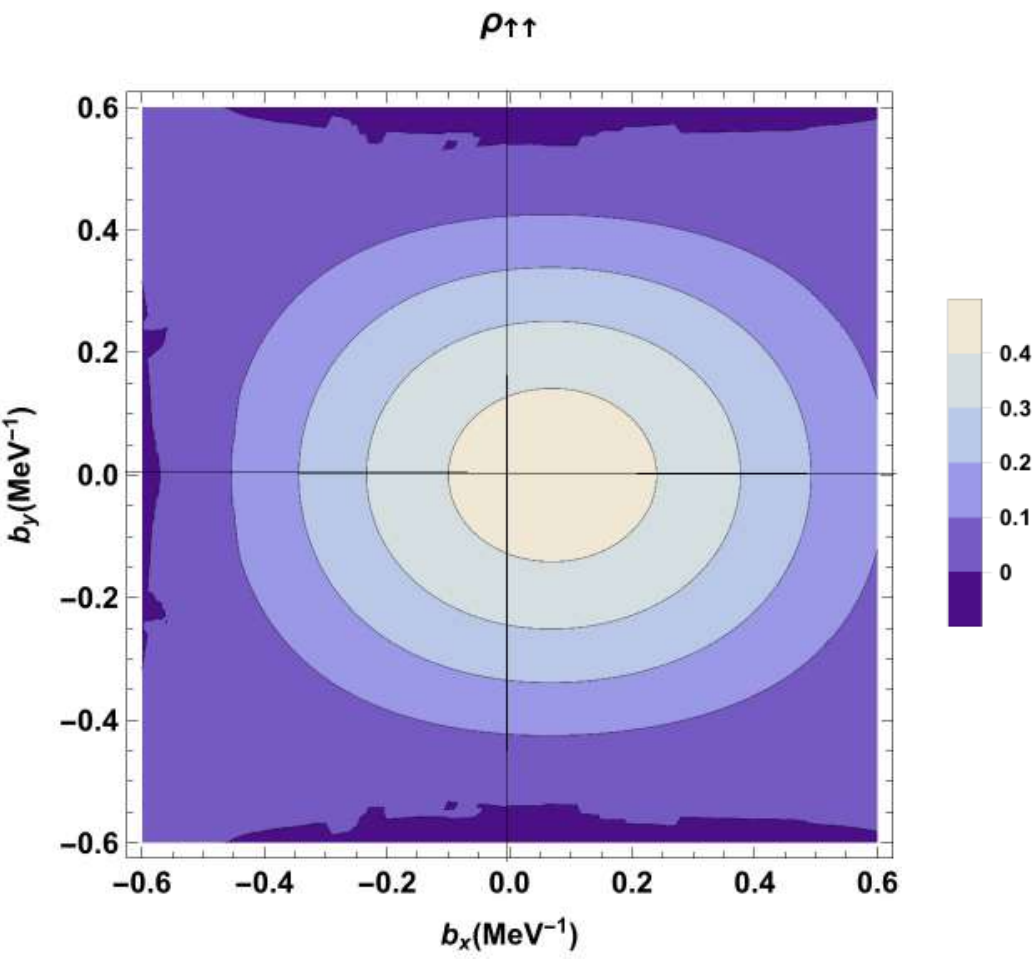}
\hspace{0.1cm}%
\small{(e)}\includegraphics[width=4.5cm,clip]{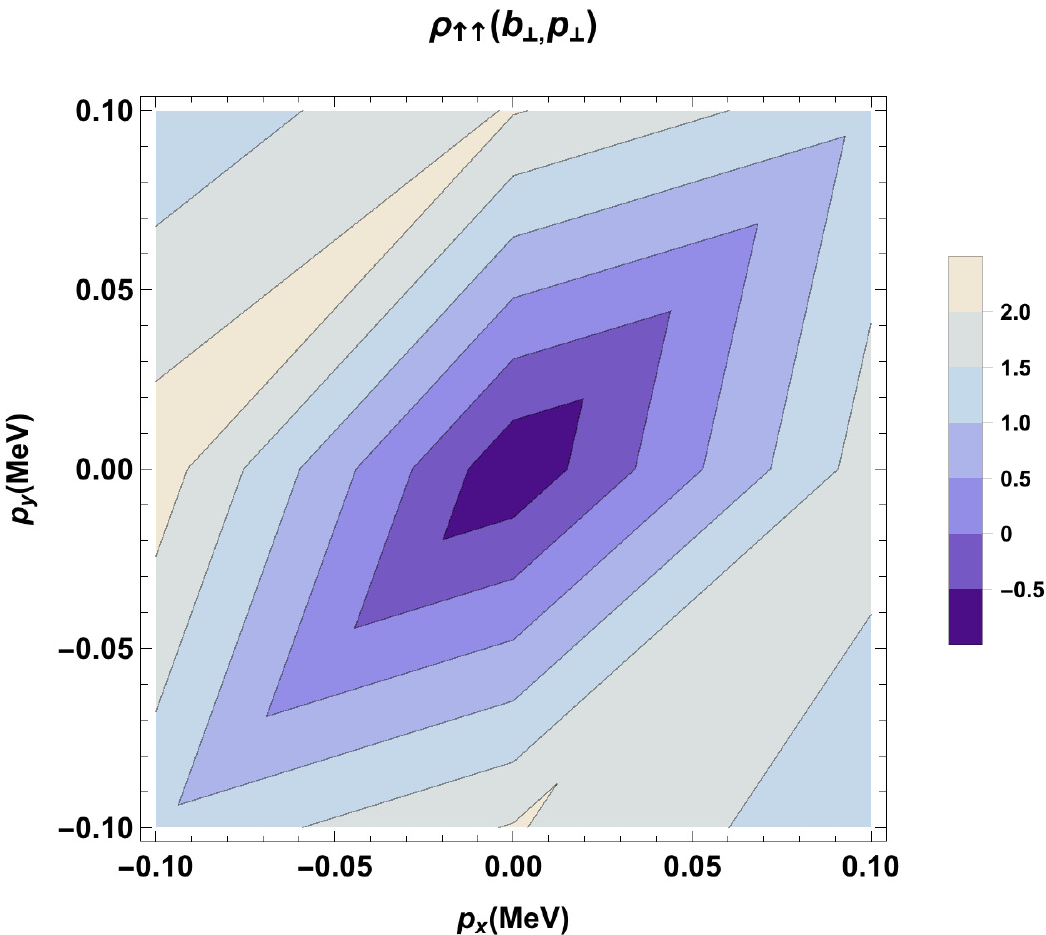}
\hspace{0.1cm}%
\small{(f)}\includegraphics[width=4.5cm,clip]{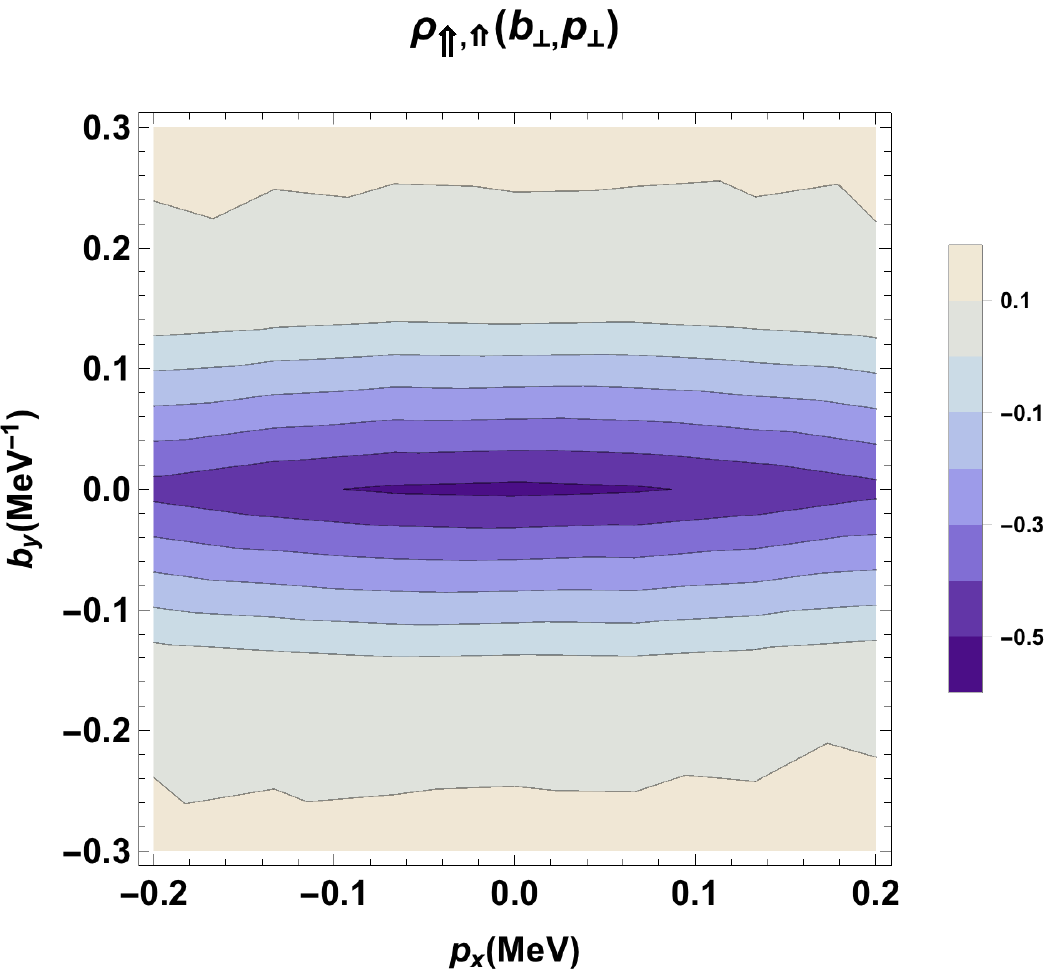}
\end{minipage}
\caption{(color online)  Plots of Wigner distribution $\rho_{\uparrow \uparrow}(\bfb,\bfp)$ for physical electron in impact-parameter space with fixed transverse momentum ${\bf p}_\perp= 0.8 ~MeV$ $\hat{e}_x$ (left panel), in momentum space with fixed impact-parameter ${\bf b}_\perp= 0.8 ~MeV^{-1}$ $\hat{e}_x$ (middle panel) and in mixed plane (right panel).}
  \label{rhoUp_up_long}
\end{figure}
\begin{figure}
\centering
\begin{minipage}[c]{0.98\textwidth}
\small{(a)}
\includegraphics[width=4.5cm,clip]{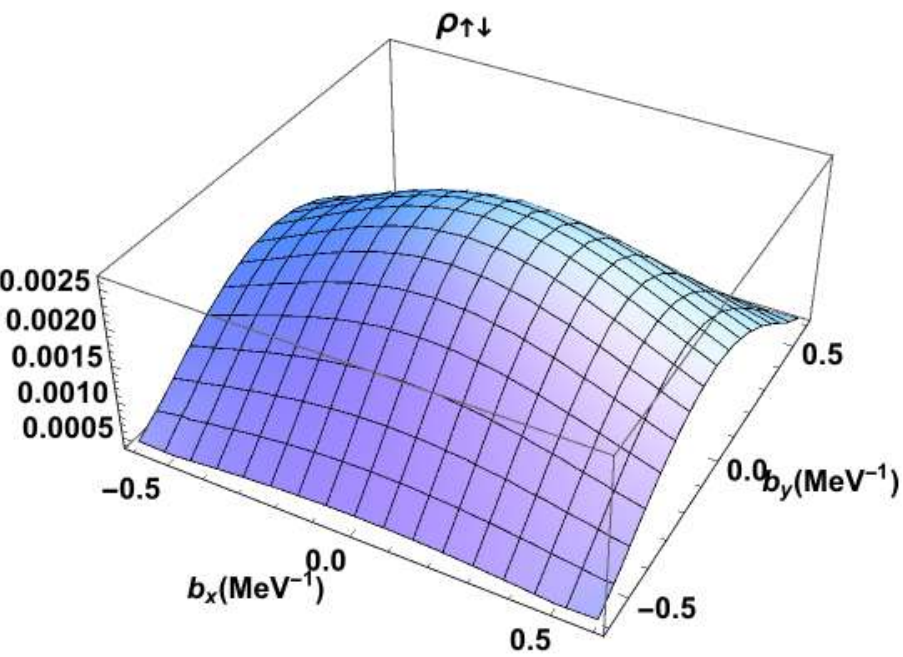}
\hspace{0.1cm}%
\small{(b)}\includegraphics[width=4.5cm,clip]{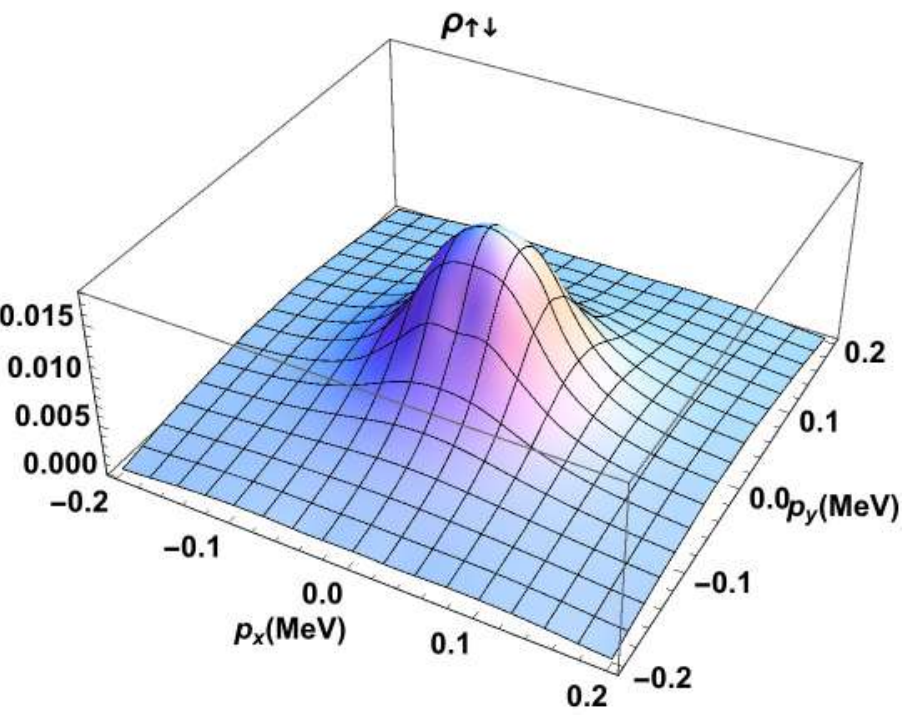}
\hspace{0.1cm}%
\small{(c)}\includegraphics[width=4.5cm,clip]{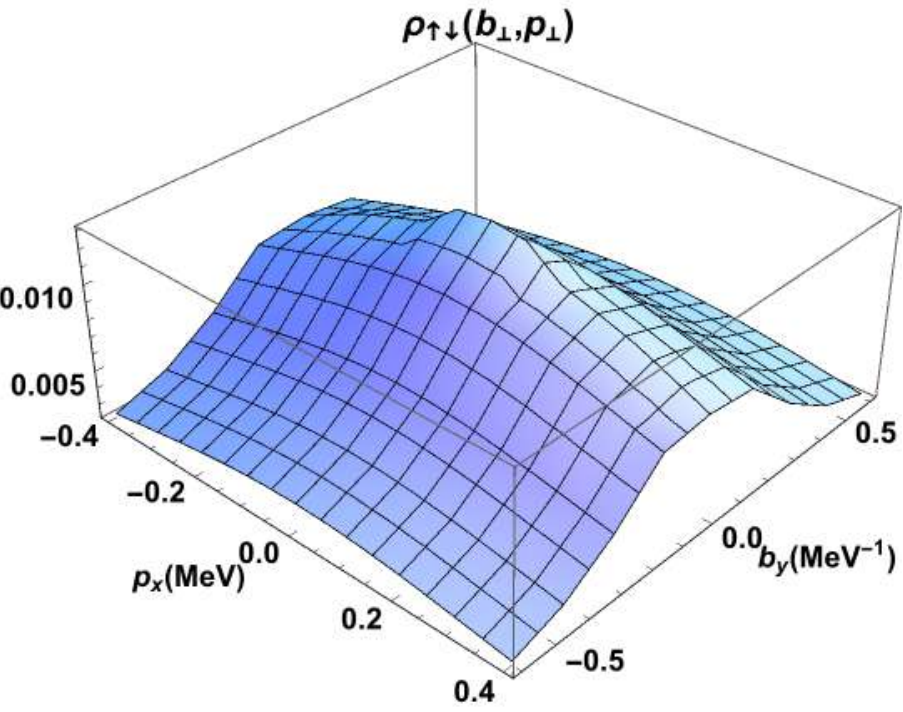}
\\
\small{(d)}\includegraphics[width=4.5cm,clip]{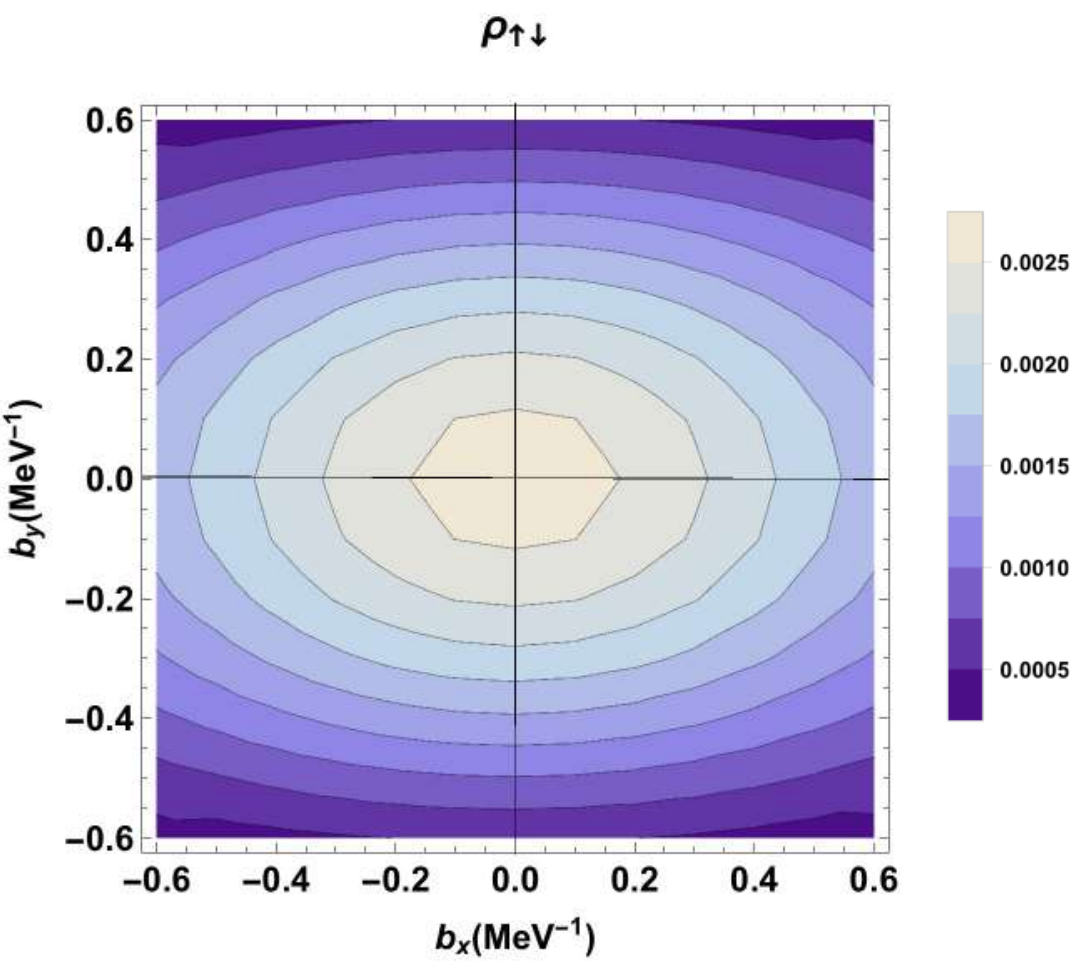}\hfill
\small{(e)}\includegraphics[width=4.5cm,clip]{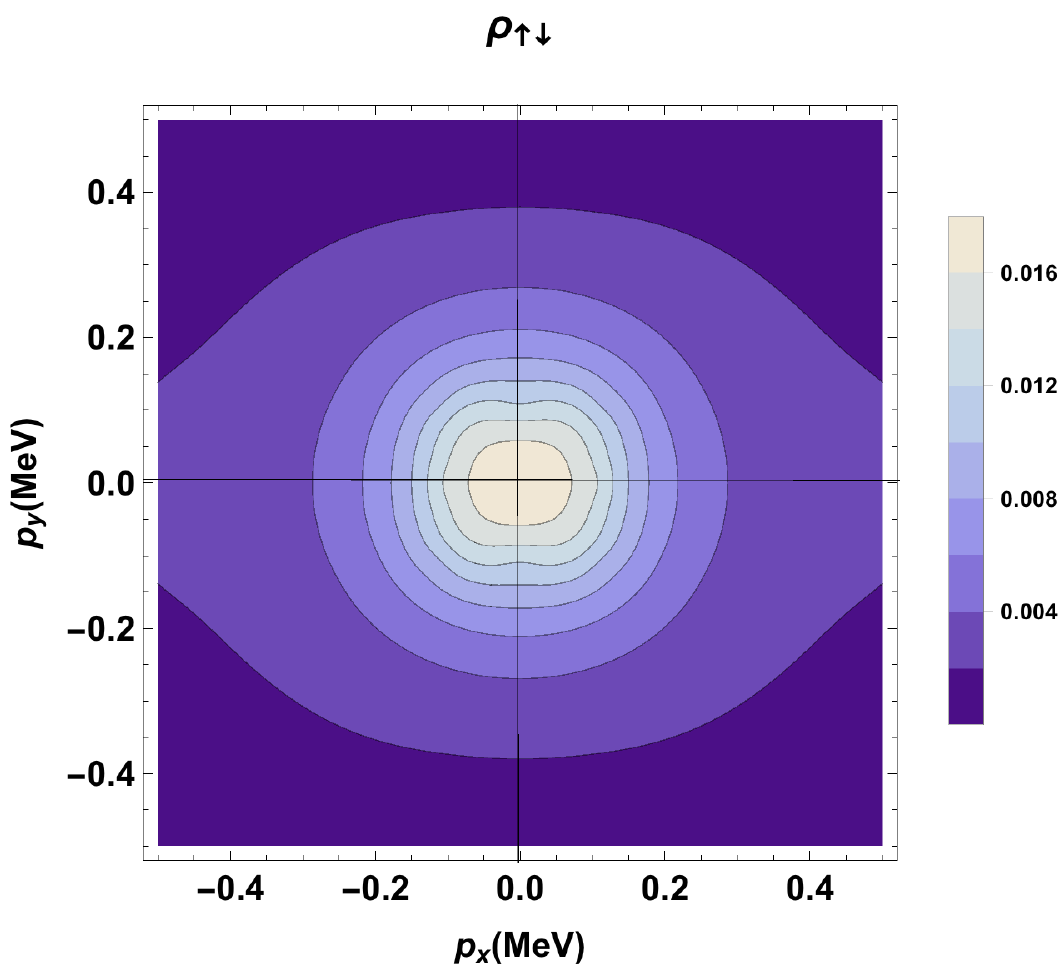}\hfill
\small{(f)}\includegraphics[width=4.5cm,clip]{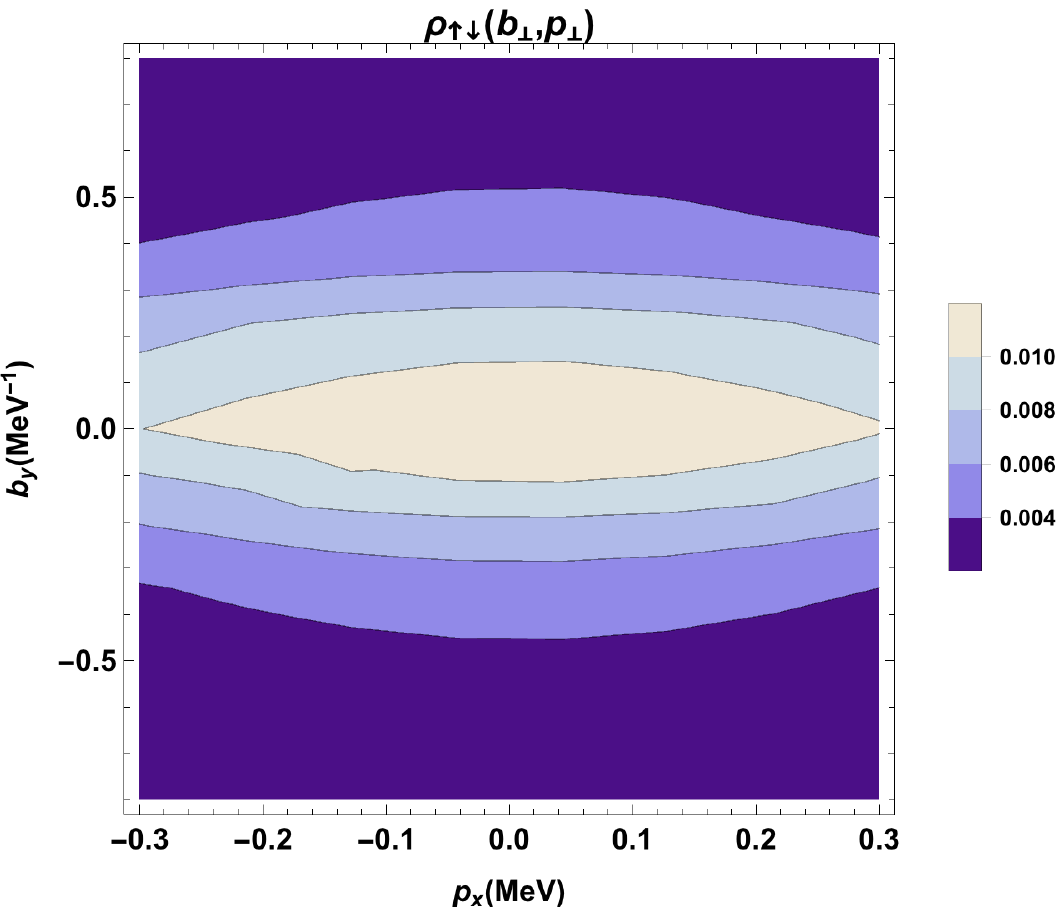}
\end{minipage}
\caption{(color online) Plots of Wigner distribution $\rho_{\uparrow \downarrow}(\bfb,\bfp)$ for physical electron in impact-parameter space with fixed transverse momentum ${\bf p}_\perp= 0.8 ~MeV$ $\hat{e}_x$ (left panel), in momentum plane with fixed impact-parameter ${\bf b}_\perp= 0.8 ~MeV^{-1}$ $\hat{e}_x$ (middle panel) and in mixed plane (right panel).}
\label{rhoUp_down_long}
\end{figure}
In Fig. \ref{rhoLT_TL}, we present the longitudinal-transverse $\rho_{LT}$ Wigner distributions.
$\rho_{LT}$ describes the correlation between transverse spin of bare electron and longitudinal spin of physical electron. This distribution vanishes when the transverse momentum of the bare electron is perpendicular to its polarization. This illustrates that the transverse spin of bare electron has strong correlation only with the transverse momentum parallel to its polarization. In impact-parameter plane, we observe a peak at $b_x=b_y=0$ whereas in momentum plane distribution is dipolar in nature. In mixed-plane, it again shows a dipolar structure. An analogous result has also been observed in a QCD spectator model \cite{Liu:2015eqa} and in dressed quark model \cite{More:2017zqq}.
\section{ Spin spin correlations}\label{spin_spin}
The Wigner distribution with the composite system helicity $\Lambda$ and fermion constituent helicity $\lambda$ is defined for $\Gamma=\gamma^+ \frac{1+\lambda\gamma^5}{2} $ and $\vec{S}=\Lambda \hat{S}_z$ as \cite{Lorce:2011kd}
\begin{eqnarray}
\rho_{\Lambda \lambda}(\bfb,\bfp,x)&=&\frac{1}{2}[\rho^{[\gamma^+]}(\bfb,\bfp,x;\Lambda \hat{S}_z) + \lambda \ \rho^{[\gamma^+\gamma^5]}(\bfb,\bfp,x;\Lambda \hat{S}_z)],
\label{rho_Lamlam0}
\end{eqnarray}
which can be decomposed as
\begin{eqnarray}
\rho_{\Lambda \lambda}(\bfb,\bfp,x)&=&\frac{1}{2}[\rho_{UU}(\bfb,\bfp,x) +\Lambda \ \rho_{LU}(\bfb,\bfp,x)+ \lambda \ \rho_{UL}(\bfb,\bfp,x) + \nonumber\\
&& \Lambda \ \lambda \ \rho_{LL}(\bfb,\bfp,x)],
\label{rho_Lamlam}
\end{eqnarray}
with $\Lambda=\uparrow, \downarrow $ and  $\lambda=\uparrow,\downarrow$ where $\uparrow$ and $\downarrow$ are corresponding to  $+1$ and $-1$ for longitudinal polarizations respectively. In this model $\rho_{UL}=\rho_{LU}$, thus,
for $\Lambda=\uparrow$ and $\lambda=\uparrow$
\begin{equation}
\rho_{\uparrow \uparrow}= \frac{1}{2}[\rho_{UU}+2 \rho_{LU}+\rho_{LL}],
\end{equation}
and for $\Lambda=\uparrow$ and $\lambda=\downarrow$
\begin{equation}
\rho_{\uparrow \downarrow}=\frac{1}{2}[\rho_{UU}-\rho_{LL}].
\label{spin-up-down}
\end{equation}
Similarly, one can also consider the case when composite system has transverse polarization $\Lambda_T= \Uparrow, \Downarrow$ and the fermion constituent has transverse polarization $\lambda_T=\Uparrow, \Downarrow$. The expression of Wigner distribution for this situation can be written as \cite{Chakrabarti:2017teq}
\begin{eqnarray}
\rho_{\Lambda_T \lambda_T}&=& \frac{1}{2} [\rho^{[\gamma^+]}({\bfb},{\bfp},x;\Lambda_T \hat{e}_i) + \Lambda_T \ \rho^{[i\sigma^{j+} \gamma_5]}({\bfb},{\bfp},x;\lambda_T \hat{e}_i)],
\end{eqnarray}
which can be decomposed as
\begin{eqnarray}
\rho_{\Lambda_T \lambda_T}({\bfb},{\bfp},x)&=&\frac{1}{2} \big[\rho_{UU}({\bfb},{\bfp},x)+\Lambda_T \ \rho_{TU}({\bfb},{\bfp},x)
+\lambda_T \ \rho_{UT}({\bfb},{\bfp},x)+\nonumber\\
&& \lambda_T \ \Lambda_T \ \rho_{TT}({\bfb},{\bfp},x) \big].
\label{rho_LambdaT_lambdaT}
\end{eqnarray}
We can also define the Wigner distribution for longitudinally polarized fermion constituent in transversely polarized composite system,
$\rho_{\Lambda_T \lambda}({\bfb},{\bfp},x)$ and transversely polarized fermion constituent in
 a longitudinally polarized composite system,
$\rho_{\Lambda \lambda_T}({\bfb},{\bfp},x)$ as
\begin{eqnarray}
\rho_{\Lambda_T \lambda}({\bfb},{\bfp},x)&=&\frac{1}{2} [\rho_{UU}({\bfb},{\bfp},x)+ \Lambda_T \ \rho_{TU}({\bfb},{\bfp},x)+ \lambda \ \rho_{UL}({\bfb},{\bfp},x)+\nonumber\\
&& \Lambda_T ~\lambda \ \rho_{TL}({\bfb},{\bfp},x)],
\end{eqnarray}
\begin{eqnarray}
\rho_{\Lambda \lambda_T}({\bfb},{\bfp},x)
&=&\frac{1}{2}[\rho_{UU}({\bfb},{\bfp},x)+ \Lambda \ \rho_{LU}({\bfb},{\bfp},x)+ \lambda_T \ \rho_{UT}({\bfb},{\bfp},x)+ \nonumber\\
&& \Lambda ~\lambda_T \ \rho_{LT}({\bfb},{\bfp},x)].
\end{eqnarray}
$\rho_{\Lambda \lambda}(\bfb,\bfp,x)$ provides information about the correlations between the spin of composite system and the spin of bare electron in the longitudinal direction. The Wigner distributions, $\rho_{\Lambda \lambda}(\bfb,\bfp,x)$ with the polarization of composite system $\Lambda=\uparrow$ and bare electron polarization $\lambda=\uparrow$ are shown in Fig.\ref{rhoUp_up_long} which implies that the physical electron is align parallel to the bare electron. The lower panel represents top view of the same distributions. We observe that in impact-parameter plane, the peak of the distribution is slightly shifted sideways toward the $+b_x$ direction which is due to $\rho_{LU}$, as it is dipole in nature and therefore it distorts the distribution significantly. No significant shifting of peak is observed in  momentum plane but a distortion is observed in the mixed plane.
The results for the spin-spin correlation for the case of $\Lambda=\uparrow$ and $\lambda=\downarrow$, (which implies that  the physical electron is polarized opposite to bare electron) are shown in Fig. \ref{rhoUp_down_long}. In the present case, the effective spin-spin correlation is given by Eq.(\ref{spin-up-down}), i.e., distribution is not getting any contribution from $\rho_{LU}$, thus there is no shifting of peaks.
In the mixed plane representation (Fig. \ref{rhoUp_down_long}(b) and (e)), the peak is symmetric and there is no shifting from central axes.
It should be mentioned here that in the light-front quark diquark model $\rho_{LU}=-\rho_{UL}$ \cite{Chakrabarti:2017teq} which in result does not contribute towards the distortion in $\rho_{\uparrow\uparrow}$ distribution.
\begin{figure}
\centering
\begin{minipage}[c]{0.98\textwidth}
\small{(a)}\includegraphics[width=4.5cm]{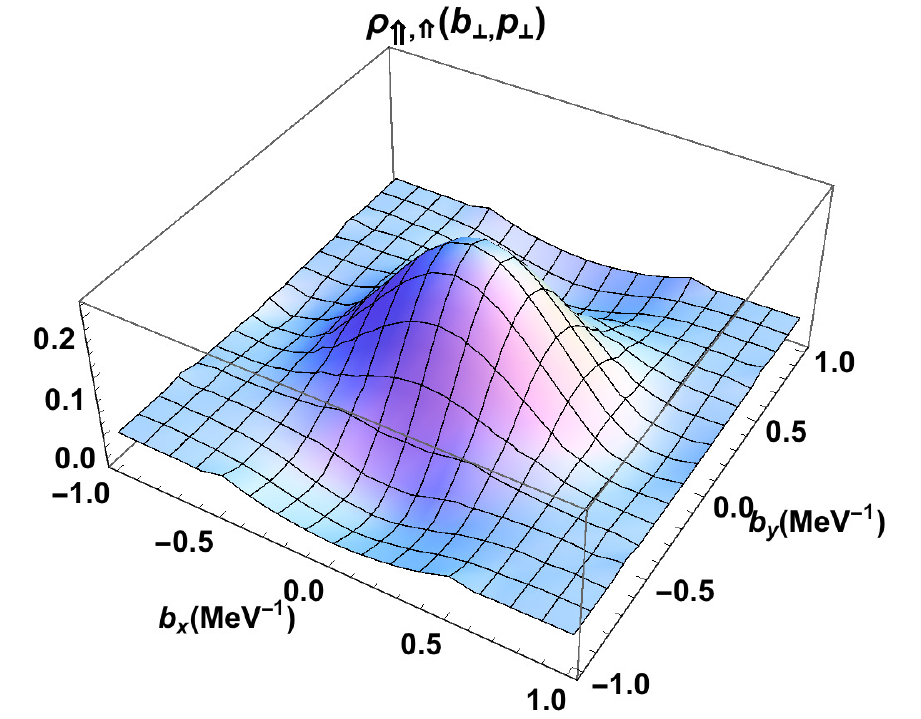}\hfill
\small{(b)}\includegraphics[width=4.5cm]{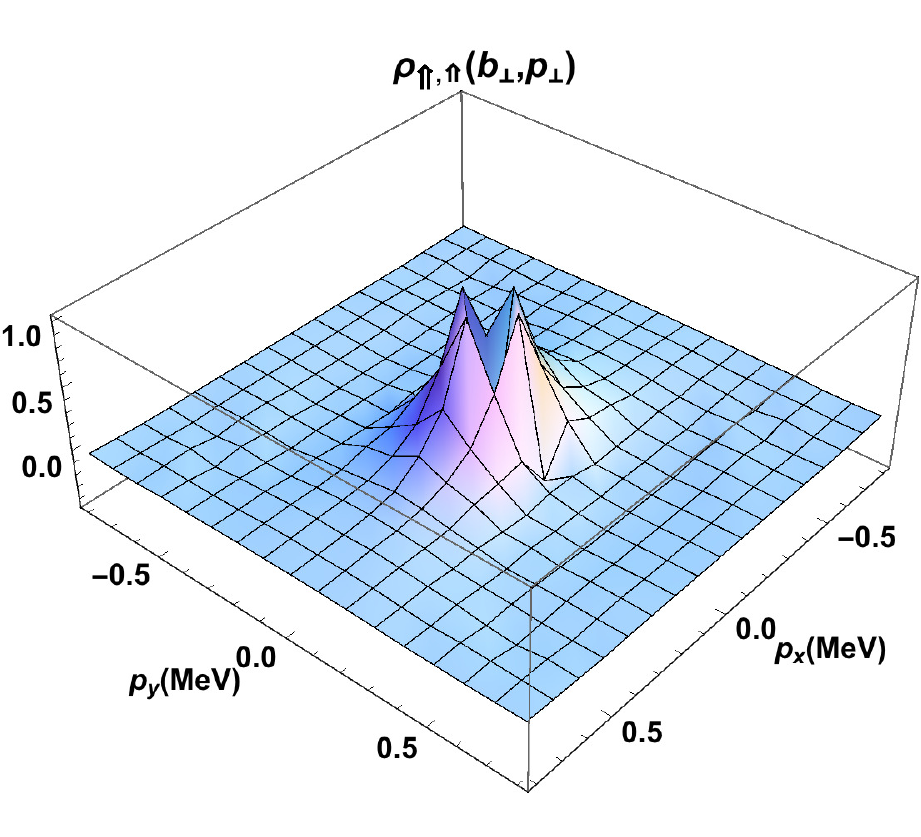}\hfill
\small{(c)}\includegraphics[width=4.5cm]{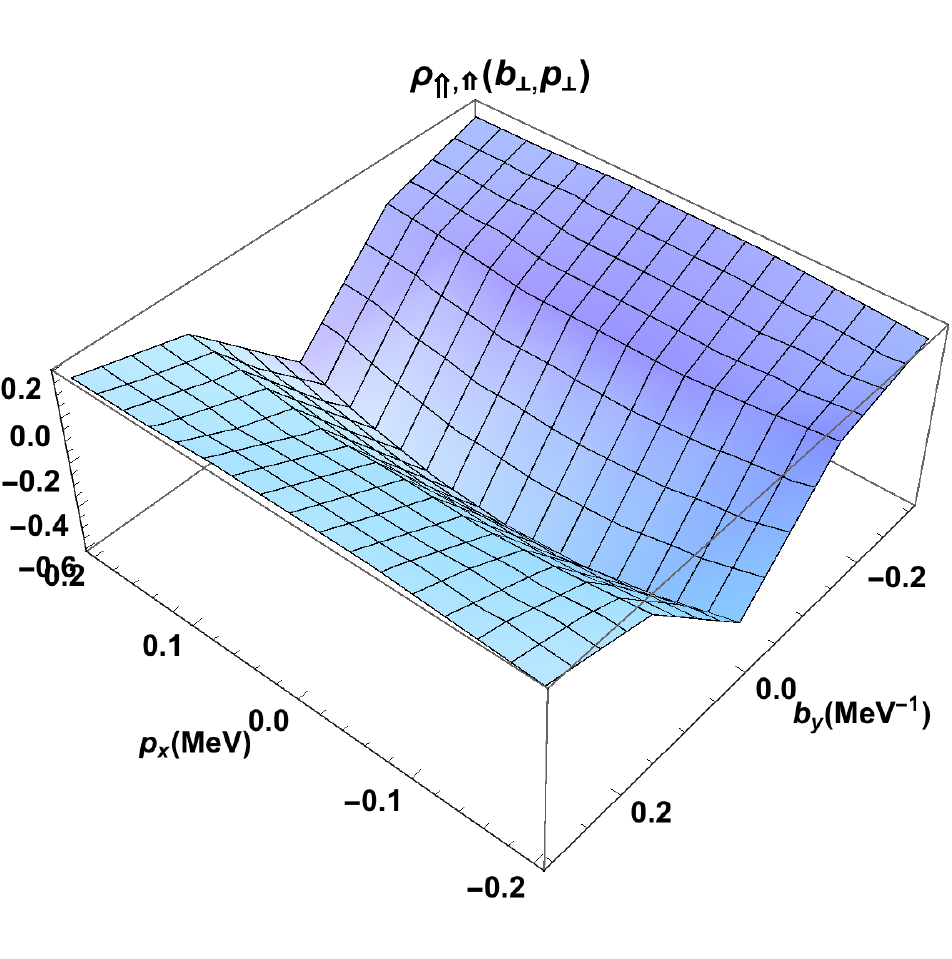}
\\
\small{(d)}\includegraphics[width=4.5cm]{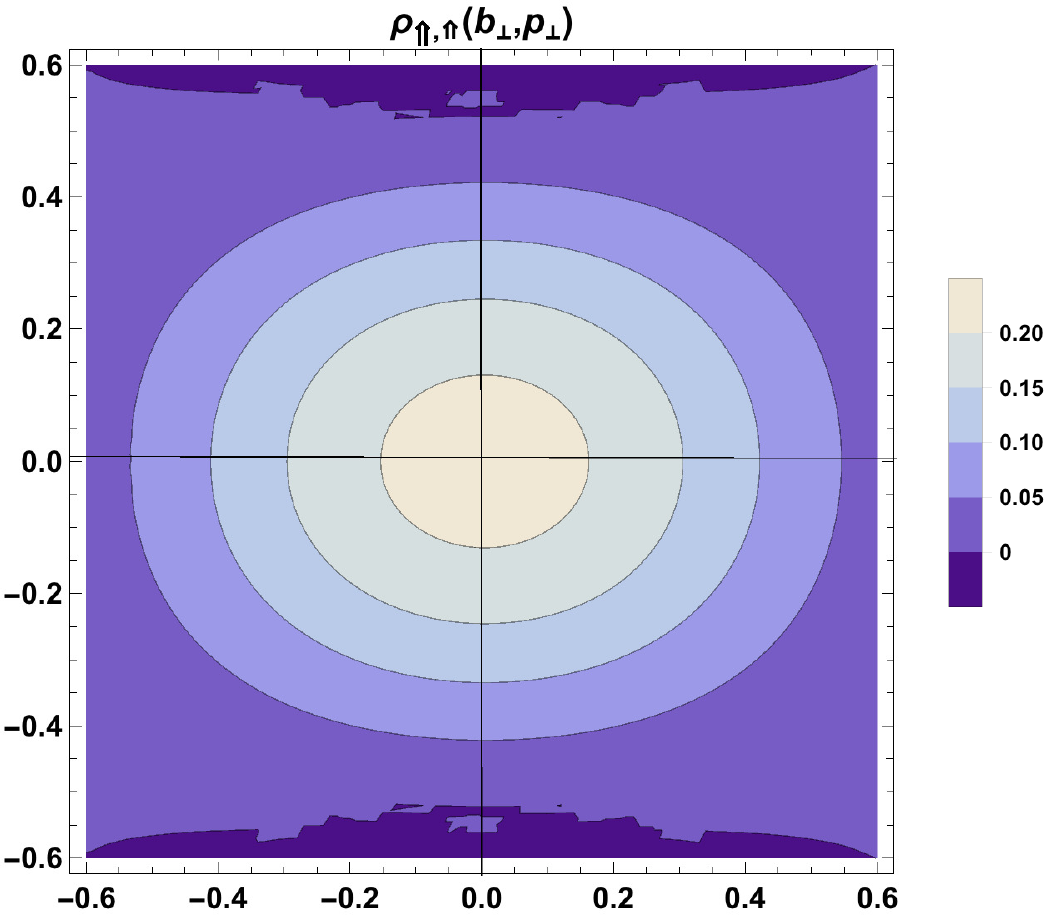}\hfill
\small{(e)}\includegraphics[width=4.5cm]{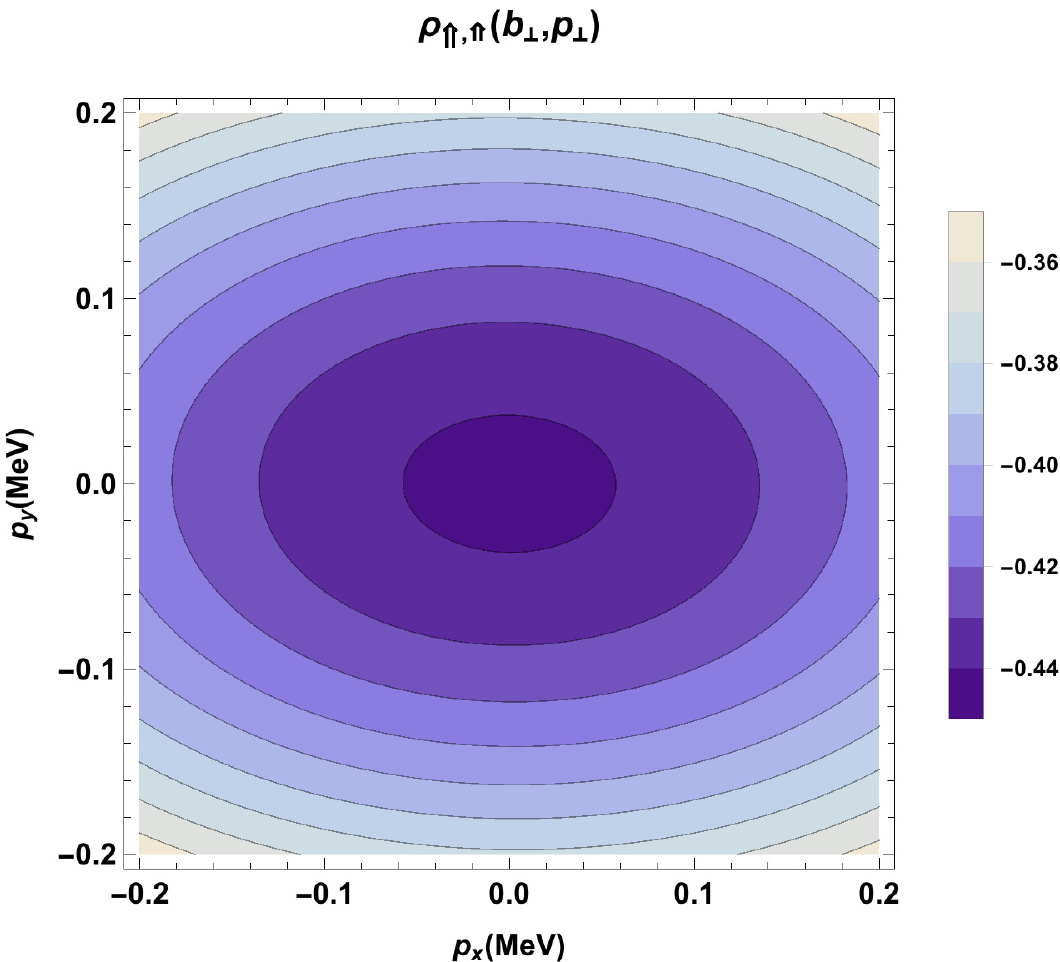}\hfill
\small{(f)}\includegraphics[width=4.5cm]{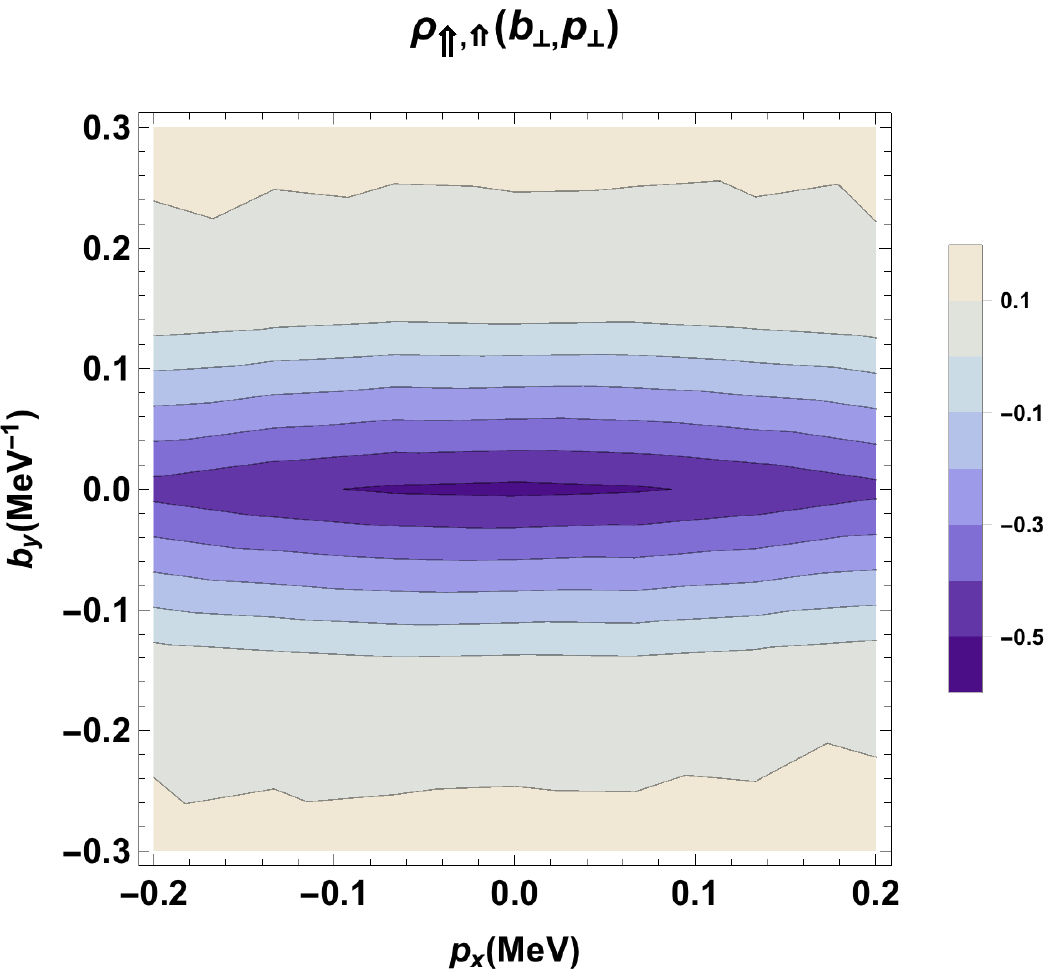}
\end{minipage}
\caption{(color online) Plots of Wigner distribution $\rho_{\Lambda_T=\Uparrow \lambda_T=\Uparrow}(\bfb,\bfp)$ for physical electron in impact-parameter plane with fixed transverse momentum ${\bf p}_\perp= 0.8 ~MeV$ $\hat{e}_x$ (left panel), in momentum plane with fixed impact-parameter ${\bf b}_\perp= 0.8 ~MeV^{-1}$ $\hat{e}_x$ (middle panel) and in mixed plane (right panel).}
\label{rhoUp_up_trans}
\end{figure}
\begin{figure}
\centering
\begin{minipage}[c]{0.98\textwidth}
\small{(a)}\includegraphics[width=4.5cm]{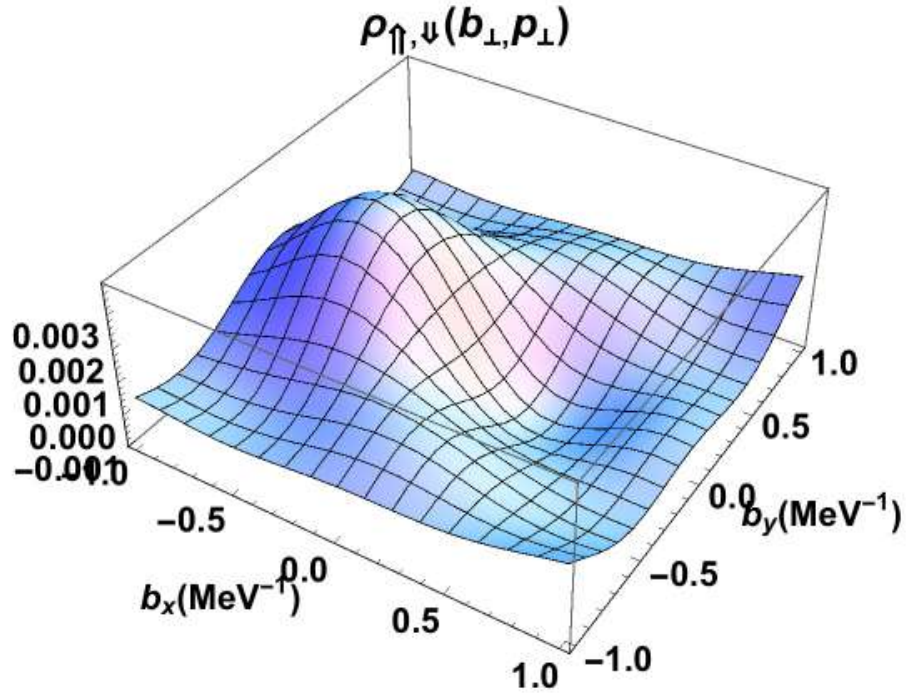}\hfill
\small{(b)}\includegraphics[width=4.5cm]{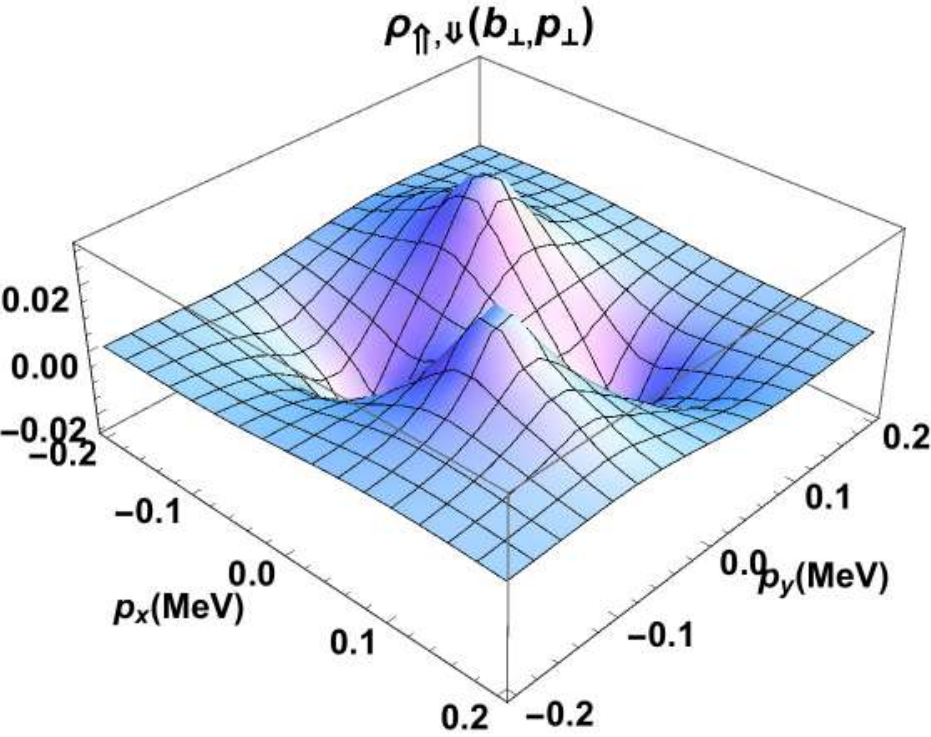}\hfill
\small{(c)}\includegraphics[width=4.5cm]{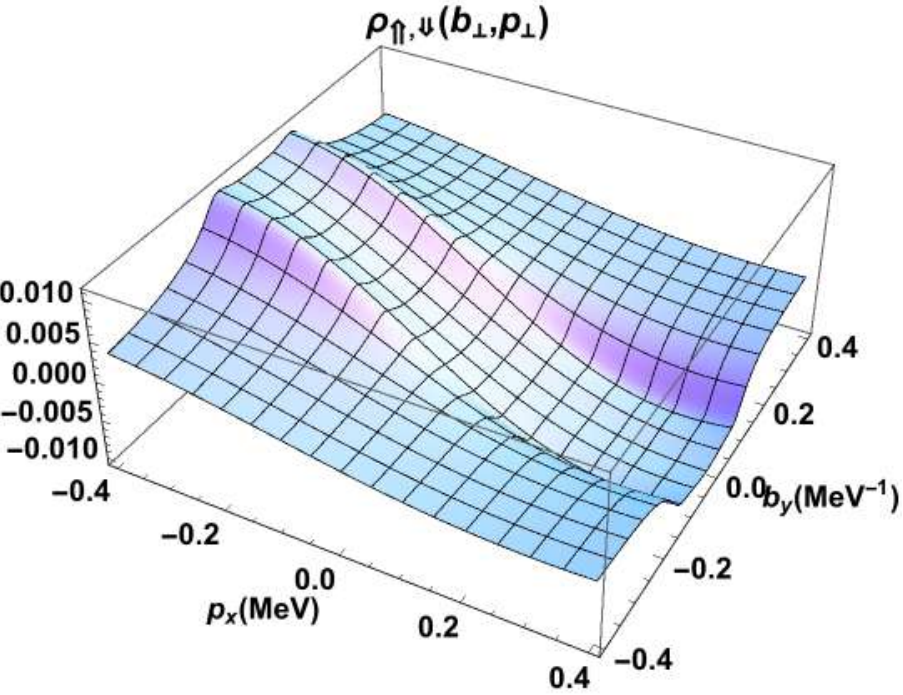}
\\
\small{(d)}\includegraphics[width=4.5cm]{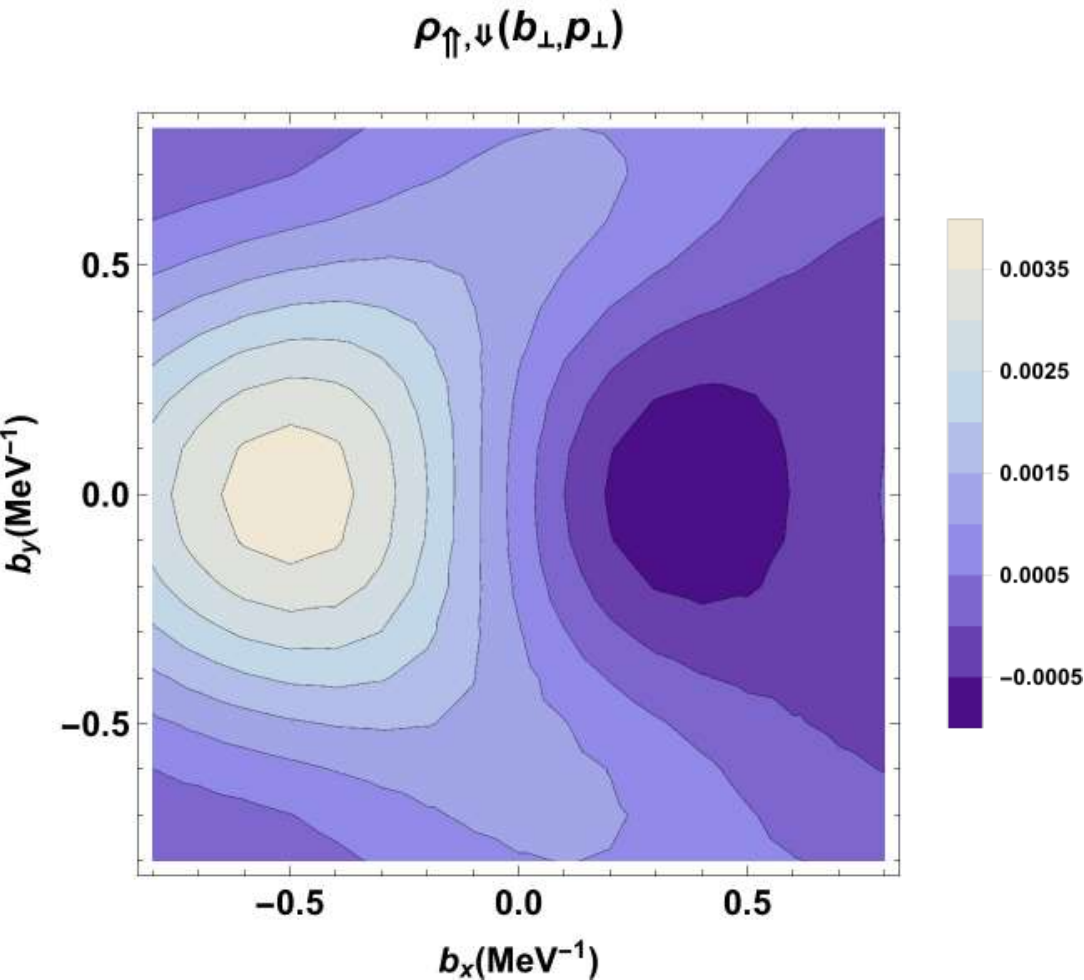}\hfill
\small{(e)}\includegraphics[width=4.5cm]{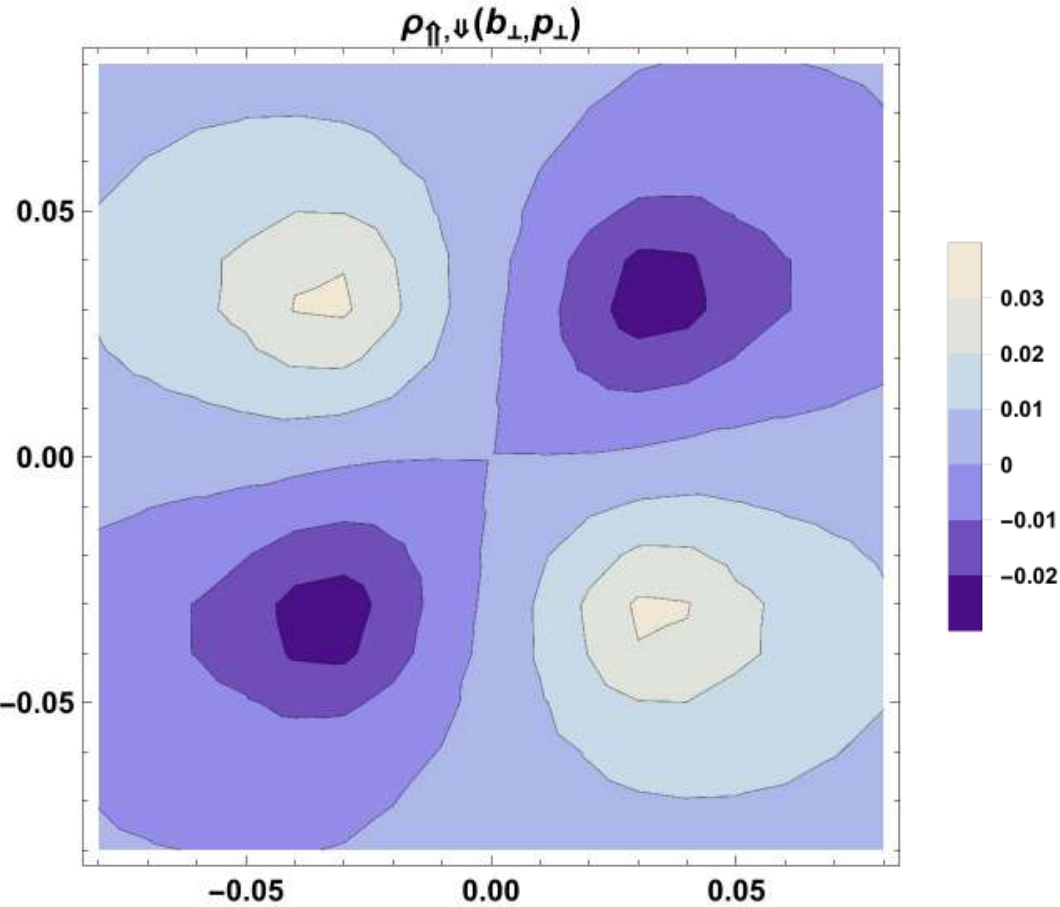}\hfill
\small{(f)}\includegraphics[width=4.5cm]{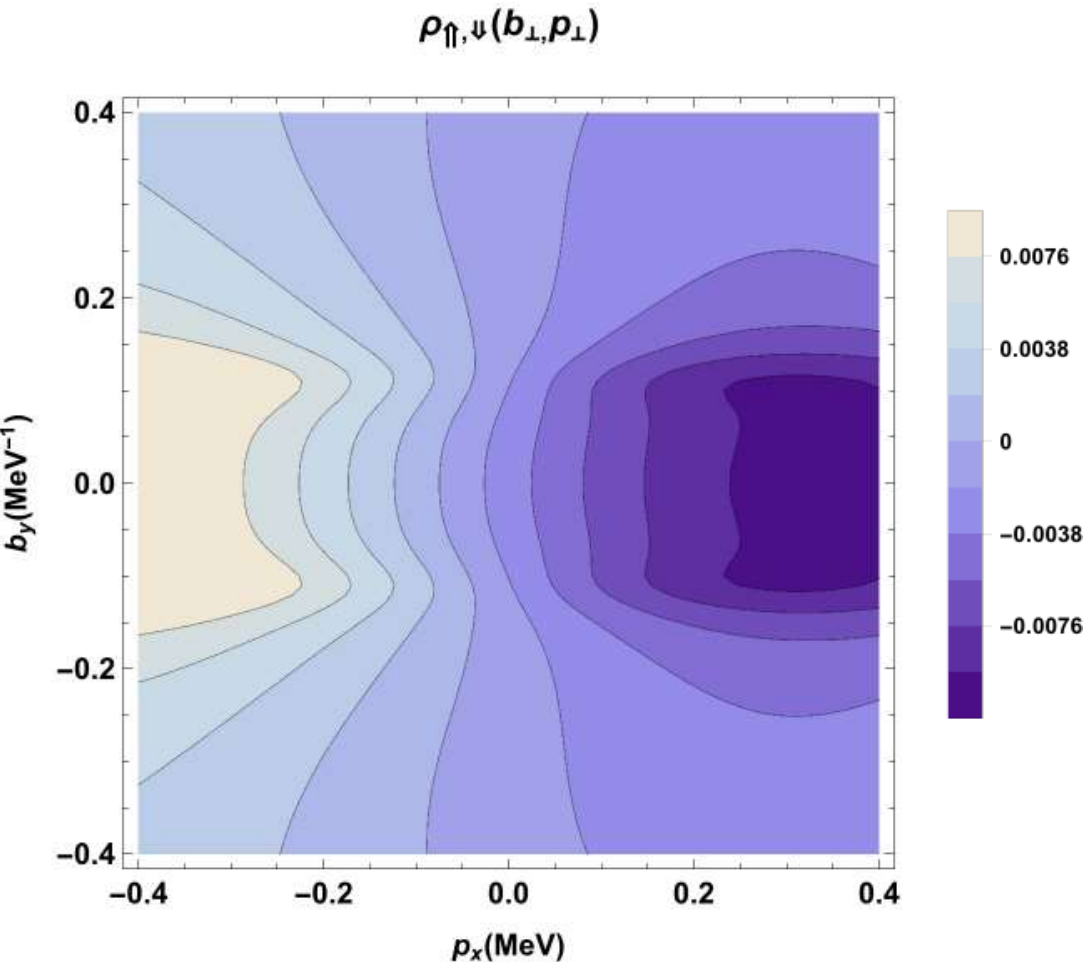}
\end{minipage}
\caption{(color online) Plots of Wigner distribution $\rho_{\Lambda_T=\Uparrow \lambda_T=\Downarrow}(\bfb,\bfp)$ for physical electron in impact-parameter plane with fixed transverse momentum ${\bf p}_\perp= 0.8 ~MeV$ $\hat{e}_x$ (left panel), in momentum plane with fixed impact-parameter ${\bf b}_\perp= 0.8 ~MeV^{-1}$ $\hat{e}_x$ (middle panel) and in mixed plane (right panel).}
\label{rhoUp_down_trans}
\end{figure}
In Fig. \ref{rhoUp_up_trans}, we plot the transverse Wigner distribution $\rho_{\Uparrow \Uparrow}$ using Eq.(\ref{rho_LambdaT_lambdaT}). It represents the spin-spin correlation of a transversely polarized fermion constituent in transversely polarized composite system. This correlation consist of contributions from $\rho_{UU}, \rho_{UT}, \rho_{TU}$ and $\rho_{TT}$. Out of these four contributions, $\rho_{TU}$ vanishes in this model. One finds that in impact-parameter plane, the distribution is symmetric about the center of the plane as both $\rho_{UU}$ and $\rho_{TT}$ are symmetric whereas $\rho_{UT}$ is also contributing which is dipolar in nature but with very small magnitude and therefore does not contribute to provide any distortion. In momentum plane the distribution has a sharp peak in negative direction but no shifting of peak has been obtained.
Effective contributions of $\rho_{UU}, \rho_{UT}$ and $\rho_{TT}$ towards the mixed distributions is shown in Fig. \ref{rhoUp_up_trans}(c).
The Wigner distribution, $\rho_{\Uparrow \Downarrow}$ which corresponds the spin-spin correlation of a transversely polarized fermion constituent with $\lambda_T=\Downarrow$ in transversely polarized composite system with $\Lambda_T=\Uparrow$ has been shown in Fig. \ref{rhoUp_down_trans}. In this case, again $\rho_{TU}$ vanishes. The qualitative nature of $\rho_{UU}$ and $\rho_{TT}$ are the same but due to the subtraction they cancel each other and therefore $\rho_{UT}$ plays vital role in providing the dipole nature of the distribution in impact-parameter plane. Howevere, $\rho_{\Uparrow \Downarrow}$ exhibit the quadrupole structure in momentum plane.  In mixed plane, we obtained dipole structure of the distribution.
\begin{figure}
\centering
\begin{minipage}[c]{0.98\textwidth}
\small{(a)}\includegraphics[width=4.5cm]{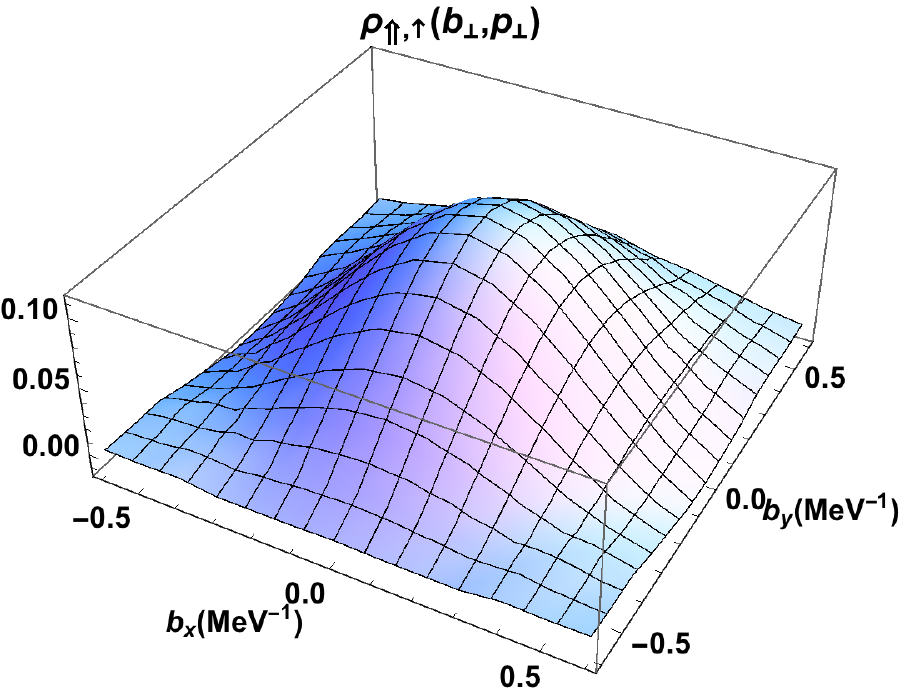}\hfill
\small{(b)}\includegraphics[width=4.5cm]{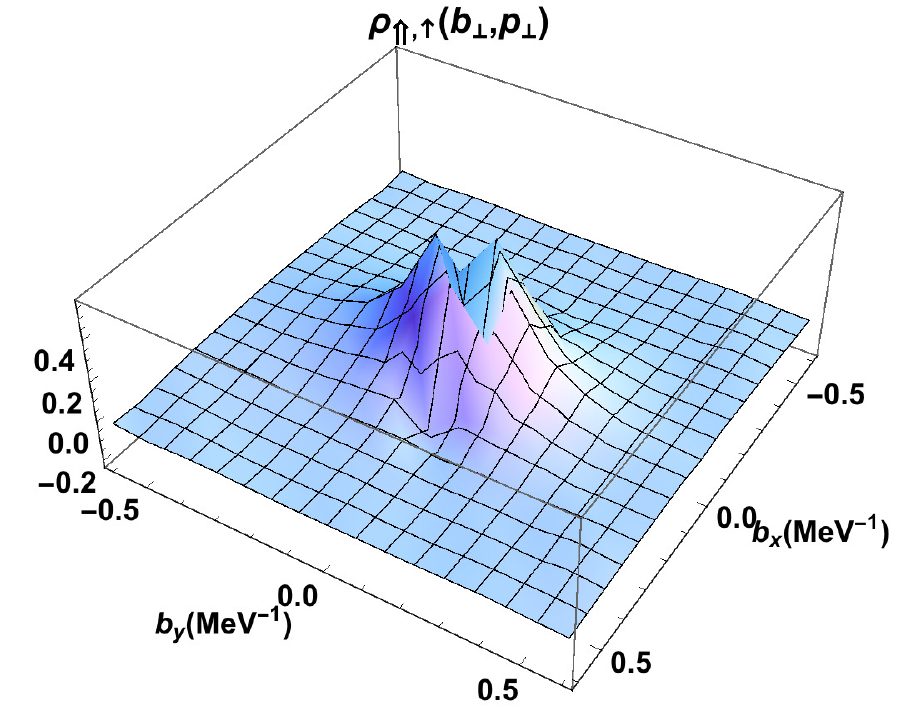}\hfill
\small{(c)}\includegraphics[width=4.5cm]{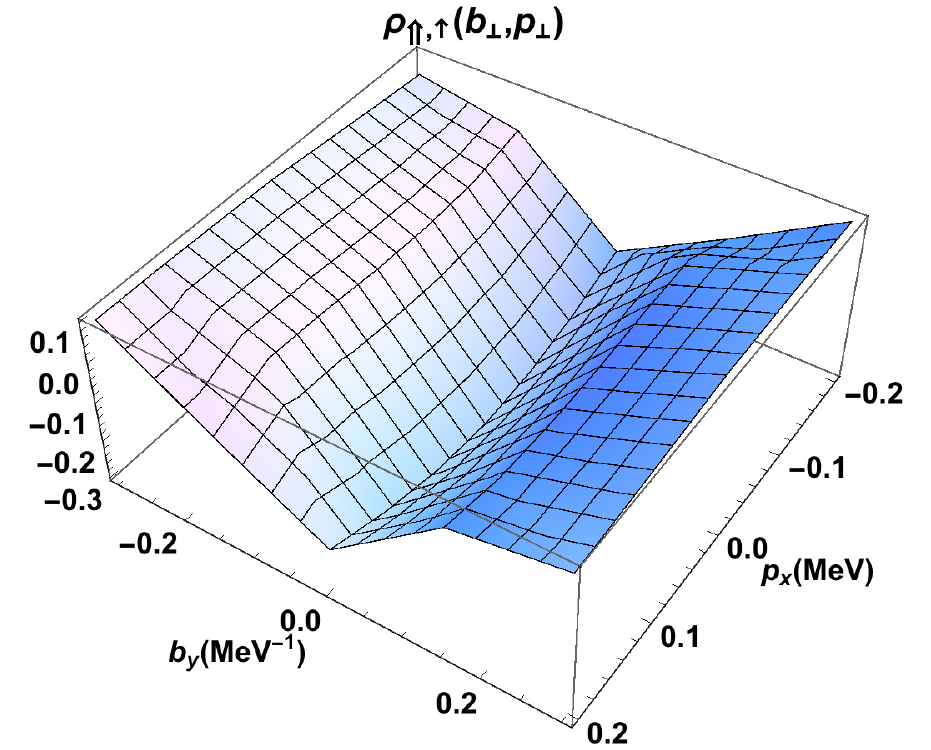}
\\
\small{(d)}\includegraphics[width=4.5cm]{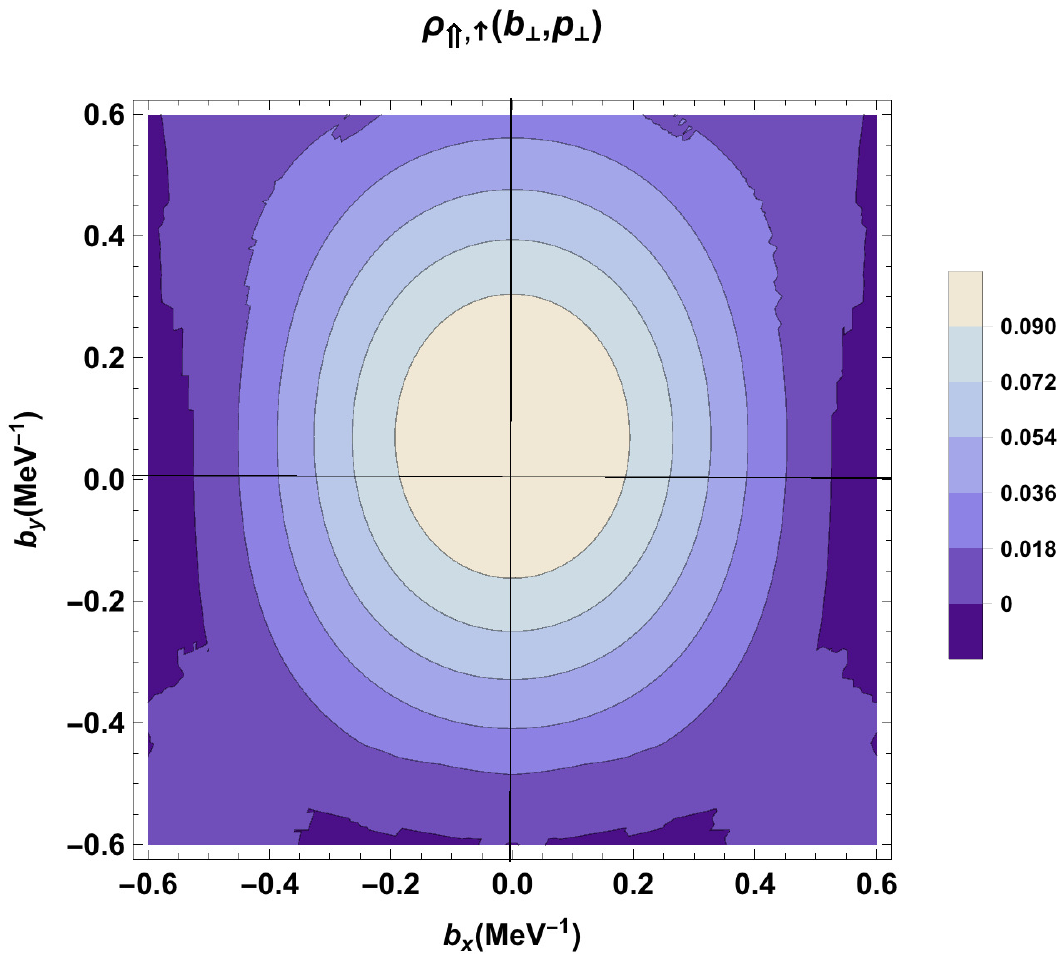}\hfill
\small{(e)}\includegraphics[width=4.5cm]{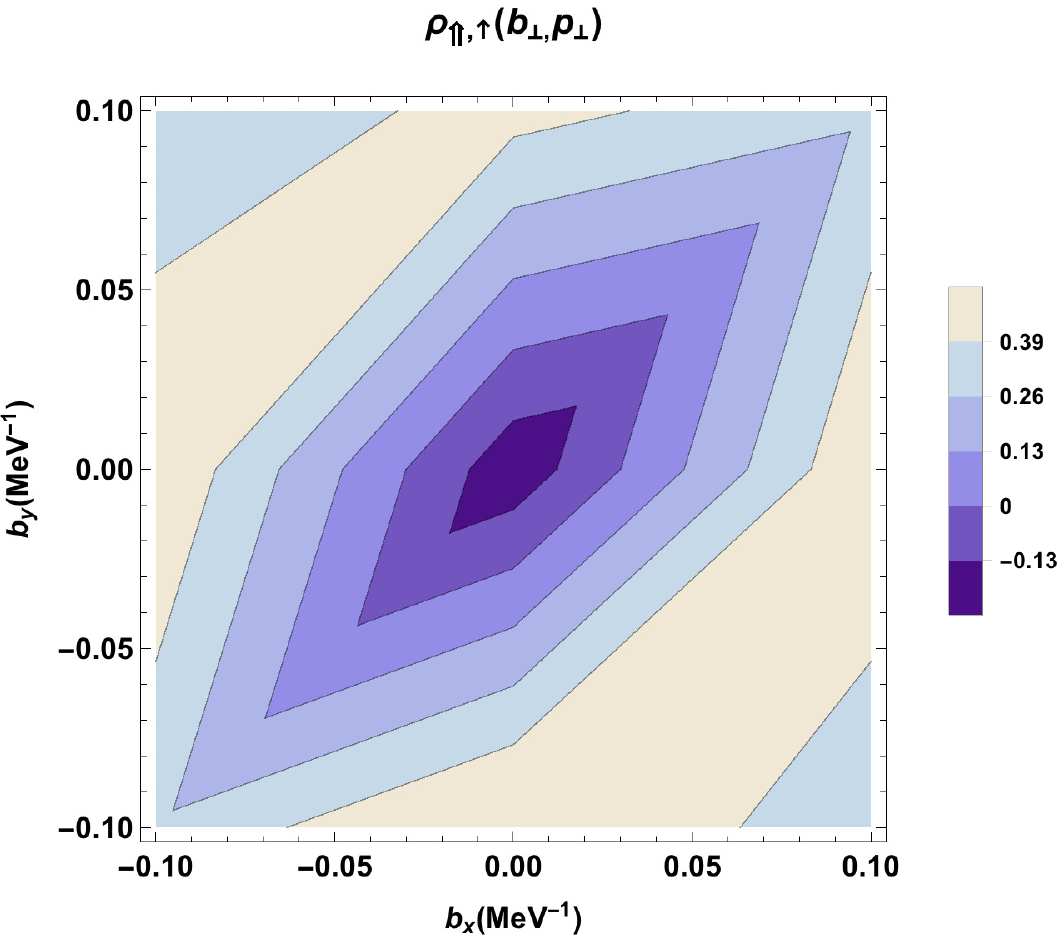}\hfill
\small{(f)}\includegraphics[width=4.5cm]{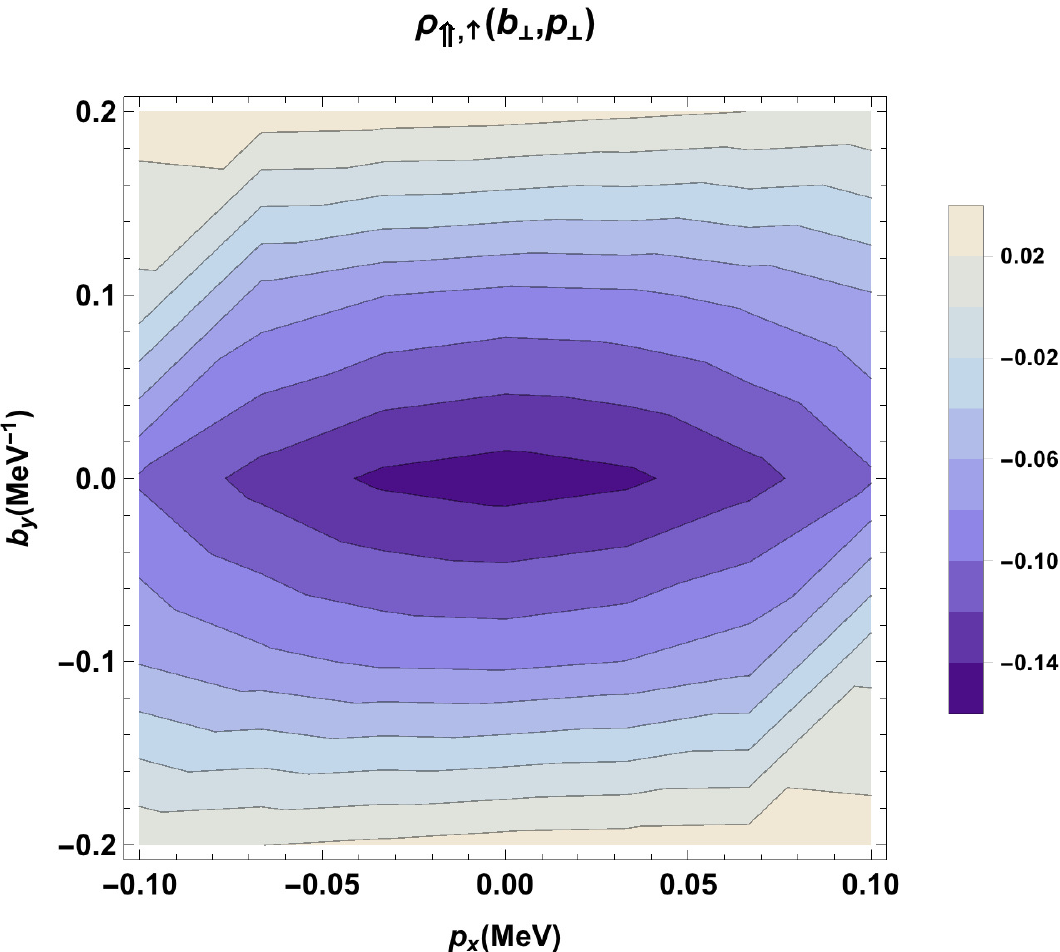}
\end{minipage}
\caption{(color online) Plots of Wigner distribution $\rho_{\Lambda_T=\Uparrow \lambda=\uparrow}(\bfb,\bfp)$ for physical electron in impact-parameter plane with fixed transverse momentum ${\bf p}_\perp= 0.8 ~MeV$ $\hat{e}_y$ (left panel), in momentum plane with fixed impact-parameter ${\bf b}_\perp= 0.8 ~MeV^{-1}$ $\hat{e}_y$ (middle panel) and in mixed plane (right panel).}
\label{rhoUp_up_trans_long}
\end{figure}
\begin{figure}
\centering
\begin{minipage}[c]{0.98\textwidth}
\small{(a)}\includegraphics[width=4.5cm]{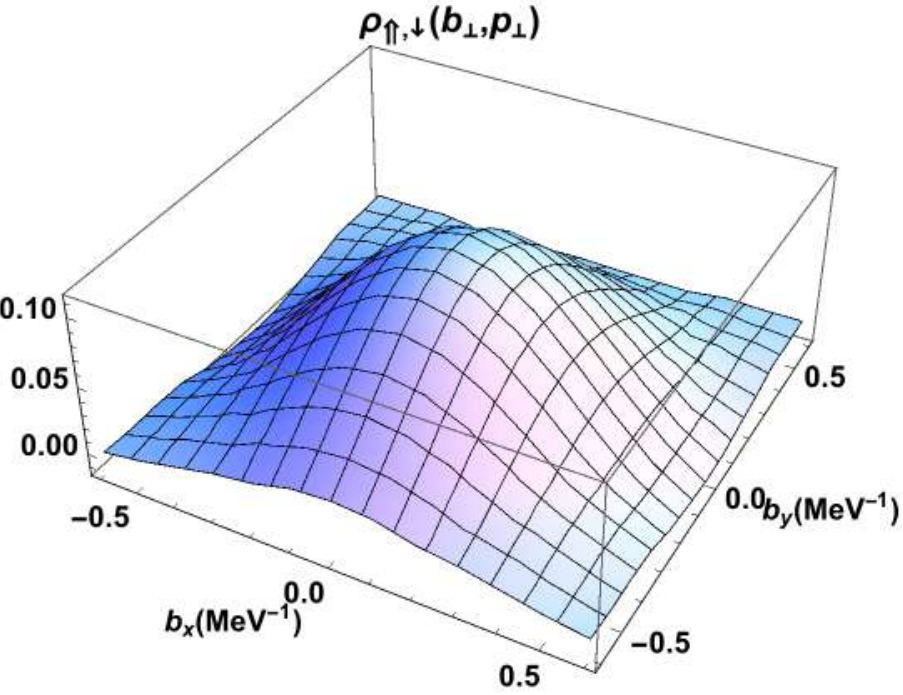}\hfill
\small{(b)}\includegraphics[width=4.5cm]{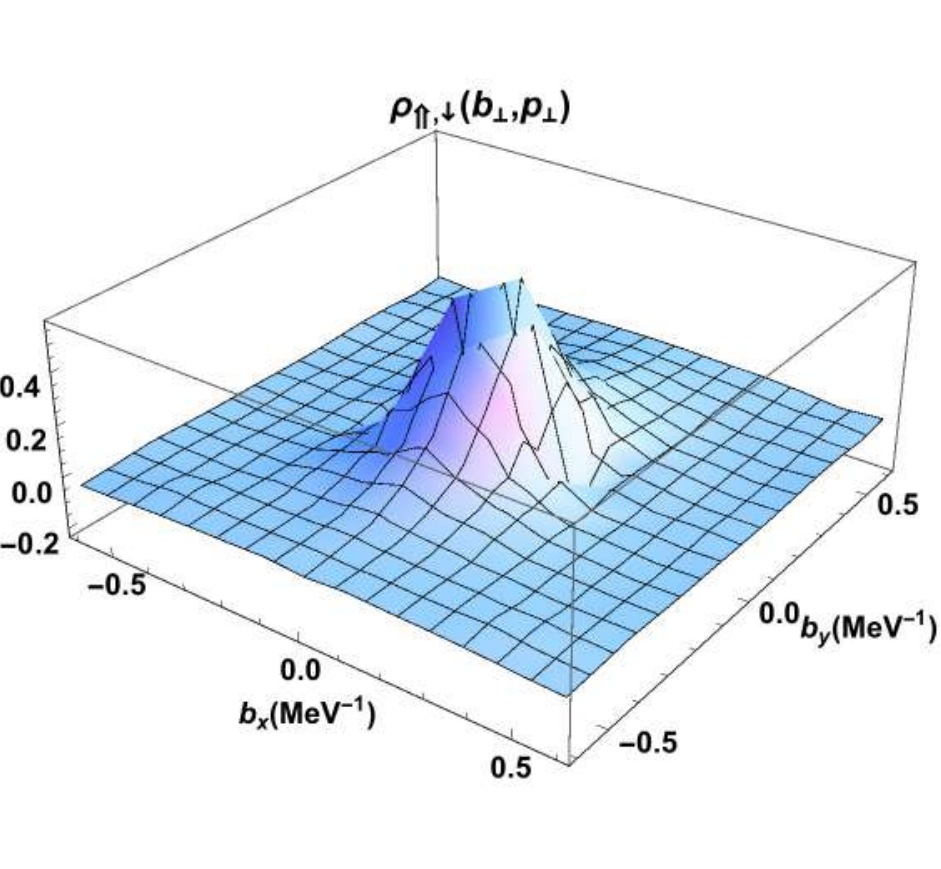}\hfill
\small{(c)}\includegraphics[width=4.5cm]{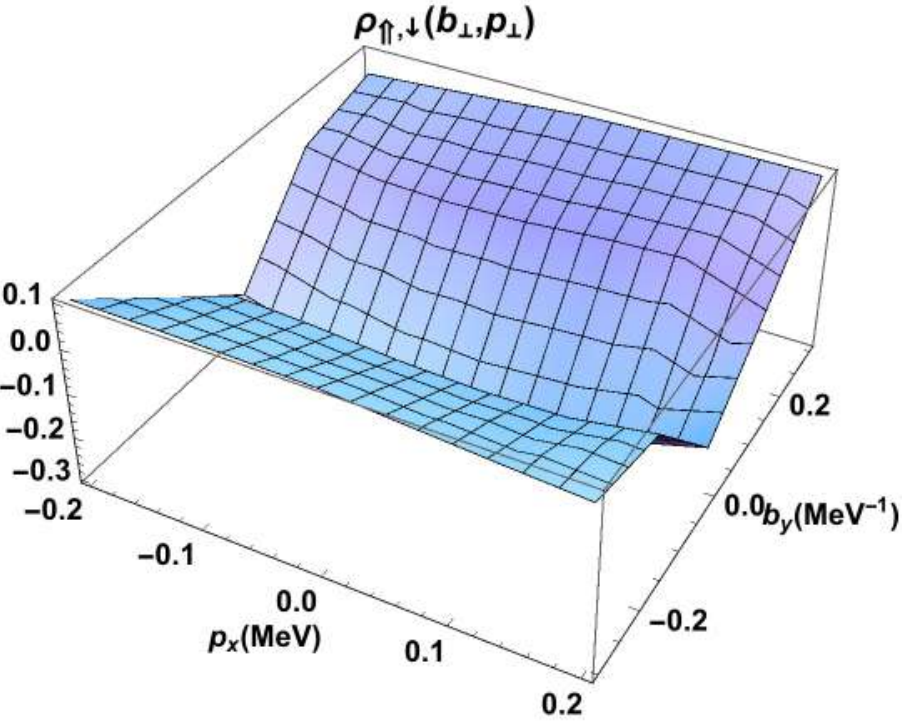}
\\
\small{(d)}\includegraphics[width=4.5cm]{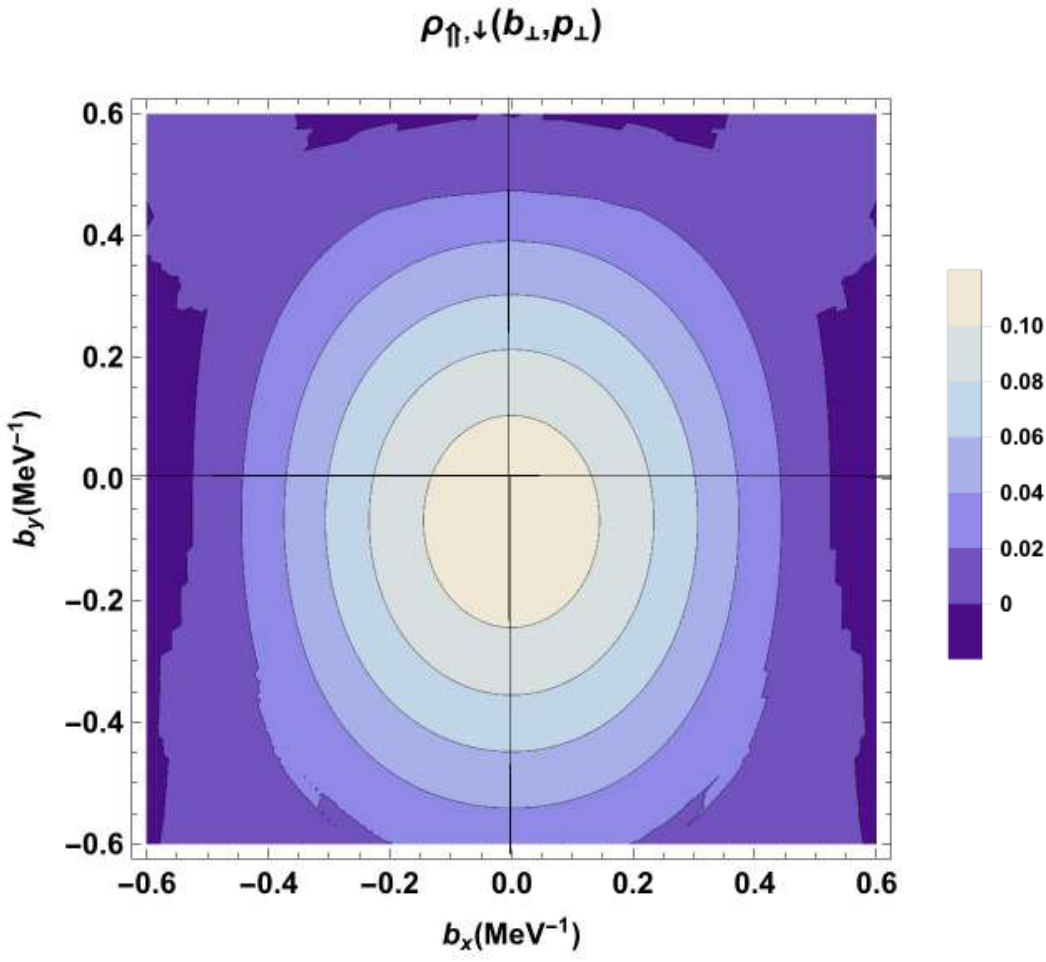}\hfill
\small{(e)}\includegraphics[width=4.5cm]{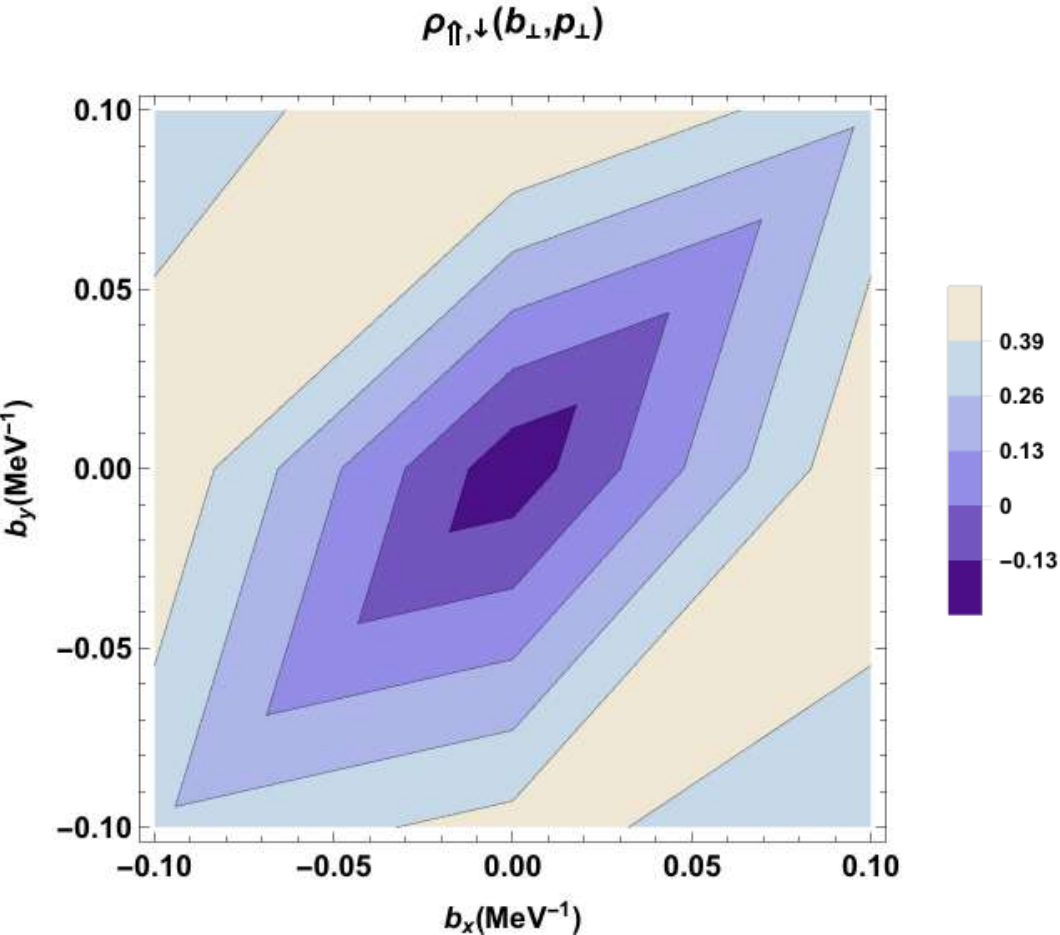}\hfill
\small{(f)}\includegraphics[width=4.5cm]{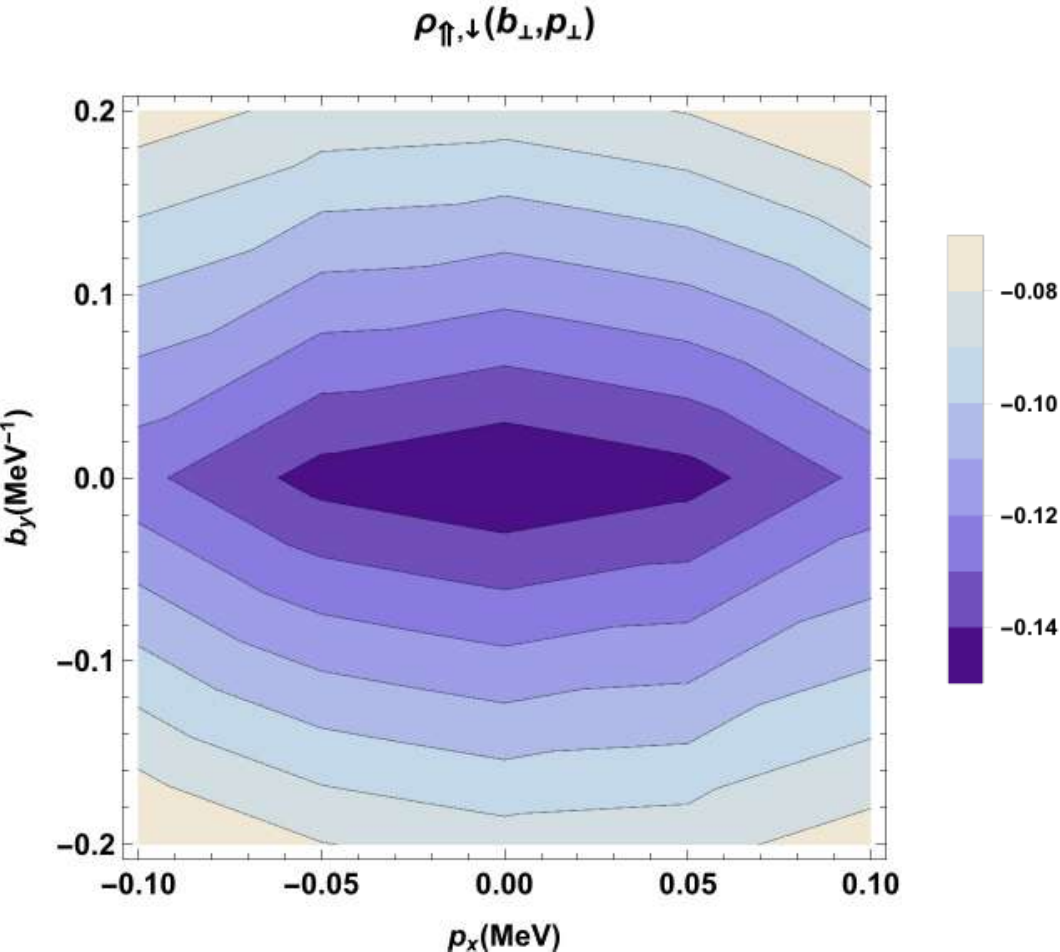}
\end{minipage}
\caption{(color online) Plots of Wigner distribution $\rho_{\Lambda_T=\Uparrow \lambda=\downarrow}(\bfb,\bfp)$ for physical electron in impact-parameter plane with fixed transverse momentum ${\bf p}_\perp= 0.8 ~MeV$ $\hat{e}_y$ (left panel), in momentum plane with fixed impact-parameter ${\bf b}_\perp= 0.8 ~MeV^{-1}$ $\hat{e}_y$ (middle panel) and in mixed plane (right panel).}
\label{rhoUp_down_trans_long}
\end{figure}
\begin{figure}
\centering
\begin{minipage}[c]{0.98\textwidth}
\small{(a)}\includegraphics[width=4.5cm]{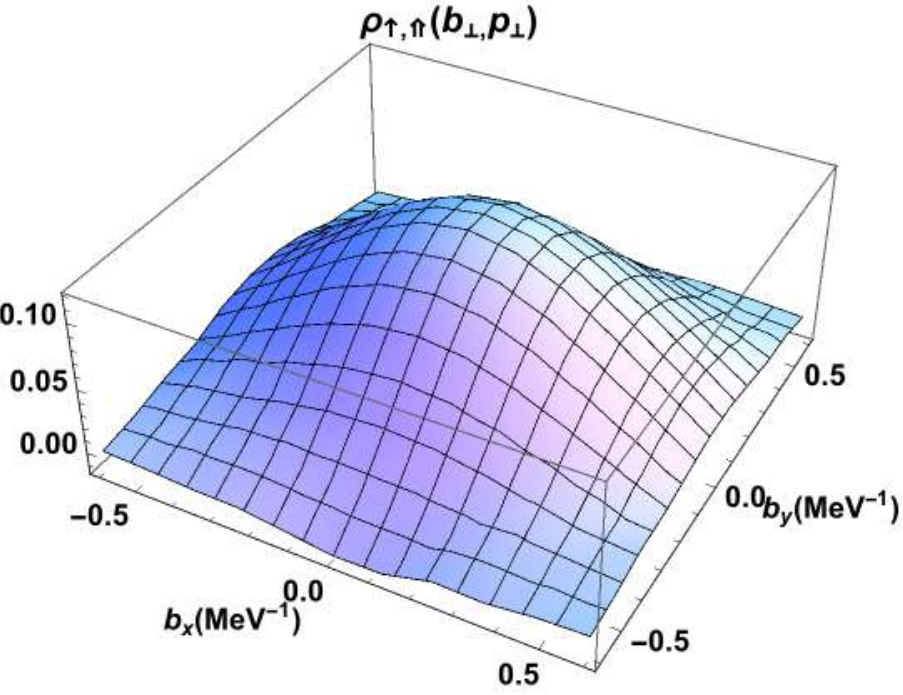}\hfill
\small{(b)}\includegraphics[width=4.5cm]{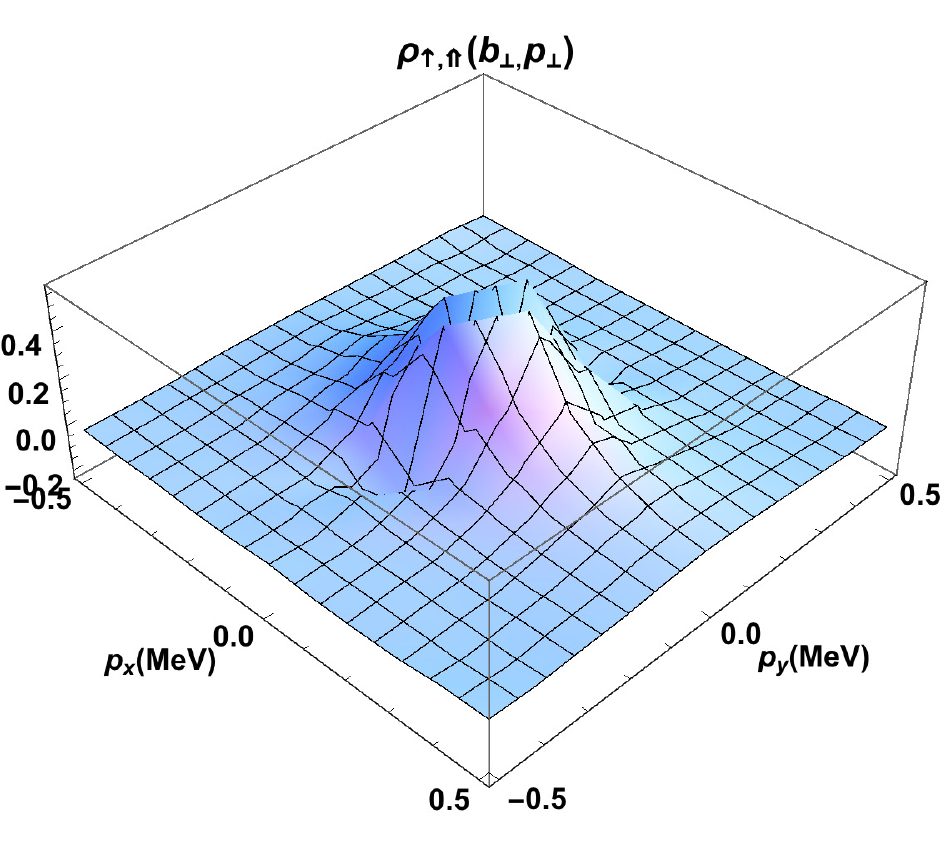}\hfill
\small{(c)}\includegraphics[width=4.5cm]{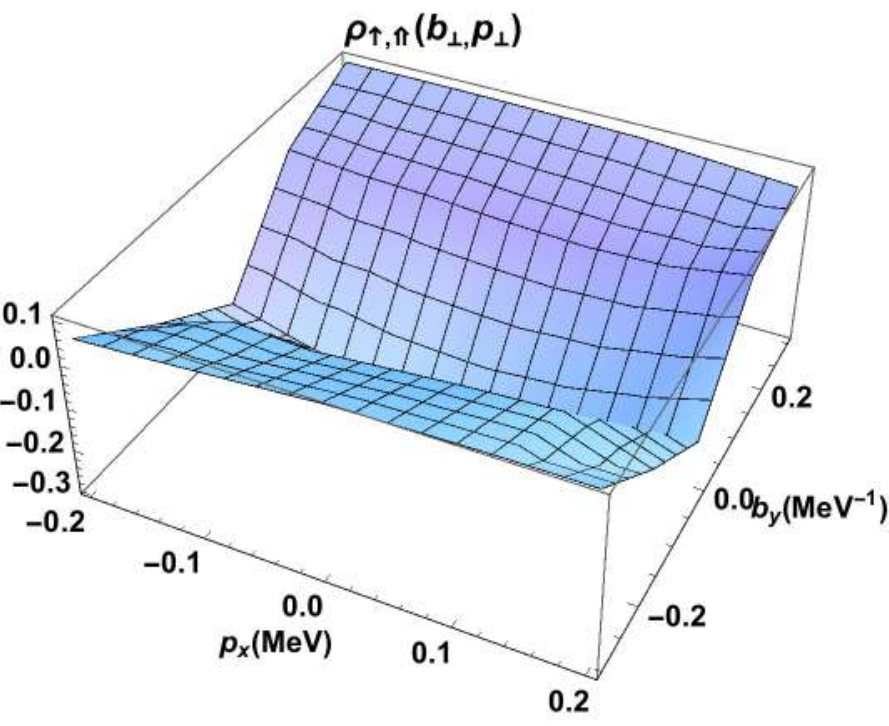}
\\
\small{(d)}\includegraphics[width=4.5cm]{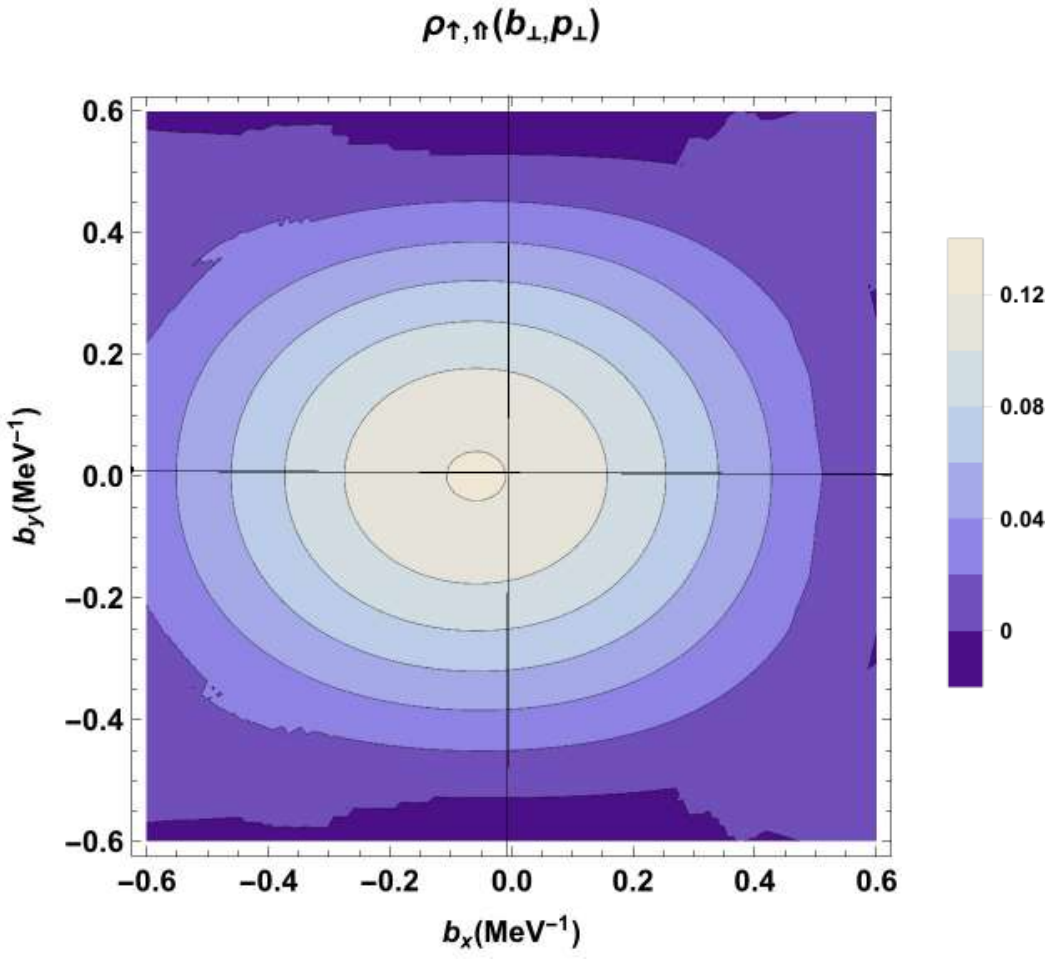}\hfill
\small{(e)}\includegraphics[width=4.5cm]{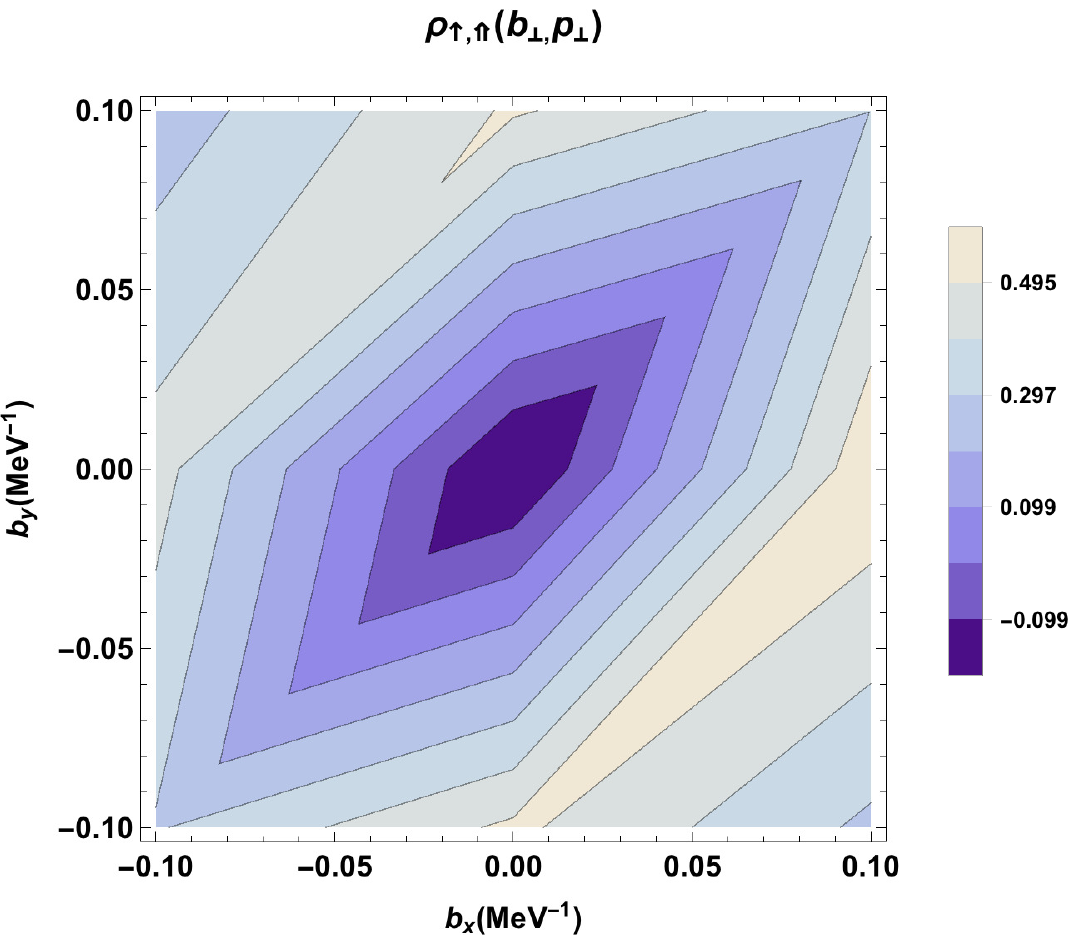}\hfill
\small{(f)}\includegraphics[width=4.5cm]{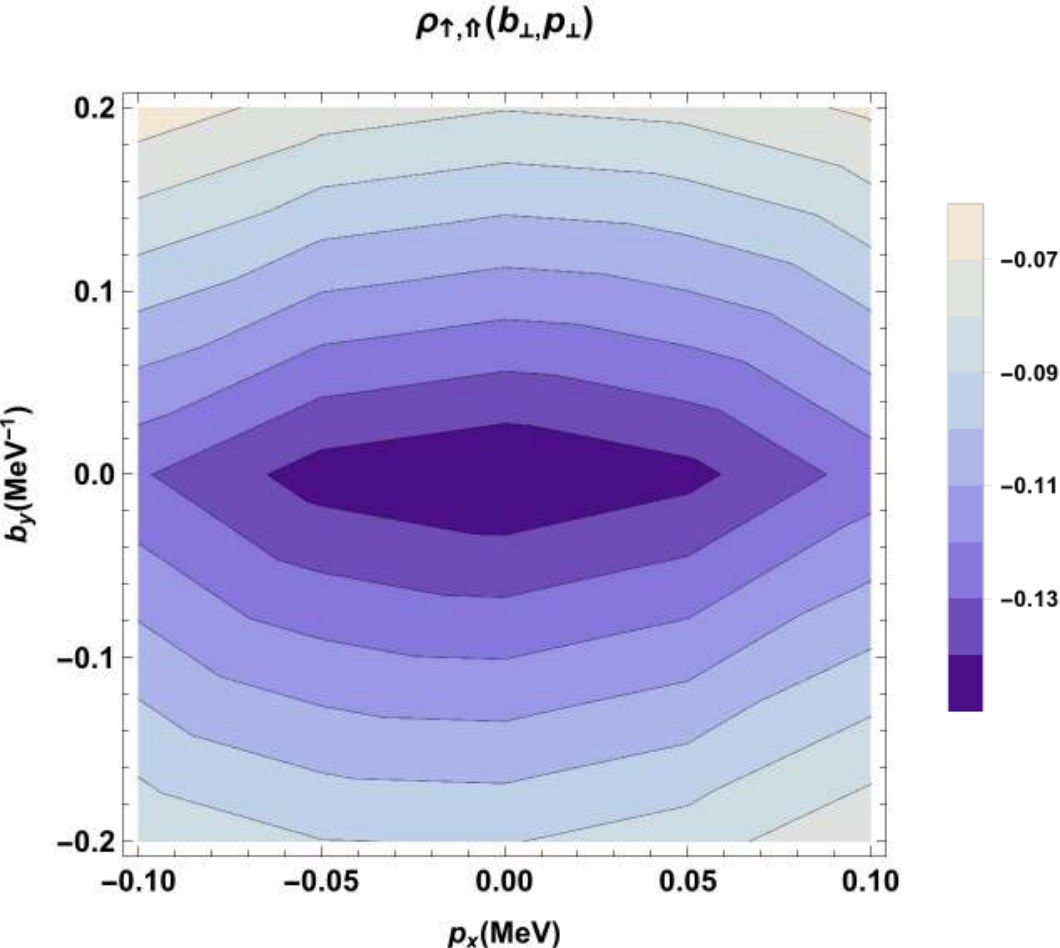}
\end{minipage}
\caption{(color online) Plots of Wigner distribution $\rho_{\Lambda=\uparrow \lambda_T=\Uparrow}(\bfb,\bfp)$ for physical electron in impact-parameter plane with fixed transverse momentum ${\bf p}_\perp= 0.8 ~MeV$ $\hat{e}_y$ (left panel), in momentum plane with fixed impact-parameter ${\bf b}_\perp= 0.8 ~MeV^{-1}$ $\hat{e}_y$ (middle panel) and in mixed plane (right panel).}
\label{rhoUp_up_long_trans}
\end{figure}
\begin{figure}
\centering
\begin{minipage}[c]{0.98\textwidth}
\small{(a)}\includegraphics[width=4.5cm]{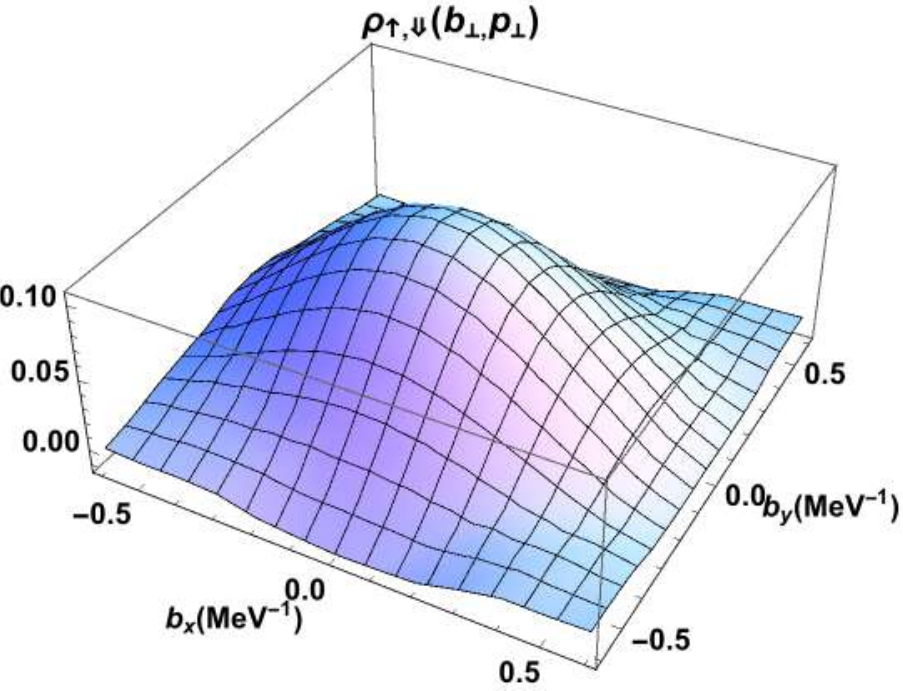}\hfill
\small{(b)}\includegraphics[width=4.5cm]{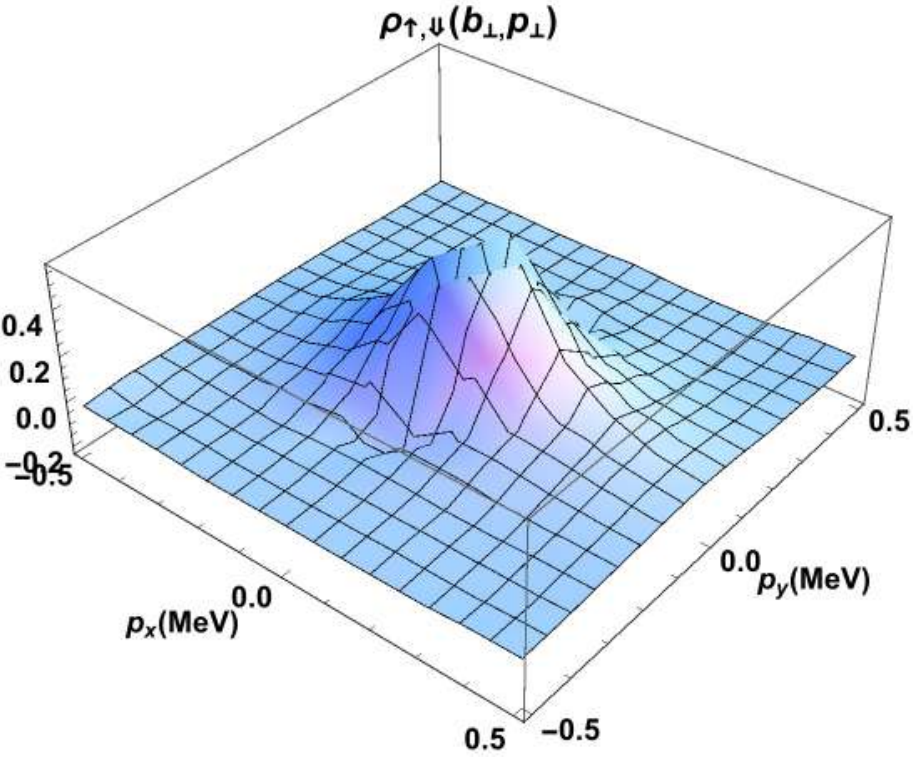}\hfill
\small{(c)}\includegraphics[width=4.5cm]{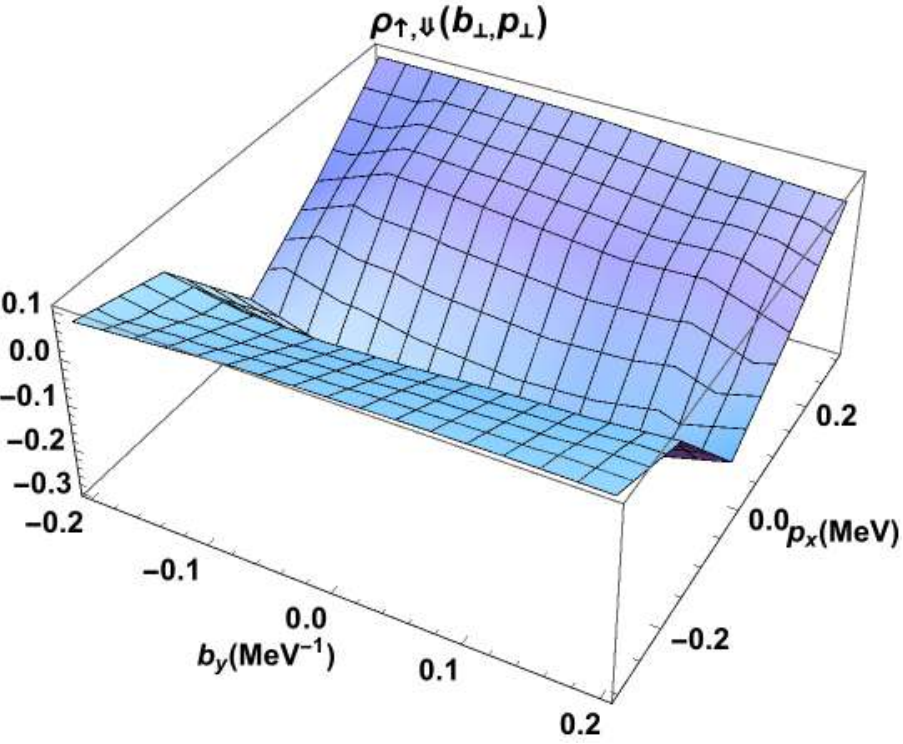}
\\
\small{(d)}\includegraphics[width=4.5cm]{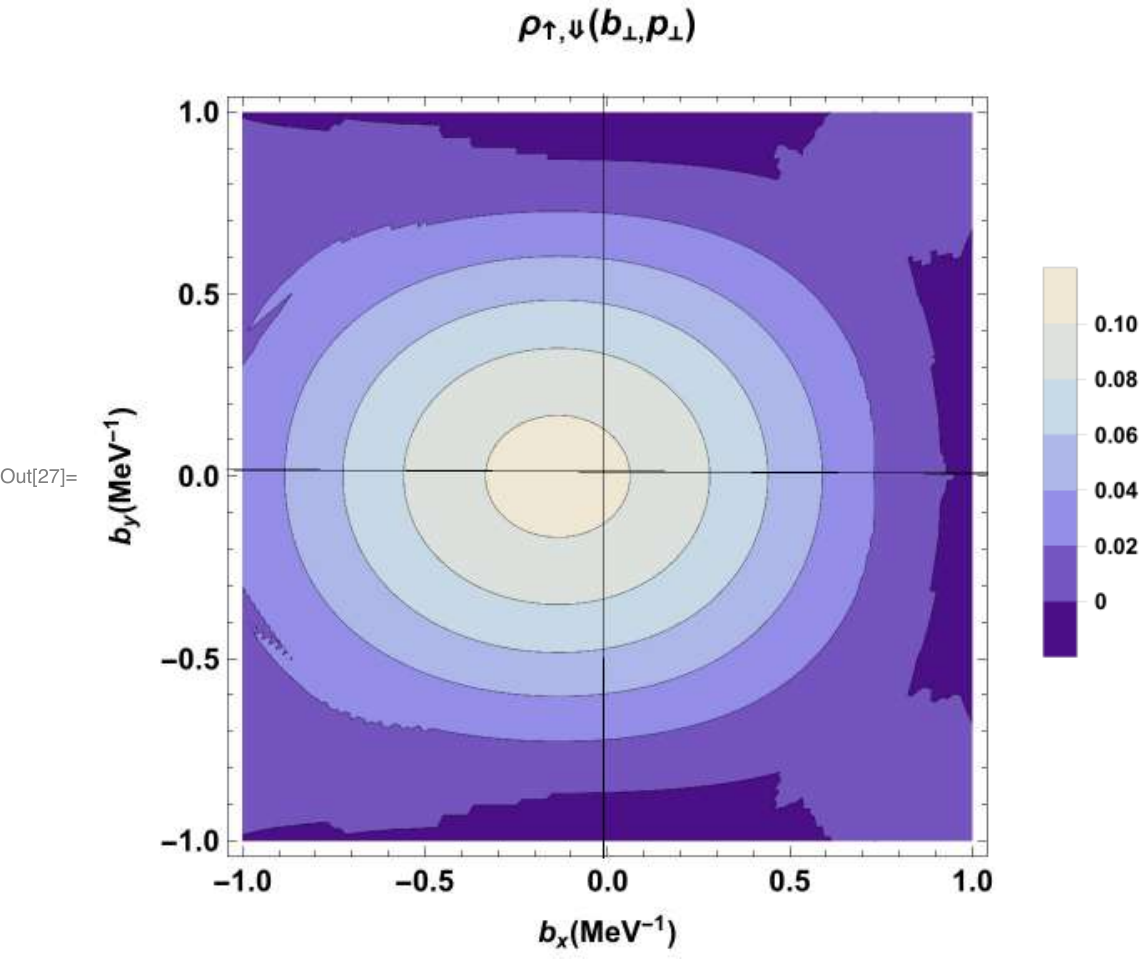}\hfill
\small{(e)}\includegraphics[width=4.5cm]{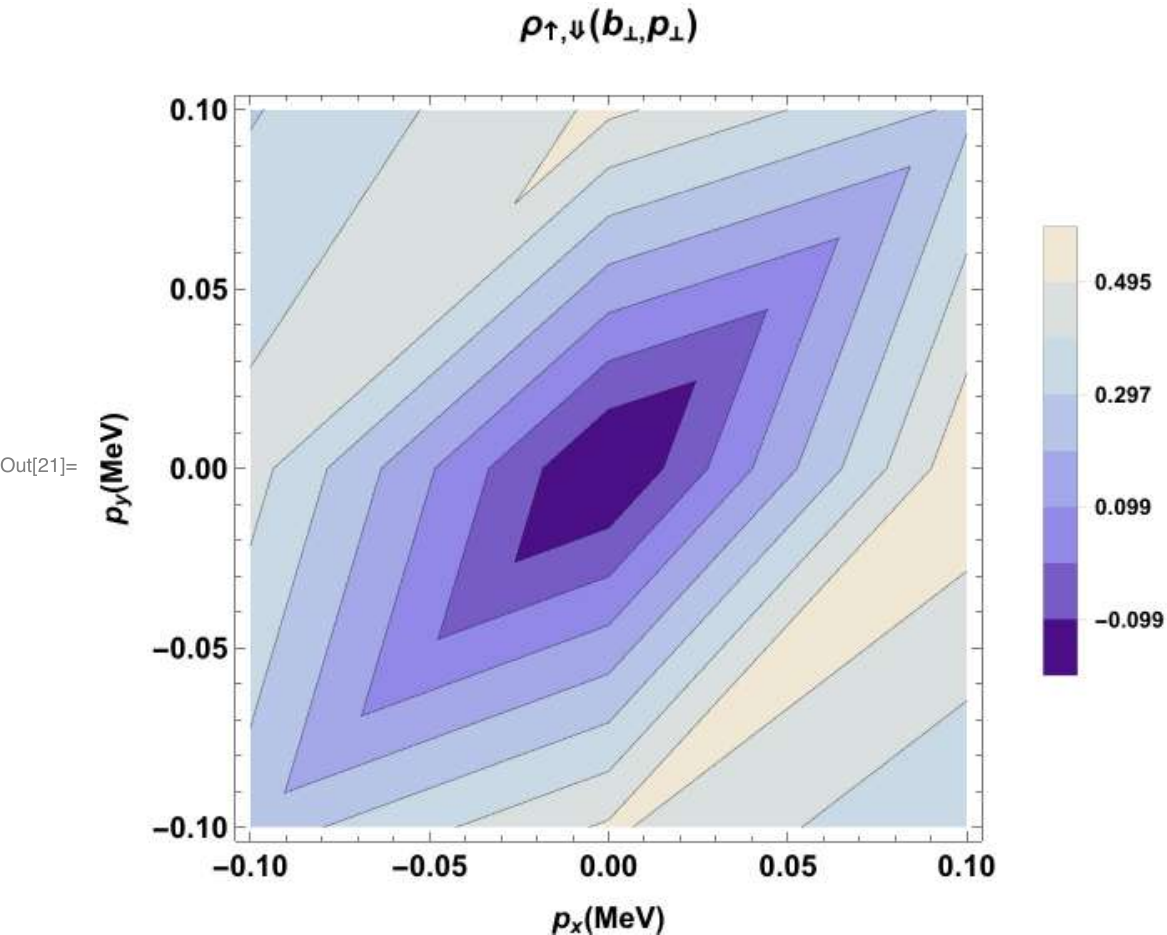}\hfill
\small{(f)}\includegraphics[width=4.5cm]{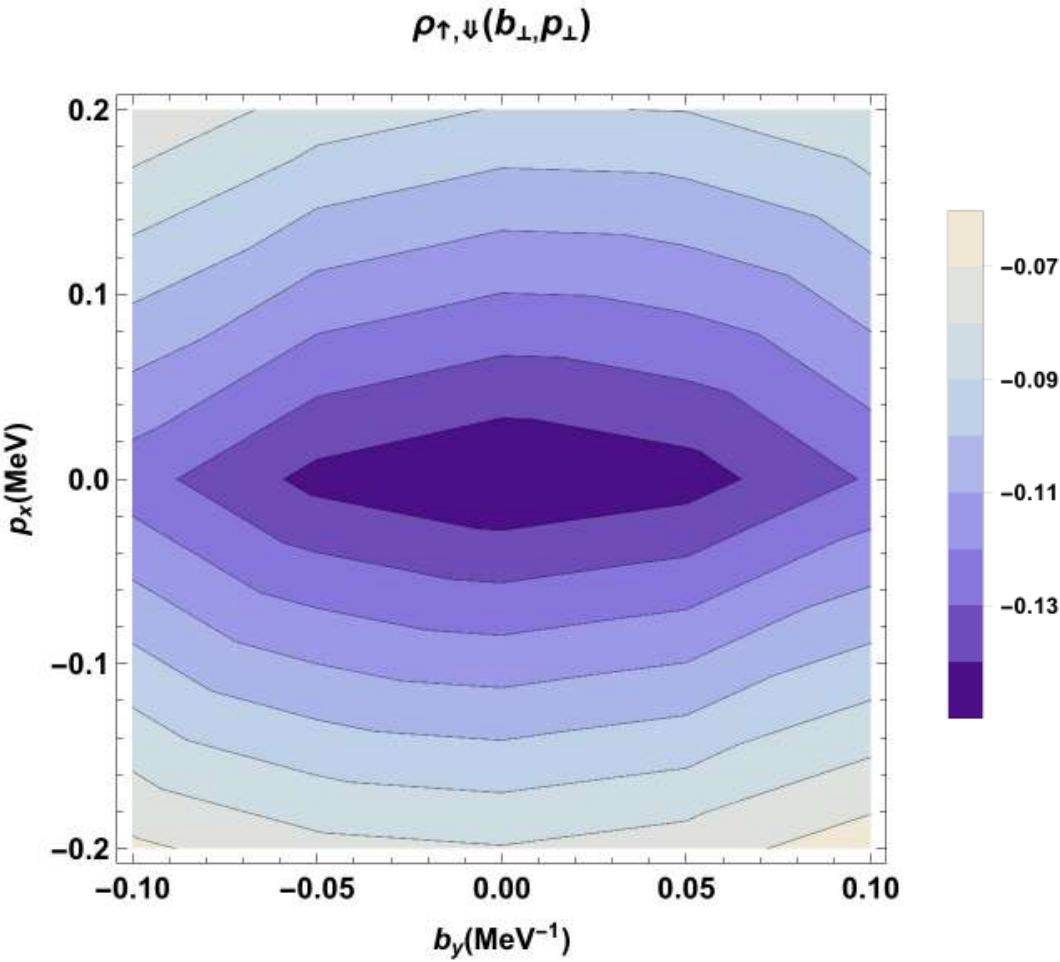}
\end{minipage}
\caption{(color online) Plots of Wigner distribution $\rho_{\Lambda=\uparrow \lambda_T=\Downarrow}(\bfb,\bfp)$ for physical electron in impact-parameter plane with fixed transverse momentum ${\bf p}_\perp= 0.8 ~MeV$ $\hat{e}_x$ (left panel), in momentum plane with fixed impact-parameter ${\bf b}_\perp= 0.8 ~MeV^{-1}$ $\hat{e}_x$ (middle panel) and in mixed plane (right panel).}
\label{rhoUp_down_long_trans}
\end{figure}
\begin{figure}[http]
\begin{minipage}[r]{0.98\textwidth}
(a)\includegraphics[width=6cm]{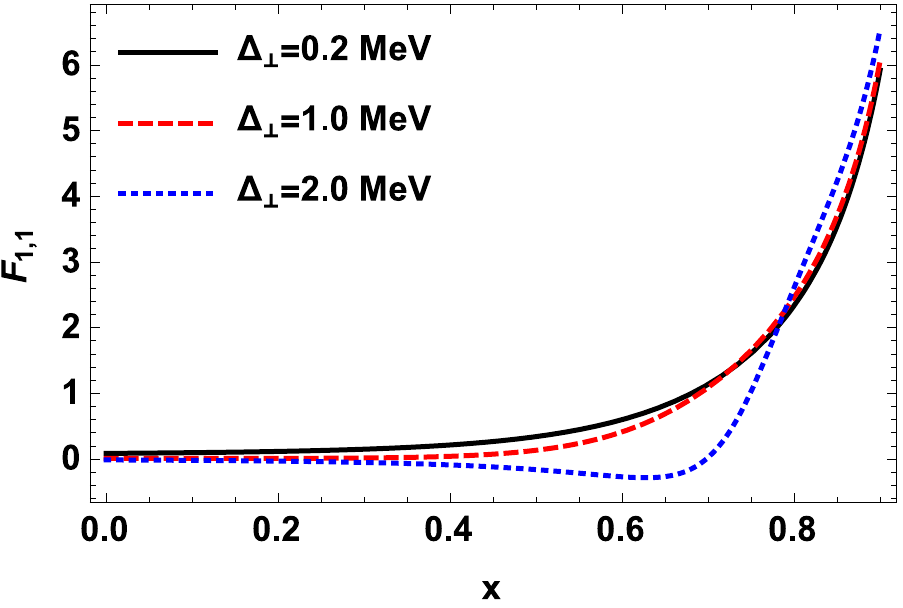}
(b)\includegraphics[width=6cm]{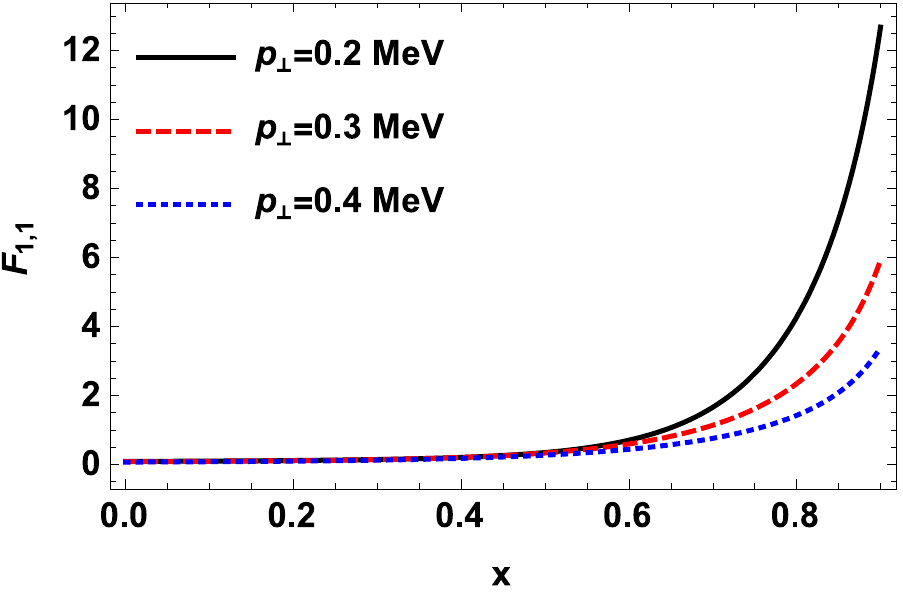}
\end{minipage}
\begin{minipage}[r]{0.98\textwidth}
\vspace{0.2cm}
(c)\includegraphics[width=6cm]{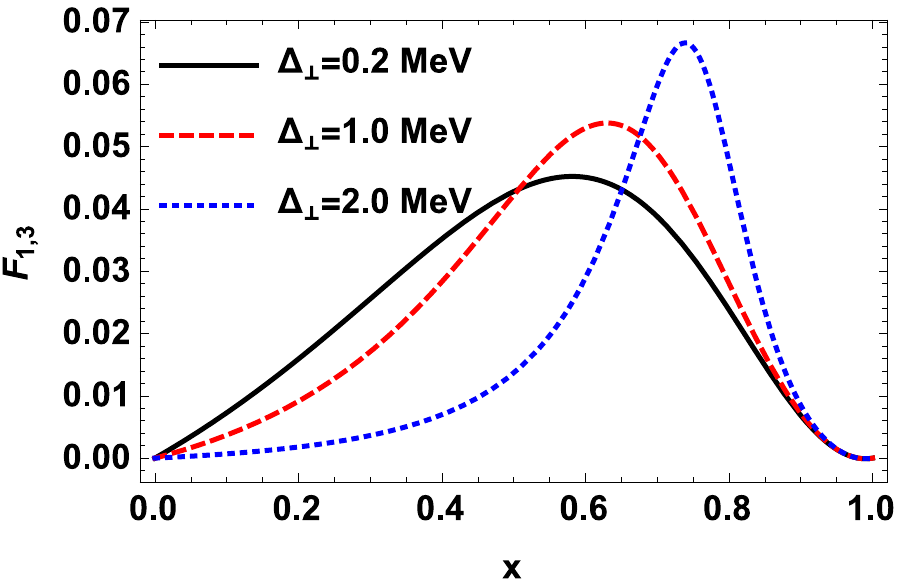}
(d)\includegraphics[width=6cm]{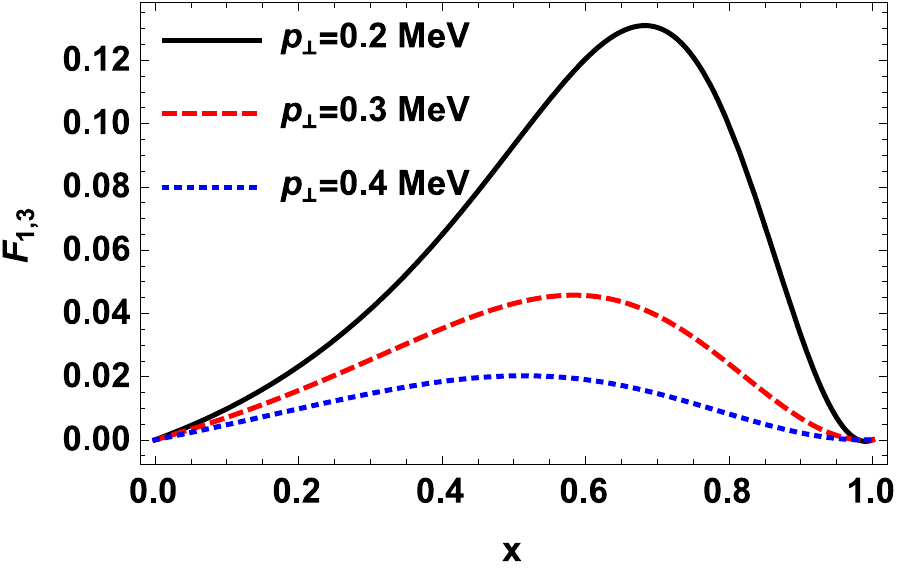}
\end{minipage}
\begin{minipage}[r]{0.98\textwidth}
\vspace{0.2cm}
(e)\includegraphics[width=6cm]{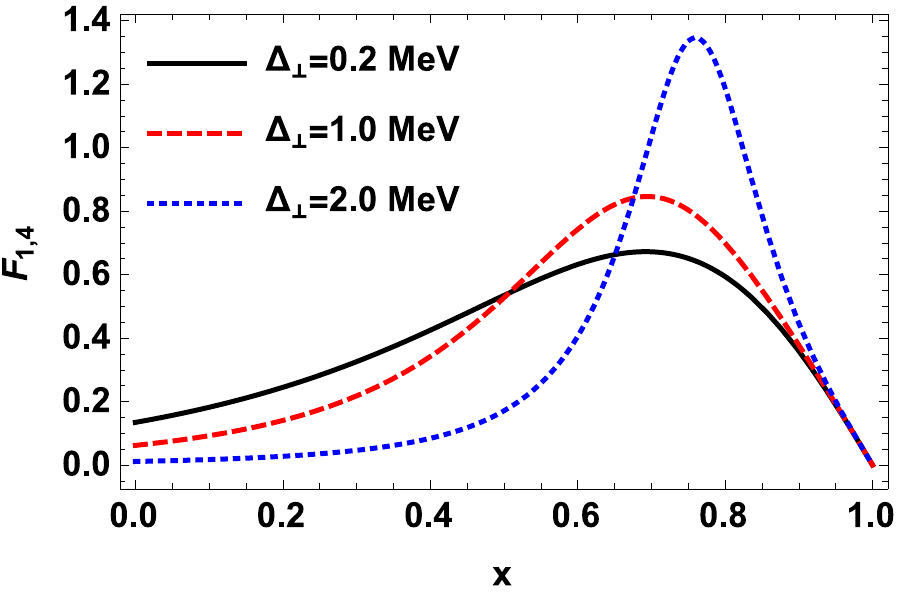}
(f)\includegraphics[width=6cm]{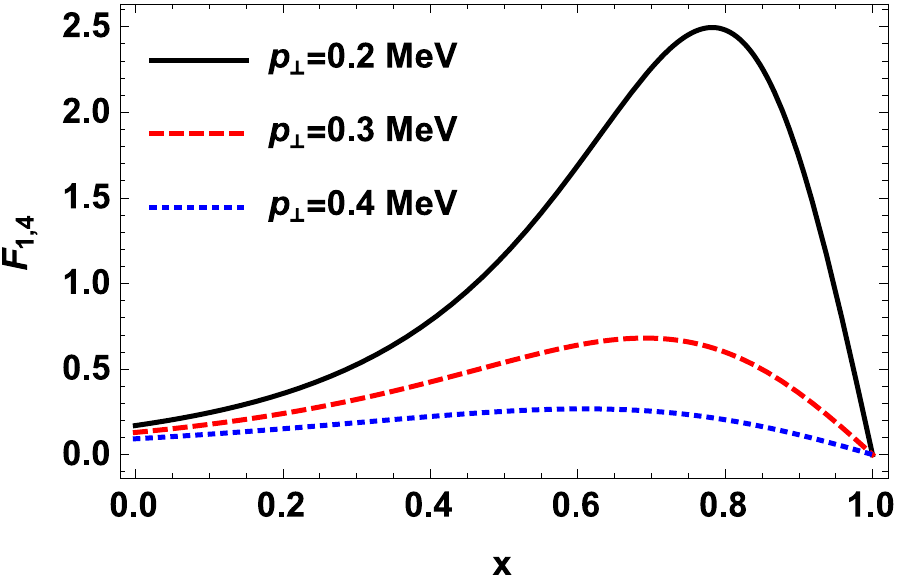}
\end{minipage}
\caption{(color online) Plots of GTMDs $F_{1,1},F_{1,3}$ and $F_{1,4}$ with fixed value of ${\bf p}_\perp=0.3 MeV$ but with different values of ${\bf \Delta}_\perp$ (left panel) and fixed value of ${\bf \Delta}_\perp=0.3 MeV $ with different values of ${\bf p}_\perp$ (right panel).}
\label{F_GTMD}
\end{figure}
\begin{figure}[http]
\begin{minipage}[r]{0.98\textwidth}
(a)\includegraphics[width=6cm]{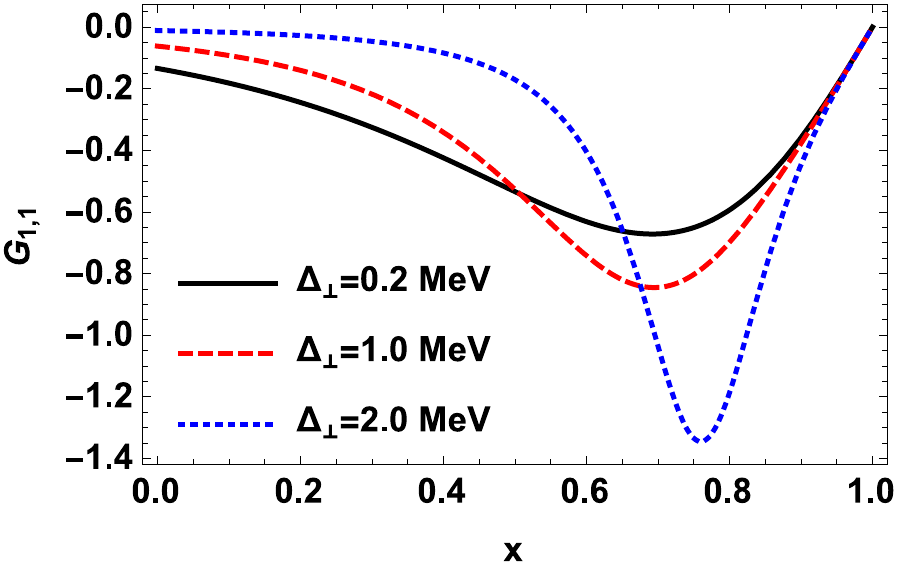}
(b)\includegraphics[width=6cm]{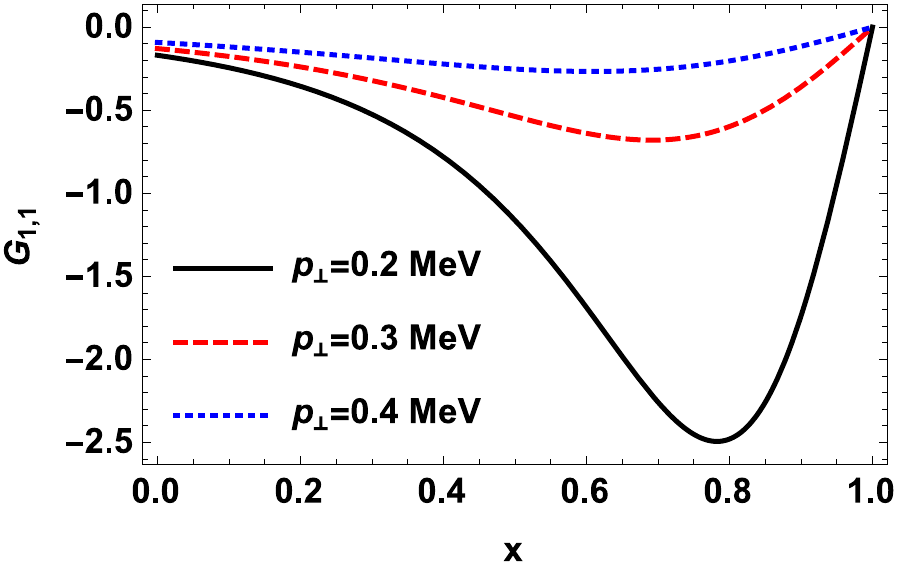}
\end{minipage}
\begin{minipage}[r]{0.98\textwidth}
\vspace{0.2cm}
(c)\includegraphics[width=6cm]{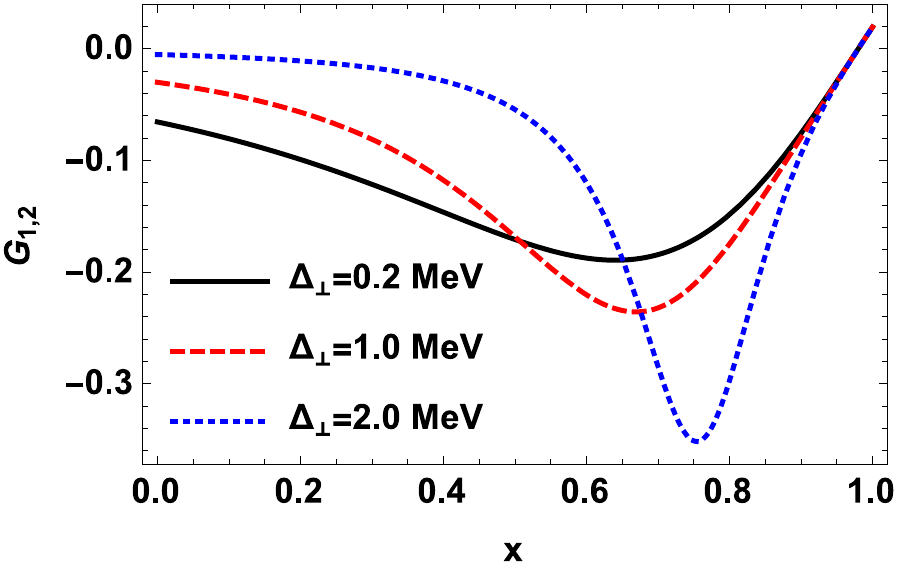}
(d)\includegraphics[width=6cm]{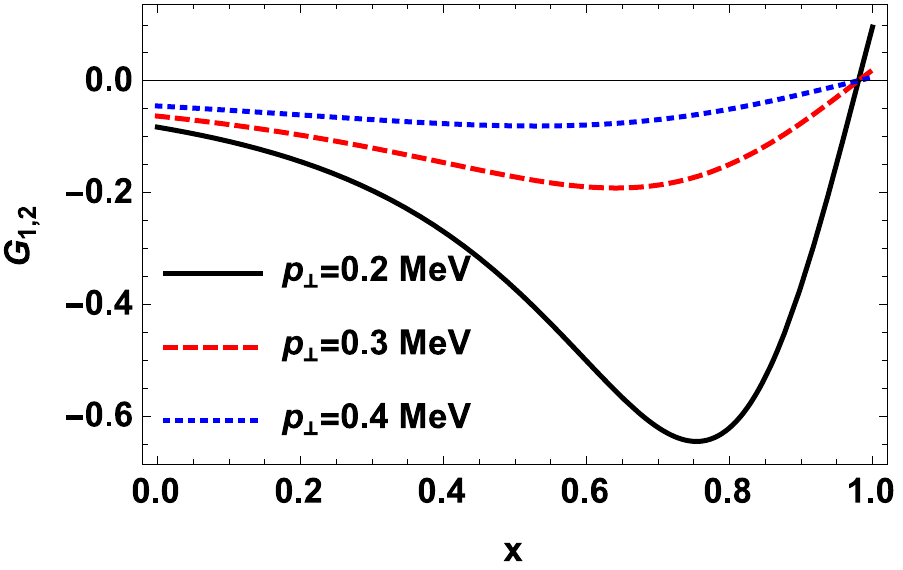}
\end{minipage}
\begin{minipage}[r]{0.98\textwidth}
\vspace{0.2cm}
(e)\includegraphics[width=6cm]{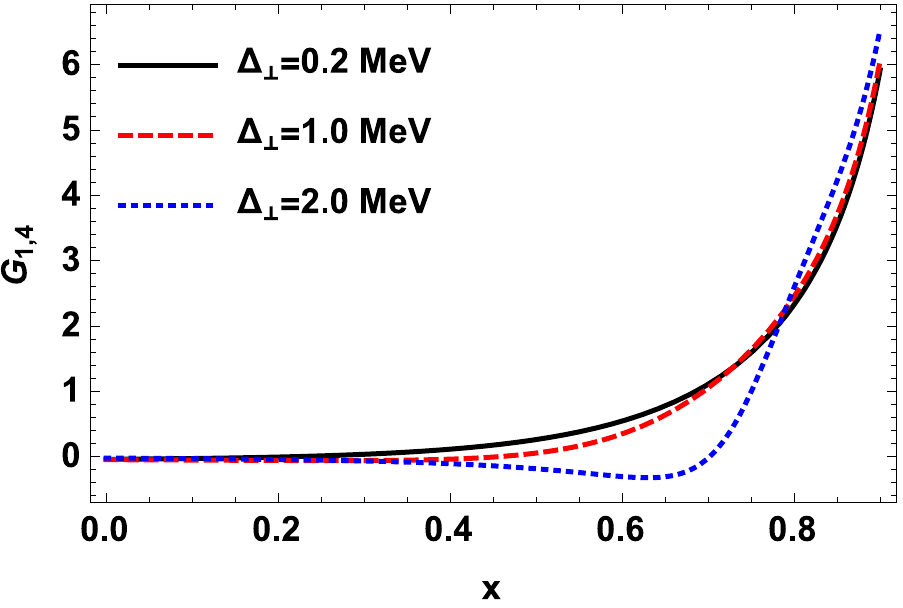}
(f)\includegraphics[width=6cm]{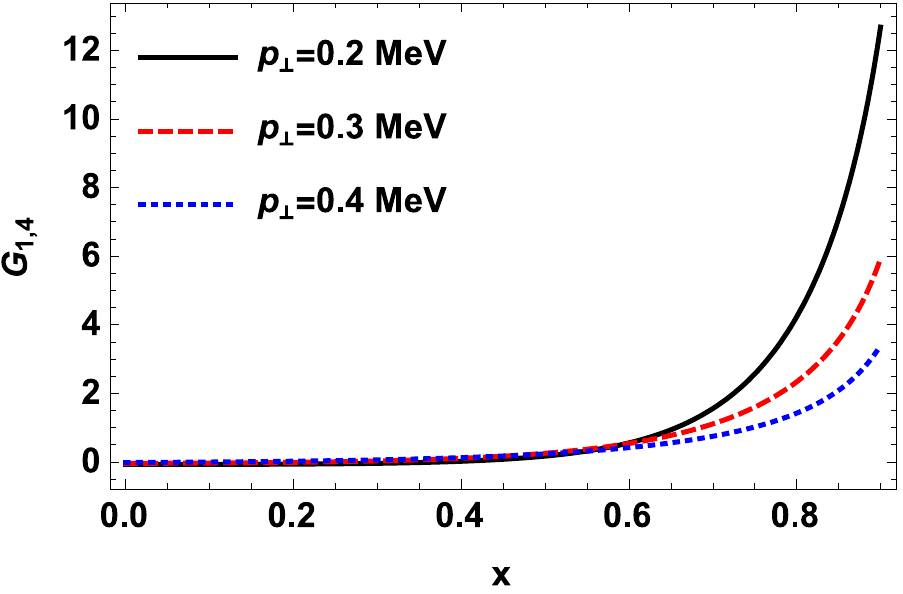}
\end{minipage}
\caption{(color online) Plots of GTMDs $G_{1,1},G_{1,2}$ and $G_{1,4}$ with fixed value of ${\bf p}_\perp=0.3 MeV$ but with different values of ${\bf \Delta}_\perp$ (left panel) and fixed value of ${\bf \Delta}_\perp=0.3 MeV $ with different values of ${\bf p}_\perp$ (right panel).}
\label{G_GTMD}
\end{figure}

Results for the $\rho_{\Lambda_T \lambda}$ which describes the spin-spin correlation between a longitudinally polarized fermion constituent with polarization $\lambda$ and transversely polarized composite system with polarization $\Lambda_T$ are presented in Fig. \ref{rhoUp_up_trans_long} and Fig. \ref{rhoUp_down_trans_long} for $\Lambda_T=\Uparrow,\lambda=\uparrow$ and $\Lambda_T=\Uparrow,\lambda=\downarrow$ respectively. Since, $\rho_{TU}$ and $\rho_{TL}$ vanish in present QED model therefore effective contributions are only from $\rho_{UU}$ and $ \rho_{UL}$. In impact-parameter plane, we observe that $\rho_{\Uparrow\uparrow}$ shifts towards $+b_y$ direction whereas $\rho_{\Uparrow\downarrow}$ gets shifted in opposite direction. It is due to the fact that for $\Lambda_T=\Uparrow \lambda=\uparrow$, the contribution from $\rho_{LU}$ distorts the symmetric nature of $\rho_{UU}$. Similarly when $\Lambda_T=\Uparrow \lambda=\downarrow$ one obtains the same distortion in opposite direction.  However in momentum plane representation, no distortion has been observed for both cases.  

In Fig. \ref{rhoUp_up_long_trans} and Fig. \ref{rhoUp_down_long_trans}, we show the distribution $\rho_{\Lambda\lambda_T}$ where the spin-spin correlation is considered of a transversely polarized fermion constituent with polarization $\lambda_T$ in a longitudinally polarized composite system with the  polarization $\Lambda$. Fig. \ref{rhoUp_up_long_trans} is for $\Lambda=\uparrow,\lambda_T=\Uparrow$ and Fig. \ref{rhoUp_down_long_trans} describes the case when $\Lambda=\uparrow,\lambda_T=\Downarrow$. In both the cases, unlike $\rho_{\Lambda_T\lambda}$, in impact-parameter plane $\rho_{\Lambda\lambda_T}$ gets distorted along the $-b_x$ direction. $\rho_{\Lambda\lambda_T}$ is getting contribution from $\rho_{LU}, \rho_{UT}$ and $\rho_{LT}$ which break the symmetric nature of $\rho_{UU}$.

Spin-spin correlation distributions immediately reduce to spin densities in impact-parameter plane \cite{Pasquini:2007xz,Diehl:2005jf,Kumar:2015yta} when one integrates them over $\bfp$. In general, spin density $\rho (x, \bfb,\lambda,\Lambda)$ and $\rho(x, \bfb, \lambda_T, \Lambda_T)$ gives the probability of finding a parton with momentum fraction $x$ and transverse position $\bfb$ with light-cone helicity $\lambda$ in a composite system with longitudinal polarization $\Lambda$ or with a transverse spin $\lambda_T$ in the composite system with transverse spin $\Lambda_T$. We find that the longitudinal spin density $\rho (x, \bfb,\lambda,\Lambda)$ is related to the Wigner distribution $\rho_{\Lambda \lambda}(\bfb,\bfp,x)$ defined in Eq.(\ref{rho_Lamlam}) and the transverse spin density is connected with the transverse spin-spin correlation $\rho_{\Lambda_T \lambda_T}({\bfb},{\bfp},x)$ defined in Eq.(\ref{rho_LambdaT_lambdaT}). $\rho (x, \bfb,\lambda,\Lambda)$ involves the GPDs $H(x,b^2)$ and $\tilde{H}(x,b^2)$, thus comparing with Eq.(\ref{rho_Lamlam}), one can easily find out that $\rho_{UU}$ and $\rho_{LL}$ reduce $H(x,b^2)$ and $\tilde{H}(x,b^2)$ respectively and $\rho_{UL}$ and $\rho_{LU}$ give zero when we integrate them over $\bfp$. Similarly, in the case of transverse polarization $\rho_{TU}$ is connected with $ E'(x,b^2)$, $\rho_{UT}$ is connected with $(E_T'(x,b^2)+2 \widetilde{H}_T'(x,b^2))$ and $\rho_{TT}$ is related with $(H_T(x,b^2)-\frac{\Delta_b}{4M^2} \widetilde{H}_T(x,b^2))$, where
$
f' = \frac{\partial}{\partial b^2}\, f ,
~
f''= \Big( \frac{\partial}{\partial b^2} \Big)^2 f,
~
\Delta_b f
= 4\, \frac{\partial}{\partial b^2}
    \Big( b^2 \frac{\partial}{\partial b^2} \Big) f .$
 On the other hand, integrating over $\bfb$ the correlation distributions for electron of definite longitudinal or transverse polarizations, one can obtain spin densities in the momentum plane \cite{Bacchetta:2015qka,Pasquini:2007xz} which are again expressed in terms of TMDs.
Hence with these connections, the correlations between the electron distributions in impact-parameter plane and longitudinal
momentum as well as the three-dimensional information about the strength of different spin-spin and spin-orbit correlations in the momentum plane for different polarizations of bare electron and physical electron can be obtained from the Wigner distributions.
\section{GTMDs for electron}\label{gtmd}
Generalized transverse momentum distributions \cite{Goeke:2005hb,Meissner:2007rx,Meissner:2009ww} which are known as the mother distributions of GPDs and TMDs can be extracted from different Wigner distributions. For the leading twist, the Wigner correlator, Eq. \ref{wigner-operator}, can be parametrized in terms of GTMDs as \cite{Meissner:2009ww}
\begin{eqnarray}
 W_{\lambda \lambda'}^{[\gamma^+]}
&=&\frac{1}{2M} \, \bar{u}(p', \lambda') \, \bigg[
      F_{1,1}
      + \frac{i\sigma^{i+} k_\perp^i}{P^+} \, F_{1,2} + \frac{i\sigma^{i+} \Delta_\perp^i}{P^+}
   F_{1,3}  + \frac{i\sigma^{ij} k_\perp^i \Delta_\perp^j}{M^2} \, F_{1,4}
     \bigg] \, u(p, \lambda)
     \,, \label{e:gtmd_1}\\
W_{\lambda \lambda'}^{[\gamma^+\gamma_5]}&=& \frac{1}{2M} \, \bar{u}(p', \lambda') \, \bigg[
      - \frac{i\varepsilon_\perp^{ij} k_\perp^i \Delta_\perp^j}{M^2} \, G_{1,1}
      + \frac{i\sigma^{i+}\gamma_5 k_\perp^i}{P^+} G_{1,2}+ \frac{i\sigma^{i+}\gamma_5 \Delta_\perp^i}{P^+} \, G_{1,3}+
    i\sigma^{+-}\gamma_5 \, G_{1,4}
     \bigg]\nonumber\\ &&u(p, \lambda)
     \,, \label{e:gtmd_2}\\
W_{\lambda \lambda'}^{[i\sigma^{j+}\gamma_5]}&=&\frac{1}{2M} \, \bar{u}(p', \lambda') \, \bigg[
      - \frac{i\varepsilon_\perp^{ij} k_\perp^i}{M} \, H_{1,1}
     - \frac{i\varepsilon_\perp^{ij} \Delta_\perp^i}{M} \, H_{1,2} + \frac{M \, i\sigma^{j+}\gamma_5}{P^+} \, H_{1,3}+ \frac{k_\perp^j \, i\sigma^{k+}\gamma_5 k_\perp^k}{M \, P^+} \, H_{1,4} \nonumber\\
&&+ \frac{\Delta_\perp^j \, i\sigma^{k+}\gamma_5 k_\perp^k}{M \, P^+} \, H_{1,5}
      + \frac{\Delta_\perp^j \, i\sigma^{k+}\gamma_5 \Delta_\perp^k}{M \, P^+} \, H_{1,6}+ \frac{k_\perp^j \, i\sigma^{+-}\gamma_5}{M} \, H_{1,7}
      +\frac{\Delta_\perp^j \, i\sigma^{+-}\gamma_5}{M} \, H_{1,8}
     \bigg]\nonumber\\
     &&  u(p, \lambda),
     \, \label{e:gtmd_3}
\end{eqnarray}
where $\varepsilon_\perp^{ij}$ is the antisymmetric tensor.
At the leading twist there are 16 GTMDs and at $\Delta_\perp=0$, GTMDs reduces to TMDs which are function of transverse momentum $\bf{p}_\perp$ and longitudinal momentum fraction $x$. There are 8 TMDs at the leading twist. The GTMDs $F_{1,4}$ and $G_{1,1}$ give contributions to spin-orbit angular momentum correlation. In this model, we obtain $10$ nonzero GTMDs for the physical electron. The explicit expressions of the GTMDs are given by
\begin{eqnarray}
F_{1,1}(x,{\bf \Delta_\perp},{\bf p_\perp})&= &\frac{4 e^2}{2(16 \pi^3)} \Big[\frac{1+x^2}{x^2(1-x)^2}\Big({\bf p}_\perp^2-\frac{(1-x)^2}{4}  {\bf \Delta}_\perp^2 \Big)\nonumber\\
&&+ \left(M-\frac{m}{x}\right)^2 \Big]\varphi^\dagger({\bf p''_\perp}) \varphi({\bf p'_\perp}),\\
F_{1,2}(x,{\bf \Delta_\perp},{\bf p_\perp})&=&0,\\
F_{1,3}(x,{\bf \Delta_\perp},{\bf p_\perp})&=& \frac{F_{1,1}}{2}-\frac{4 e^2}{2(16 \pi^3)} M \left(M-\frac{m}{x}\right) \varphi^\dagger(\bf{p''_\perp}) \varphi(\bf{p'_\perp}),\\
F_{1,4}(x,{\bf \Delta_\perp},{\bf p_\perp})&=&\frac{4 e^2}{2(16 \pi^3)} M^2 \frac{(1-x)}{(1-x)^2} \frac{(1-x^2)}{x^2} \varphi^\dagger(\bf{p''_\perp}) \varphi(\bf{p'_\perp}),\\
G_{1,1}(x,{\bf \Delta_\perp},{\bf p_\perp})&=& -\frac{4 e^2}{2(16 \pi^3)} M^2 \frac{(1-x^2)(1-x)}{x^2(1-x)^2}\varphi^\dagger(\bf{p''_\perp}) \varphi(\bf{p'_\perp}),\\
G_{1,2}(x,{\bf \Delta_\perp},{\bf p_\perp})&=& -\frac{4 e^2}{2(16 \pi^3)}M \left(M-\frac{m}{x}\right)\varphi^\dagger(\bf{p''_\perp}) \varphi(\bf{p'_\perp}),\\
G_{1,3}(x,{\bf \Delta_\perp},{\bf p_\perp})&=&0,\\
G_{1,4}(x,{\bf \Delta_\perp},{\bf p_\perp})&=&\frac{4 e^2}{2(16 \pi^3)}\bigg[\frac{1+x^2}{x^2(1-x)^2} \Big({\bf p}_\perp^2-\frac{(1-x)^2}{4}  {\bf \Delta}_\perp^2 \Big)\nonumber\\
&&+ \left(M-\frac{m}{x}\right)^2\bigg]\varphi^\dagger(\bf{p''_\perp}) \varphi(\bf{p'_\perp}),
\end{eqnarray}
\begin{eqnarray}
H_{1,1}(x,{\bf \Delta_\perp},{\bf p_\perp})&=&0, \\
H_{1,2}(x,{\bf \Delta_\perp},{\bf p_\perp})&=& -\frac{4 e^2}{2(16 \pi^3)}\frac{2 M^2}{4 x}\varphi^\dagger(\bf{p''_\perp}) \varphi(\bf{p'_\perp}),\\
H_{1,3}(x,{\bf \Delta_\perp},{\bf p_\perp})&=& \frac{4 e^2}{2(16 \pi^3)} \frac{\bf{p}_\perp^2}{x (1-x)^2} \varphi^\dagger(\bf{p''_\perp}) \varphi(\bf{p'_\perp}),\\
H_{1,4}(x,{\bf \Delta_\perp},{\bf p_\perp})&=&0,\\
H_{1,5}(x,{\bf \Delta_\perp},{\bf p_\perp})&=&0,\\
H_{1,6}(x,{\bf \Delta_\perp},{\bf p_\perp})&=&-\frac{4 e^2}{2(16 \pi^3)}\frac{M^2}{4 x}\varphi^\dagger(\bf{p''_\perp}) \varphi(\bf{p'_\perp}),\\
H_{1,7}(x,{\bf \Delta_\perp},{\bf p_\perp})&=&\frac{-4 e^2}{2(16 \pi^3)}\frac{M}{x(1-x)}\left(M-\frac{m}{x}\right)\varphi^\dagger(\bf{p''_\perp}) \varphi(\bf{p'_\perp}),\\
H_{1,8}(x,{\bf \Delta_\perp},{\bf p_\perp})&=&0.
\end{eqnarray}
\begin{figure}[htbp]
\begin{minipage}[r]{0.98\textwidth}
(a)\includegraphics[width=6cm]{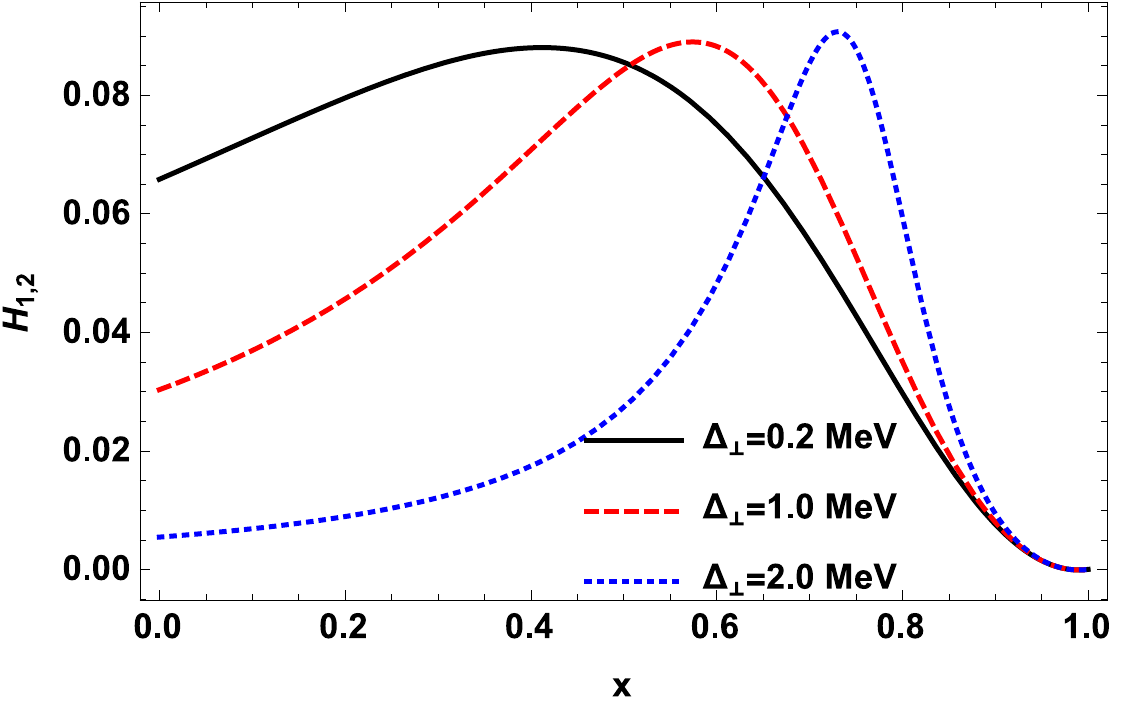}
(b)\includegraphics[width=6cm]{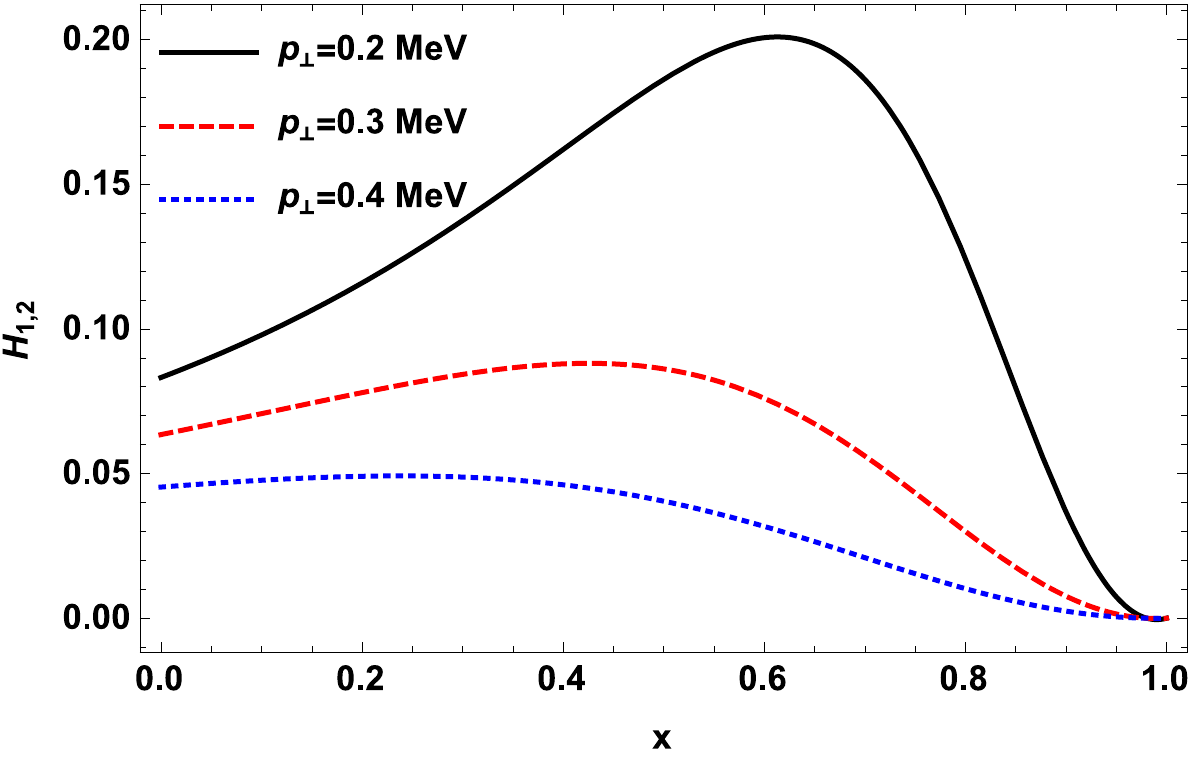}
\end{minipage}
\begin{minipage}[r]{0.98\textwidth}
\vspace{0.2cm}
(c)\includegraphics[width=6cm]{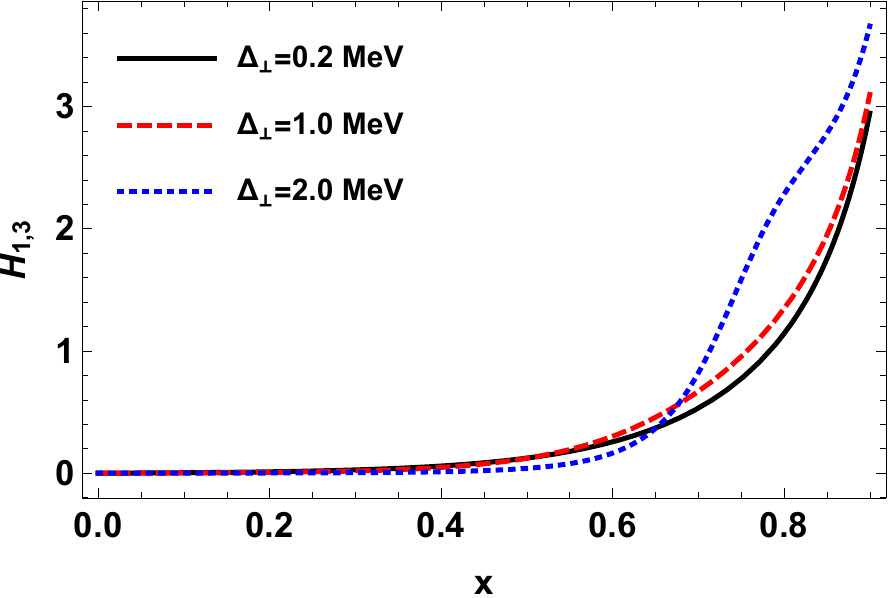}
(d)\includegraphics[width=6cm]{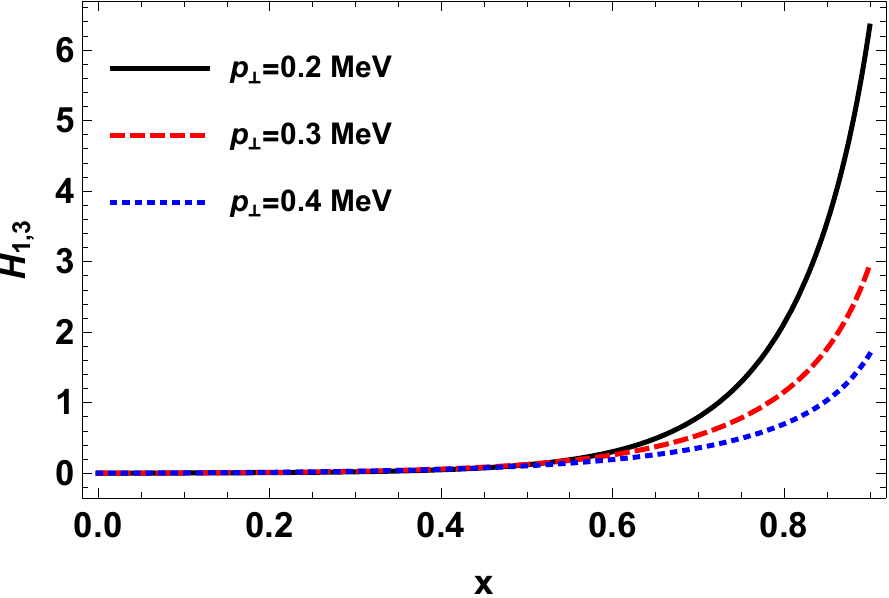}
\end{minipage}
\begin{minipage}[r]{0.98\textwidth}
\vspace{0.2cm}
(e)\includegraphics[width=6cm]{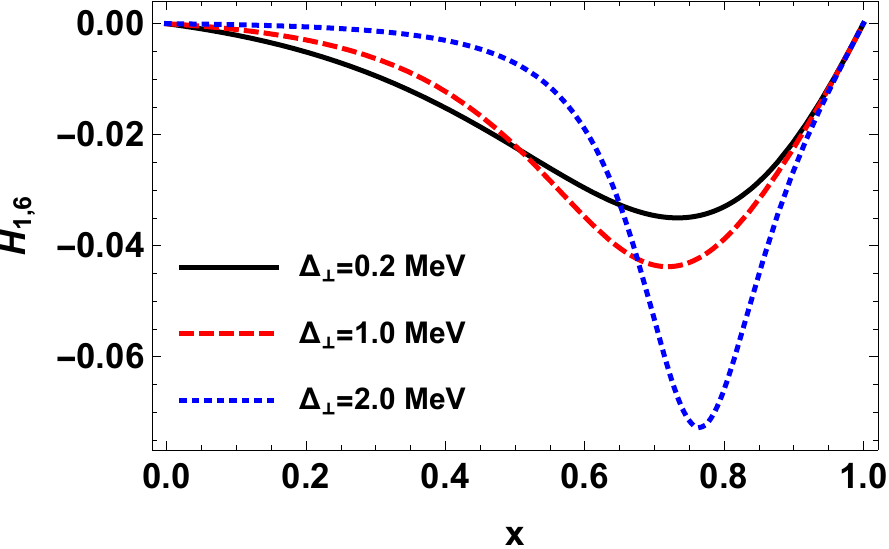}
(f)\includegraphics[width=6cm]{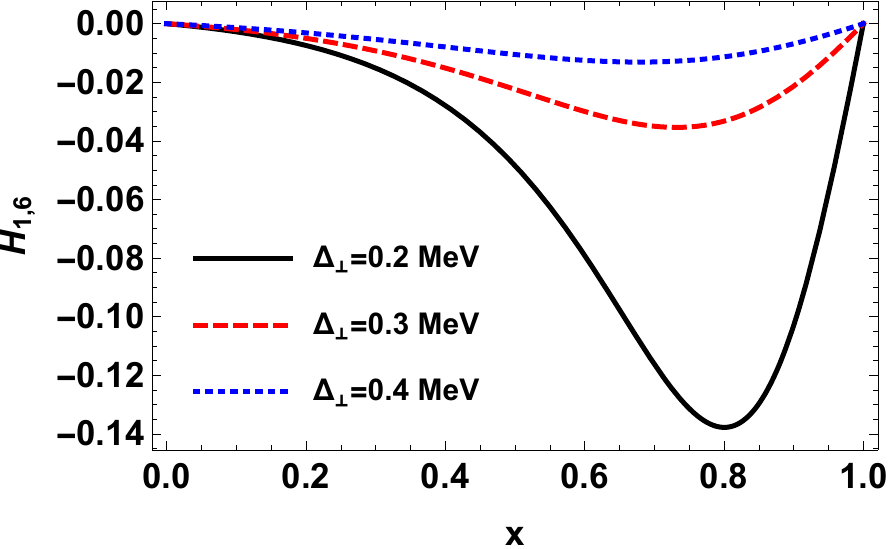}
\end{minipage}
\begin{minipage}[r]{0.98\textwidth}
\vspace{0.2cm}
(g)\includegraphics[width=6cm]{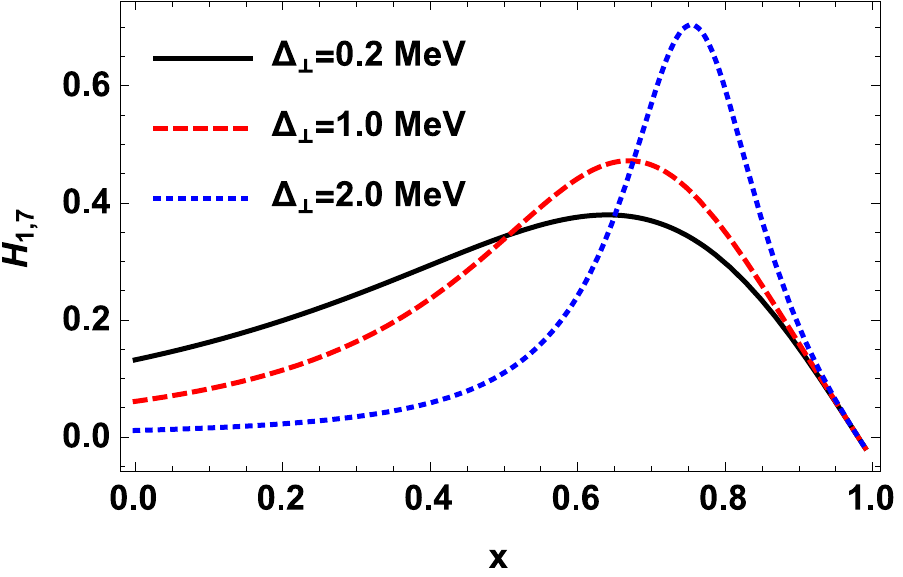}
(h)\includegraphics[width=6cm]{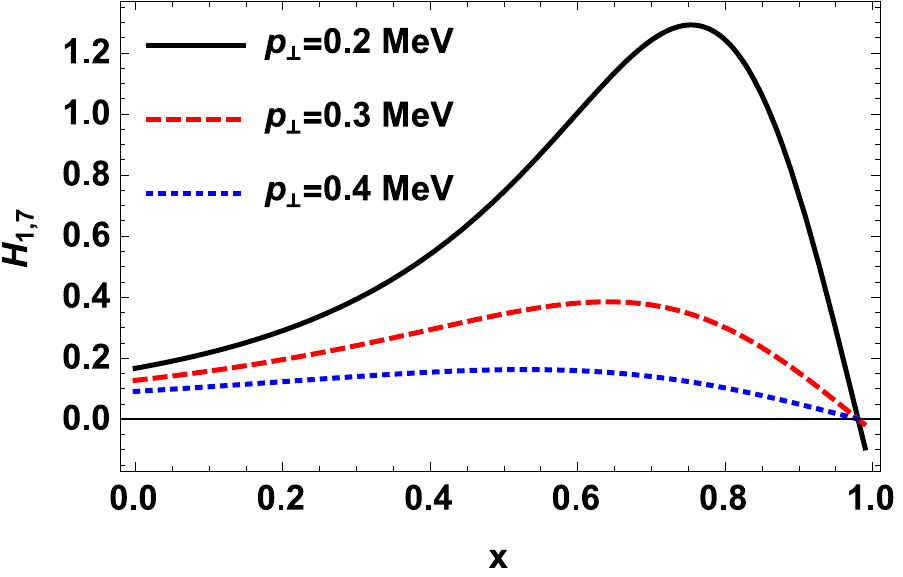}
\end{minipage}
\caption{(color online) Plots of GTMDs $H_{1,2}, H_{1,3}, H_{1,6}$ and $H_{1,7}$ with fixed value of ${\bf p}_\perp=0.3 MeV$ but with different values of ${\bf \Delta}_\perp$ (left panel) and fixed value of ${\bf \Delta}_\perp=0.3 MeV$ with different values of ${\bf p}_\perp$ (right panel).}
\label{H_GTMD}
\end{figure}

In Fig. \ref{F_GTMD}, we show the GTMDs $F_{1,1},F_{1,3}$ and $F_{1,4}$ with fixed value of $p_\perp=0.3 MeV$ but with different values of $\Delta_\perp$ (left panel) and fixed value of $\Delta_\perp=0.3 MeV$ with different values of $p_\perp$ (right panel) respectively. 
It can be noticed that $F_{1,1}$ and $F_{1,3}$ diverges at $x=1$. This is due to the fact that we consider only the quantum fluctuation
into the $|e\gamma\rangle$ Fock's sector which encodes the information on the structure of the physical electron, but if one considers the state of a physical electron as $|e_{phy}\rangle\rightarrow|e\rangle+|e\gamma\rangle$, the single particle$(|e\rangle)$ also contributes to the $F_{1,1}$ at exactly $x = 1$ which, with proper normalization, cancels the divergence. Since the GPD $H$ is related to the GTMD $F_{1,1}$, a similar behavior of $F_{1,1}$ has also been observed in $H$ for physical electron \cite{Chakrabarti:2014cwa}.
We find $F_{1,2}=0$ in this model. For fixed $p_\perp$ with increasing values of $\Delta_\perp$ the peaks in $F_{1,4}$ shift towards higher values of $x$ and the height of the peaks increase. But for fixed $\Delta_\perp$ with increasing values of $p_\perp$, peaks move towards lower values of $x$ however magnitude of GTMDs decreases.
In Fig. \ref{G_GTMD}, we present the results for $G_{1,1}, ~G_{1,2}$ and $G_{1,4}$ GTMDs with same parameters used in Fig. \ref{F_GTMD}. We observe that $G_{1,1}=-F_{1,4}$ and $G_{1,3}=0$. We show the GTMDs $H_{1,2},~H_{1,3}$,~$H_{1,6}$and $H_{1,7}$ in Fig. \ref{H_GTMD}. The GTMDs $H_{1,1},~H_{1,4},~H_{1,5}$ and $~H_{1,8}$ are zero in this model.
\begin{figure}[H]
\begin{center}
\includegraphics[width=7cm]{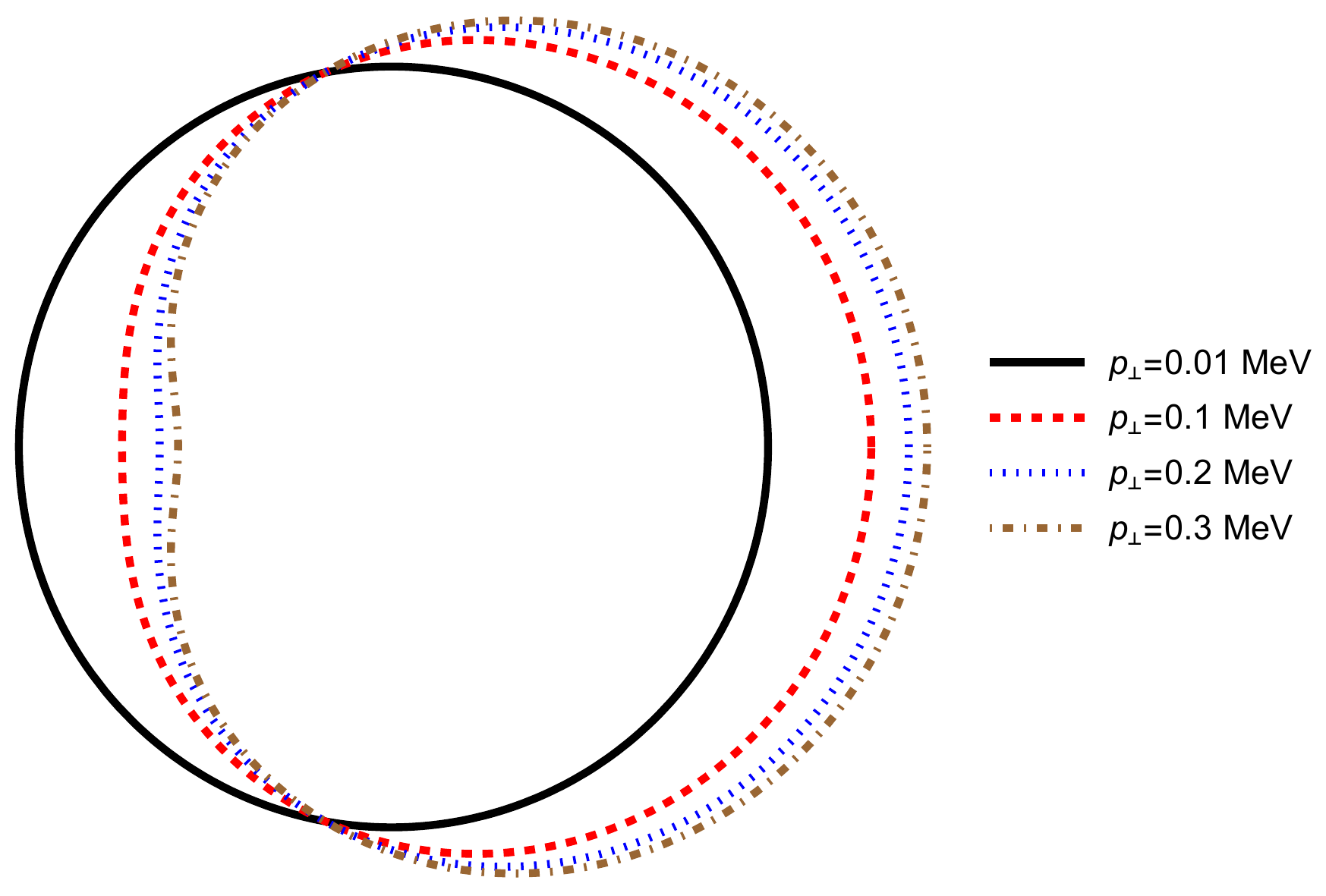}
\end{center}
\caption{Transverse shape of electron for different values of $\bfp$. The shapes are denoted with different lines.}
\label{electron_shape}
\end{figure}
 \section{Transverse shape of electron}\label{shape-electron}
The shapes of an electron can be obtained by using the following relation which was first introduced in \cite{Miller:2007ae},
\begin{eqnarray}
\frac{\hat{\rho}_{RELT}({\bfp},{\bfn})/M}{\tilde{f}_1({\bfp}^2)}&=&1+ \frac{\tilde{h_1}(\bfp^2)}{\tilde{f_1}(\bfp^2)} \cos \phi_n + \frac{\bfp^2}{2 M^2}\cos(2 \phi- \phi_n)
\frac{\tilde{h}_{1T}^\perp(\bfp^2)}{\tilde{f}_1(\bfp^2)}
\end{eqnarray}
where $\phi$ is the angle between  ${\bfp}$ and ${\bfS}$ and $\phi_n$ is the angle between ${\bfn}$ and $\bfS$. ${\bfn}$ is the unit vector which describes the arbitrary spin of the particle in a fixed direction. $\bfS$ is the physical electron polarization in the transverse direction. Further, $f_1$, $h_1$ and $h_{1 T}^{\perp}$ are the unpolarized electron distribution, transversity, and pretzelous distributions respectively and $\tilde{f} (\bfp^2)= \int dx f(x,\bfp^2)$. The definition of the TMDs ($f_1, h_1, h_{1 T}^{\perp}$) can be found in Ref.\cite{Bacchetta:2015qka} and their expressions in the present QED model are given by
\begin{eqnarray}
f_1&=& \frac{e^2}{16 \pi^3} \Bigg[\frac{2 \ \bfp^2 (1+x^2)}{x^2(1-x)^2}+2 \Bigg(M-\frac{m}{x}\Bigg)^2\Bigg] y, \\
h_1&=&\frac{e^2}{16 \pi^3} \frac{4 \bfp^2}{x(1-x)^2} y, \\
h_{1T}^{\perp}&=& 0,
\end{eqnarray}
where
\begin{eqnarray}
y&=&\frac{x^2(1-x)}{(\bfp^2-M^2 x (1-x)+m^2 (1-x)+\lambda^2 x)^2}.\nonumber
\end{eqnarray}
Since the pretzelous distribution $h_{1T}^{\perp}$ is zero in this model, 
thus the shape of an electron explicitly depends upon $f_1$ and $h_1$. In Fig. \ref{electron_shape} we present the shape of electron by considering different angle $\phi_n$. We show that how the shape of an electron emerges when we take the different angles between $\bfn$ and $\bfS$.
We consider the value of $\phi_n$ ranging from 0 to $2 \pi$ ($\phi_n=0$ or $2\pi$ reflects that $\bfn$ is parallel to $\bfS$ and $\phi_n= \pi$ reflects that $\bfn$ is anti-parallel to $\bfS$) and increase the value of $\bfp$. We observe that with increasing the value of $\bfp$, the shape gets distorted (moving from black color$\rightarrow$ brown color). Since we are excluding the $\bfp=0$ point, we present the results starting from $\bfp=0.01 MeV$ which comes out to be spherical and thereafter by increasing the value of $\bfp$, we observe that distortion increases in the shape of electron.
\section{Conclusions}\label{con}
We discussed all the leading twist Wigner distributions of a physical electron which provide the multi-dimensional images of electron. Here, we used the perturbative QED model where one can consider physical electron as an effective composite system of electron and photon. Using the overlap of LFWFs the Wigner distributions have been evaluated for the different polarization configurations e.g. unpolarized, longitudinally polarized and transversely polarized composite system and fermion constituent. The spin-spin correlations have been investigated for the system. We have also evaluated the GTMDs of electron. In this model we have found $10$ nonzero leading twist GTMDs. Results provide rich and interesting information on the distribution of QED partons in impact-parameter, momentum plane and mixed planes representation. We have also discussed the shape of electron.
\section{Acknowledgements}
Authors would like to thank Jai More (IIT Bombay) for technical help and insightful discussions. Authors are highly thankful to Dipankar Chakrabarti (IIT Kanpur) for critically reading the manuscript and giving valuable suggestions. N.K. is also thankful to Oleg Teryaev (JINR, Dubna) for helpful discussion.
N.K. acknowledge financial support received from Science and Engineering Research Board a statutory board under
Department of Science and Technology, Government of India (Grant No. PDF/2016/000722) under National Post-Doctoral Fellowship. This work of CM is supported by the funding from the China Post-Doctoral Science Foundation under the Grant No. 2017M623279.


\section*{References}
\begin{thebibliography}{60}
\bibitem{Ji:2003ak} X. Ji, Phys. Rev. Lett. {\bf 91} 062001 (2003). 
\bibitem{Diehl:2003ny} M. Diehl, Phys. Rept. {\bf 388}, 41 (2003).                  					\bibitem{Ji:1998pc} X. Ji, J. Phys. G {\bf 24} 1181 (1998).
\bibitem{Goeke:2001tz} K. Goeke, M. Polyakov, and M. Vanderhaeghen, Prog. Part. Nucl. Phys. {\bf 47} 401 (2001).
\bibitem{Mulders:1995dh} P. Mulders and R. Tangerman, Nucl. Phys. B {\bf 461}, 197 (1996). 
\bibitem{Boer:1997nt} D. Boer and P. J. Mulders, Phys. Rev. D {\bf 57}, 5780 (1998).
\bibitem{Burkardt:2000za}  M. Burkardt, Phys. Rev. D {\bf 62}, 071503 (2000); Phys. Rev. D {\bf 66}, 119903 (2002).
\bibitem{Burkardt:2002hr} M. Burkardt, Int. Jou. of Mod. Phys. A {\bf 18}, 173 (2003).
\bibitem{Burkardt:2005td} M. Burkardt, Int. Jou. of Mod. Phys. A {\bf  21}, 926 (2006).
\bibitem{Diehl:2005jf} M. Diehl and Ph. Haegler, Eur. Phys. J. C {\bf 44}, 87 (2005).
\bibitem{Pasquini:2007xz} B. Pasquini and S. Boffi, Phys. Lett. B {\bf 653}, 23 (2007).
\bibitem{Goeke:2005hb} K. Goeke, A. Metz, and M. Schlegel, Phys. Lett. B {\bf 618}, 90 (2005).
\bibitem{Meissner:2007rx} S. Meissner, A. Metz, and K. Goeke, Phys. Rev. D {\bf 76}, 034002 (2007).
\bibitem{Lorce:2011zta} C. Lorc$\acute{e}$ and B. Pasquini, Phys. Rev. D {\bf 84}, 034039 (2011).
\bibitem{Balazs:1983hk} N. Balazs and B. Jennings, Phys. Rep. {\bf 104}, 347 (1984).
\bibitem{Hillery:1983ms} M. Hillery, R. O' Connell, M. Scully, and E. Wigner, Phys. Rep. {\bf 106}, 121 (1984).
\bibitem{LEE1995147} H.-W. Lee, Phys. Rep. {\bf 259}, 147 (1995).
\bibitem{Banaszek:1999ya} K. Banaszek, C. Radzewicz, K. W$\acute{o}$dkiewicz, and J. S. Krasi$\acute{n}$ski, Phys. Rev. A {\bf 60}, 674 (1999).
\bibitem{Smithey:1993zz} D. T. Smithey, M. Beck, M. G. Raymer, and A. Faridani, Phys. Rev. Lett. {\bf 70}, 1244 (1993).
\bibitem{Vogel:1989zz} K. Vogel and H. Risken, Phys. Rev. A {\bf 40}, 2847 (1989).
\bibitem{Lorce:2011kd}  C. Lorc$\acute{e}$ and B. Pasquini, Phys. Rev. D {\bf 84}, 014015 (2011).
\bibitem{Mukherjee:2014nya} A. Mukherjee, S. Nair, and V. K. Ojha, Phys. Rev. D {\bf 90}, 014024 (2014).
\bibitem{Liu:2015eqa} T. Liu and B.-Q. Ma, Phys. Rev. D {\bf 91}, 034019 (2015).
\bibitem{Chakrabarti:2016yuw} D. Chakrabarti, T. Maji, C. Mondal, and A. Mukherjee, Eur. Phys. Jou. C {\bf 76}, 409 (2016).
\bibitem{Chakrabarti:2017teq} D. Chakrabarti, T. Maji, C. Mondal, and A. Mukherjee, Phys. Rev. D {\bf 95} 074028 (2017).
\bibitem{Meissner:2009ww} S. Meissner, A. Metz, and M. Schlegel, J. High Energy Phys. {\bf 08}, 056 (2009).
\bibitem{Lorce:2011dv}  C. Lorc$\acute{e}$, B. Pasquini, and M. Vanderhaeghen, J. High Energy Phys. {\bf 2011}, 41 (2011).
\bibitem{Echevarria:2016mrc} M. G. Echevarria {\it et. al.,} Phys. Lett. B {\bf 759}, 336 (2016).
\bibitem{Kanazawa:2015zpa} K. Kanazawa {\it et. al.,} Int. Jou. of Mod. Phys.: Conference Series {\bf 37}, 1560037 (2015).
\bibitem{Miller:2014vla} G. A. Miller, Phys. Rev. D {\bf 90} 113001 (2014).
\bibitem{Hoyer:2009sg}  P. Hoyer and S. Kurki, Phys. Rev. D {\bf 81} 013002 (2010).
\bibitem{Hatta:2016aoc} Y. Hatta, Y. Nakagawa, F. Yuan and Y. Zhao, Phys. Rev. D 95 , 114032 (2017).
\bibitem{Hatta:2016dxp} Y. Hatta, B. W. Xiao and F. Yuan, Phys. Rev. Lett. 116 , no. 20, 202301 (2016).
\bibitem{Ji:2016jgn}  X.-D. Ji, F. Yuan and Y. Zhao, Phys. Rev. Lett. 118, 192004 (2017). 
\bibitem{Zhou:2016rnt} J. Zhou, Phys. Rev. D 94 , 114017 (2016).
\bibitem{Hagiwara:2016kam} Y. Hagiwara, Y. Hatta and T. Ueda, Phys. Rev. D 94 , 094036 (2016).
\bibitem{Bhattacharya:2017bvs} S. Bhattacharya, A. Metz and J. Zhou, Phys. Lett. B 771, 396 (2017).
\bibitem{Accardi:2012qut} A. Accardi {\it et. al.,} Eur. Phys. Jou. A 52 268 (2016).
\bibitem{Dudek:2012vr} J. Dudek {\it et al.}, Eur. Phys. A 48, 187 (2012).
\bibitem{Mirhosseini} M. Mirhosenni {\it et al.}, Phys. Rev. Lett. 116 130402 (2016).
\bibitem{Brodsky:2000ii} S. J. Brodsky, D. S. Hwang, B.-Q. Ma, and I. Schmidt, Nucl. Phys. B {\bf 593}, 311 (2001).
\bibitem{Brodsky:2006ku}  S. J. Brodsky {\it et. al.,} Phys. Rev. D {\bf 75}, 014003 (2007).
\bibitem{Kumar:2015fta}  N. Kumar and H. Dahiya, Int. Jou. of Mod. Phys. A {\bf 30}, 1550010 (2015).
\bibitem{Kumar:2015tpa} N. Kumar and H. Dahiya, Eur. Phys. J. A {\bf 51}, 19 (2015).
\bibitem{Kumar:2014coa} N. Kumar and H. Dahiya, Phys. Rev. D {\bf 90}, 094030 (2014).
\bibitem{Bacchetta:2015qka} A. Bacchetta, L. Mantovani, and B. Pasquini, Phys. Rev. D {\bf 93}, 013005 (2016).
\bibitem{baron} J. Baron {\it et al.} [ACME Collaboration], Science 343, no. 6168, 269 (2014).
\bibitem{hudson} J. J. Hudson {\it et al.}, Nature 473 , 493 (2014)
\bibitem{wesley} Wesley C .Campbell {\it et. al.}, EPJ Web of Conferences {\bf 57,} 02004 (2013)
\bibitem{Meissner:2008ay} S. Meissner, K. Goeke, A. Metz, and M. Schlegel, J. High Energy Phys. {\bf 2008}, 038 (2008).
\bibitem{Lorce:2011ni}  C. Lorc$\acute{e}$, B. Pasquini, X. Xiong, and F. Yuan, Phys. Rev. D {\bf 85}, 114006 (2012).
\bibitem{levin} D. Levin, Math. Comput. {\bf 38}, 531 (1982).
\bibitem{levin1} D. Levin, J. Comput. Appl. Math. {\bf 67}, 95 (1996).
\bibitem{levin2}  D. Levin, J. Comput. Appl. Math. {\bf 78}, 131 (1997).
\bibitem{More:2017zqq} J. More, A. Mukherjee and S. Nair, Phys. Rev D {\bf 95} 074039 (2017).
\bibitem{Kumar:2015yta} N. Kumar and H. Dahiya, Phys. Rev. D {\bf 91}, 114031 (2015).
\bibitem{Chakrabarti:2014cwa} D. Chakrabarti {\it et al.}, Phys. Rev. D {\bf 89}, 116004 (2014).
\bibitem{Miller:2007ae} G. A. Miller, Phys. Rev. C {\bf 76}, 065209 (2007).

   \end{thebibliography}
\end{document}